\documentclass[journal,comsoc]{IEEEtran}

\usepackage{xurl}
\usepackage{caption}
\DeclareCaptionLabelSeparator{period}{. }
\usepackage{booktabs} 
\usepackage[normalem]{ulem}
\usepackage[table]{xcolor}
\usepackage{xcolor}
\usepackage{lipsum}
\usepackage{color,soul}
\usepackage{xspace}
\usepackage[utf8x]{inputenc} 
\usepackage[LGR,T1]{fontenc}
\usepackage{textalpha}
\usepackage{algorithm}
\usepackage{algpseudocode}
\usepackage{url}
\usepackage{amssymb,amsfonts,amsmath}
\usepackage{pifont}
\usepackage{tikz}
\usepackage{cite} 
\usetikzlibrary{arrows,shapes}
\usepackage[gen]{eurosym}

\usepackage{multicol}
\usepackage{lscape}
\usepackage{afterpage}
\usepackage{graphicx}
\usepackage{varwidth}
\usepackage[tight]{subfigure}
\usepackage{microtype}

\usepackage{tabularray}

\usepackage{rotating}
\usepackage{xcolor}
\usepackage{multirow}
\usepackage{rotating}
\usepackage{makecell}

\usepackage{array,multirow,graphicx}
\usepackage{float}
\usepackage{tabularx}
\usepackage{lscape}
\usepackage{longtable}
\usepackage{supertabular,booktabs}

\usepackage{amssymb}
\usepackage{pifont}
\newcommand{\cmark}{\ding{51}}%
\newcommand{\xmark}{\ding{55}}%
\newcommand{\hardwareC}{\ding{108}}%
\newcommand{\softwareC}{\ding{115}}%

\usepackage{fontawesome}
\usepackage{marvosym}

\algrenewcommand\algorithmicrequire{\textbf{Input:}}
\algrenewcommand\algorithmicensure{\textbf{Output:}}

\usepackage[T1]{fontenc}

\usepackage{array,multirow,makecell}

\usepackage[linguistics]{forest}

\usepackage{stfloats}

\usepackage{enumitem}
\setitemize{itemsep=0pt,topsep=2pt,parsep=0pt,partopsep=0pt}

\usepackage{amsmath}

\graphicspath{{figure/}{figures/}}

\usepackage{etoolbox}
\usepackage{xstring}
\DeclareListParser{\doslashlist}{/}
\newcounter{ndnNameComponentCounter}%
\newcommand{\name}[1]{{%
  \setcounter{ndnNameComponentCounter}{0}%
  \renewcommand{\do}[1]{{%
    \ifnumgreater{\value{ndnNameComponentCounter}}{0}{\allowbreak/}{}%
    \ifnumodd{\value{ndnNameComponentCounter}}{}{}%
    ##1}%
    \stepcounter{ndnNameComponentCounter}}%
``{\fontfamily{cmtt}\small\selectfont\IfBeginWith{#1}{/}{/}{}\doslashlist{#1}}''%
}}

\usepackage{caption} 
\usepackage{enumitem}
\usepackage{tikz}
\setitemize{itemsep=0pt,topsep=2pt,parsep=0pt,partopsep=0pt}

\usepackage{etoolbox}
\usepackage{xstring}
\usepackage{comment}

\usepackage{diagbox}
\usepackage{xr}
\usepackage{hyperref}
\floatstyle{plaintop}
\restylefloat{table}
\usepackage{wrapfig}

\forestset{font=\small}

\providecommand{\keywords}[1]
{
  \small	
  \textbf{\textit{Keywords---}} #1
}

\usepackage{enumitem}
\setlist{parsep=0pt,listparindent=\parindent}
\usepackage{placeins}
\usepackage{lscape} 

\usetikzlibrary{positioning}

\begin{document}

\title{A Comprehensive Survey of Unmanned Aerial Systems’ Risks and Mitigation Strategies} %
 \author{Sharad Shrestha~\IEEEmembership{Student Member,~IEEE},
         Mohammed Ababneh,
         Satyajayant Misra~\IEEEmembership{Member,~IEEE},
         Henry M. Cathey,~Jr., 
         Roopa Vishwanathan,
         Matt Jansen,
         Jinhong Choi,         
         Rakesh Bobba~\IEEEmembership{Member,~IEEE},
         Yeongjin Jang
 \thanks{S. Shrestha, M Ababneh, S. Misra, R. Vishwanathan are with the Department of Computer Science at New Mexico State University (NMSU); S. Misra is also with the Department of Electrical and Computer Engineering at NMSU. H.M. Cathey is with the Physical Science Laboratory at NMSU. M. Jansen, J. Choi, R. Bobba are with the Department of Computer Science at Oregon State University, Y.Jang is with Samsung Research.}
}

\IEEEpeerreviewmaketitle

\maketitle

\begin{abstract}
In the last decade, the rapid growth of Unmanned Aircraft Systems (UAS) and Unmanned Aircraft Vehicles (UAV) in communication, defense, and transportation has increased. The application of UAS will continue to increase rapidly. 
This has led researchers to examine security vulnerabilities in various facets of UAS infrastructure and UAVs, which form a part of the UAS system to reinforce these critical systems. 
This survey summarizes the cybersecurity vulnerabilities in several phases of UAV deployment, the likelihood of each vulnerability’s occurrence, the impact of attacks, and mitigation strategies that could be applied.  
We go beyond the state-of-the-art by taking a comprehensive approach to enhancing UAS security by performing an analysis of both UAS-specific and non-UAS-specific mitigation strategies that are applicable within the UAS domain to define the lessons learned. 
We also present relevant cybersecurity standards and their recommendations in the UAS context. 
Despite the significant literature in UAS security and the relevance of cyberphysical and networked systems security approaches from the past, which we identify in the survey, we find several critical research gaps that require further investigation. These form part of our discussions and recommendations for the future exploration by our research community. 
\end{abstract}


\section{\textbf{Introduction}\label{Introduction}}
Unmanned Aerial Vehicles (UAVs) are complex aircraft systems operating without a human on board; they can be programmed to operate automatically, semi-automatically, or manually. 
UAVs, also known as drones, have both civilian and military applications and are used in various domains, such as agriculture, e-commerce, surveillance, search and rescue, aerial data collection, aerial photography/videography, and emergency services~\cite{shakhatreh2019unmanned}. 
UAVs come in various sizes that range from 200~gms to 600~Kgs in some military applications~\cite{anatomy}.
UAVs are automatic if they can execute a pre-planned flight operation without human intervention. 

However, despite their advanced capabilities, UAVs' reliance on wireless communication channels makes them vulnerable to various cyber threats, some of which have resulted in significant consequences. For example, in 2012, the S-100 spy drone crashed during a test flight in South Korea, killing one and injuring two when the UAV collided with a ground station (control vehicle). At the same time, the South Korean military is reported to be investigating whether the interference of GPS signals by North Korea could have caused the crash~\cite{att}. The RQ-170 incident was a well-known attack on a Lockheed Martin RQ-170 Sentinel UAV that resulted in its capture~\cite{rq170}. The attacker forced the UAV to land by jamming the GPS satellites and spoofing GPS information. In another case, In 2011, a computer virus infiltrated U.S. Predator and Reaper drones at Creech Air Force Base, recording pilots’ keystrokes during missions. The virus spread through removable drives, bypassing the security measures separating classified and public networks. Despite repeated efforts to remove it, the virus remained, exposing serious vulnerabilities in military systems~\cite{shachtman2011computer}. These real-world cases show how important it is to have strong cybersecurity measures in place to protect UAVs from advanced threats and malicious attacks.

This point forward in the document, the term `UAV' is used to refer to unmanned aircraft, whereas `UAS' (Unmanned Aerial Systems) denotes the combination of the UAV and the requisite infrastructure for its operation together forming a system.
While the aforementioned incidents highlight the severity of cybersecurity challenges in UAV operations, they also underscore how adversaries increasingly exploit UAVs for cyberattacks, as categorized by the Cybersecurity and Infrastructure Security Agency (CISA) into hostilities, smuggling or contraband delivery, disruption of government activities, and weaponization. Despite extensive research, existing UAV cybersecurity surveys often fall short of addressing vulnerabilities comprehensively across all phases of UAV operations. Prior efforts have typically focused on isolated attack vectors, individual components, or generalized overviews, leaving critical gaps in the literature. To address these gaps, this paper investigates several key research challenges. Table~\ref{tab:UAVSurveyComparison} compares our work with state-of-the-art~\cite{leccadito2018survey, krishna2017review, gupta2015survey, choudhary2018intrusion, choudhary2018internet, manesh2019cyber, nassi2021sok, javaid2012cyber, lin2018security, wang2019survey, altawy2016security}, outlining the taxonomy of work in the area and highlighting unique features. 
We present a comprehensive ranking system that evaluates both the probability and potential impact of every attack. Our ranking system enables an understanding of the risks posed by various attacks, taking into account their likelihood of occurrence as well as the severity of their potential consequences.
As presented in this paper, some attacks on UAVs have a higher impact than others. To the best of our knowledge, this is the only paper that defines and identifies all major attacks on UAVs under each mode of operation.

\begin{table*}[t]
\footnotesize
\vspace{-5pt}
  \centering
 \begin{tabular}{||c | c | c | c | c | c | c | c ||} 
 \hline
 Survey & Attack & \shortstack{Mitigation \\ Strategy}  & \shortstack{Severity \\ Asessment} & \shortstack{UAV or UAS} & \shortstack{Hardware (\hardwareC) or \\ Software (\softwareC)} & \shortstack{Goes Beyond \\ Network Attacks} & Operational Phases \\ [0.5ex] 
 \hline\hline
 \cite{leccadito2018survey} & \cmark  & \cmark  & \xmark & UAS & \hardwareC \softwareC & \cmark & \xmark  \\ 
 \cite{krishna2017review}  & \cmark  & \xmark & \xmark & UAV & \xmark & \xmark & \xmark \\
 \cite{gupta2015survey} & \cmark  & \xmark & \xmark & UAV & \xmark & \xmark & \xmark \\
 \cite{choudhary2018intrusion} & \xmark & \xmark & \xmark & UAV & \hardwareC & \xmark & \xmark\\
 \cite{choudhary2018internet} & \cmark  & \xmark & \xmark & UAV & \softwareC & \xmark & \xmark \\ 
 \cite{manesh2019cyber} & \cmark  & \cmark  & \cmark & UAS & \hardwareC & \xmark & \xmark \\ 
 \cite{nassi2021sok} & \cmark  & \cmark  & \cmark & UAS & \hardwareC \softwareC & \cmark & \xmark \\
 \cite{javaid2012cyber} & \xmark & \xmark & \cmark & UAS & \softwareC & \cmark & \xmark\\
 \cite{lin2018security} & \xmark & \cmark  & \xmark & UAV & \xmark & \xmark & \xmark\\ 
 \cite{wang2019survey} & \xmark & \xmark & \xmark & UAS & \hardwareC \softwareC & \cmark & \xmark \\
 \cite{altawy2016security} & \cmark & \cmark & \xmark & UAS & \hardwareC \softwareC & \cmark & \xmark \\
 This Paper & \cmark & \cmark & \cmark & UAS & \hardwareC \softwareC & \cmark & \cmark\\[1ex] 
 \hline
 \end{tabular}
  \caption{Comparison among existing UAV security surveys and SoK.}
  \label{tab:UAVSurveyComparison}
    \vspace{-15pt}
\end{table*}

\begin{figure*}[b]
    \centering
    \includegraphics[height=8cm, keepaspectratio]{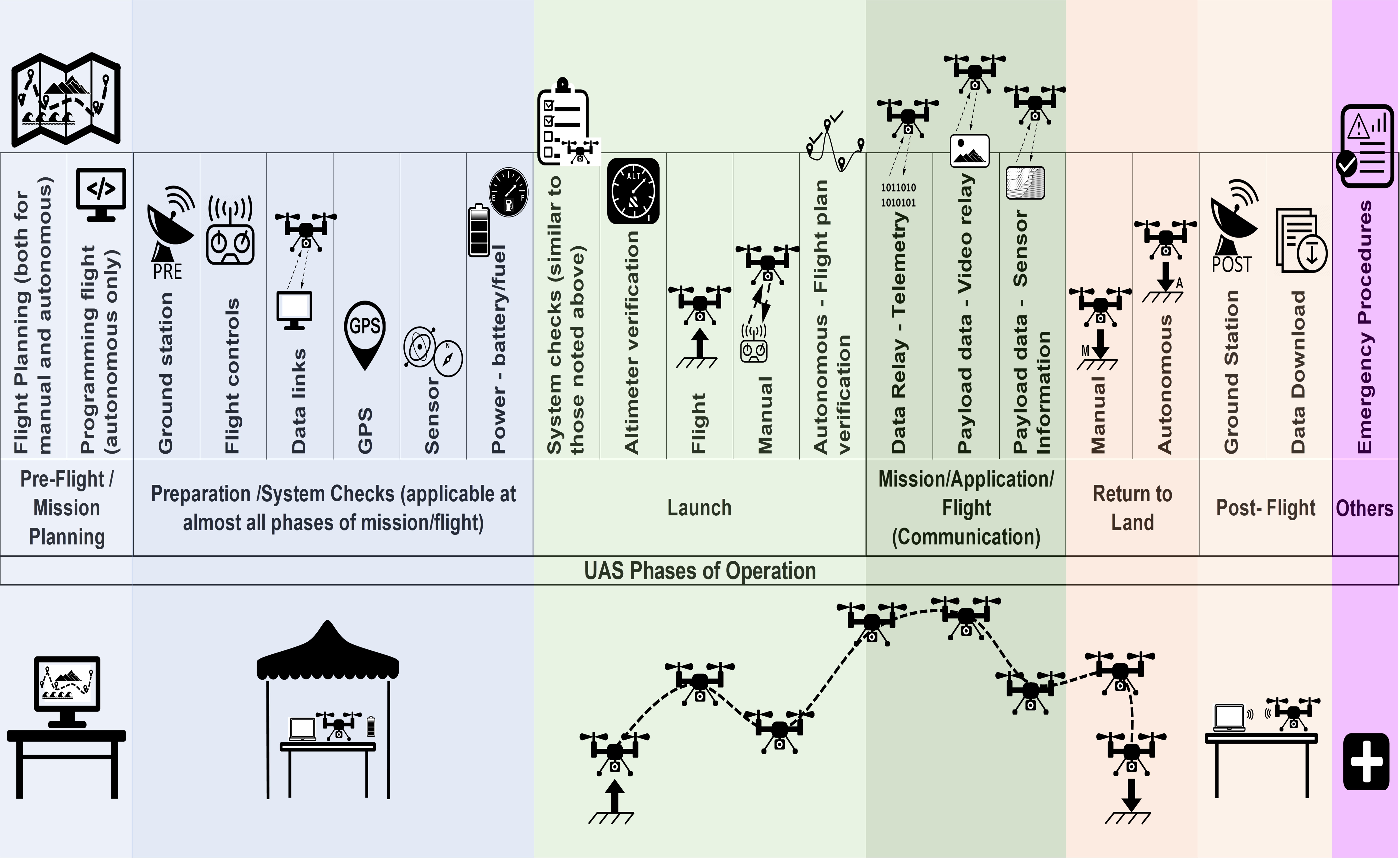}
    \caption{UAV operation can be divided into seven stages, each involving hardware, software, network, GCS, and cloud components.}
    \label{fig:phaseOfOperation}
\end{figure*}


For this manuscript, we took a systematic approach to gather the most relevant papers. We started by compiling a comprehensive list of keywords, combining UAS terms, such as `Unmanned Aircraft System', `Unmanned Aerial Vehicle', `Remotely Piloted Vehicle', `drone' and `urban air mobility' with cybersecurity terms, such as `cyber attacks', `Cyber-physical security', and `cybersecurity'. A combination of these keywords resulted in 56 distinct queries that we ran across major technical databases published in the 10~years, including Institute of Electrical and Electronics Engineers~(IEEE) Xplore (using a Python script to leverage the IEEE API), Association for Computing Machinery~(ACM), and American Institute of Aeronautics and Astronautics~(AIAA). 
We collected 6,995 articles from IEEE, 10,278 articles from the ACM digital library, and 8,117 articles from AIAA after deduplication, resulting in a total of 25,390 articles. 
For each of the articles, we assigned scores based on the number of matching keyword pairs, which helped us prioritize the most pertinent articles.
Since over 25,000 articles matched the search queries, we needed an efficient process to identify the most relevant ones. We then applied a multi-stage iterative review process: i) reviewing article abstracts to categorize relevant articles and ii) conducting a detailed technical review of those selected. 
During the abstract review, we categorized 6,833 articles with a score of 4 or higher. 
An initial set of 1,294 articles was selected for technical screening, during which each article was briefly reviewed and summarized. Following this process, 547 articles were identified for in-depth analysis for the survey.

This paper aims to thoroughly examine the cybersecurity threats associated with each phase of UAV operation, guided by three key research questions: \textbf{(RQ1)} What cybersecurity threats arise during each phase of UAV operation, and how do these threats vary in severity and associated risks? \textbf{(RQ2)} What common attack methods do adversaries typically use, and how do these methods exploit UAV vulnerabilities during different operational phases? \textbf{(RQ3)} What mitigation strategies can be effectively implemented to address the identified threats across various UAV operational stages? By exploring these questions, we intend to identify the most vulnerable phases, recognize prevalent attack vectors and their solutions. 

\textbf{Outline}: 
In Section 2, we provide rankings for the likelihood and severity of each attack, assessing the identified vulnerabilities.
Section 3 focuses on mitigation strategies. Section 4 presents the key lessons learned, and finally, Section 5 concludes our study.

\section{ \textbf{UAV Phases of Operation}\label{UAVPhasesofOperation}}

Every UAV mission or flight comprises several distinct phases. These phases are defined to organize and manage the various aspects of UAV operation, from preparation to completion. Figure~\ref{fig:phaseOfOperation} shows seven phases of UAV operations, featuring several activities and system components in hardware, software, network link, Ground Control Station (GCS), and cloud. 

\noindent \textbf{Pre-Flight/Mission Planning:}
The pre-flight phase of a UAV consists of mission planning for planning the flight path and goals. Flight and navigation plans are checked in both manual and autonomous flight systems. Clearances required for UAVs to complete the mission are achieved.  
Pilots (operators) are trained (prepared) so that they are familiar with flight path and other relevant information.
Parameters affecting flight paths in the pre-flight planning phase may include speed, size, and start time. 

Global path-finding algorithms in 3-dimensional spaces are used to find optimal and collision-free paths from launch to landing of UAVs. These include adaptive planning algorithms that calculate an optimized trajectory considering boundaries~\cite{meister2008adaptive}, or weighted shortest path~\cite{tran2021global}. Since this is done pre-flight, processing power is not limited by the battery constraint of UAVs. However, to account for unexpected obstacles or changes in environmental conditions, real-time path planning would also require that the algorithms be efficient to calculate and transmit the trajectory within a reasonable time to the UAV.

\noindent \textbf{Preparation/System Checks:}
During the preparation phase, a ground station completes the flight report, and the flight controls are configured. This enables either a pilot or a computer to control the UAV. Data link communication between the GCS and UAV is established. Various onboard sensors, such as altimeters, GPS, barometers, and compass are checked and verified to operate error-free.

Before the final launch of the UAV for the mission, the preparation of UAVs is done to check that all systems are in order. This includes individual checks for each system’s health, proper mission and software loading, and installation of necessary hardware, such as appropriate batteries, and cameras. 
Other components of the UAS should also be prepared such as getting the pilot/crew ready for manual tasks, setting up the communication links and servers, and monitoring the UAV during its mission. 
Critical processes in this phase include mechanical inspection of the airframes, 
verification of system parameters and settings, functional checks of control surfaces and communication links, and calibration of sensors,
such as barometers and GPS.

\noindent \textbf{Launch:}
The launch phase starts when the UAV takes off and ends when it reaches cruise altitude. This phase includes periodic system checks to ensure that all components are running as expected. 
The altimeter has to be constantly monitored to verify if UAV has reached a cruising altitude. 
An autonomous flight plan verification mechanism is used to verify that the UAV can accomplish its flight path during the launch phase, ensuring that the UAV's sensors and systems are functioning correctly and the UAV can safely transition to its autonomous flight plan under potentially changing environmental conditions.

\noindent \textbf{Mission/Application/Flight:}
As soon as a UAV reaches cruising altitude, it stays at that altitude for most of the flight unless there is a change in flight path due to weather changes or physical obstacles. A constant communication link between the UAV and the base station lets the pilot know that the UAV is on the correct predetermined path. Payloads should be securely attached to the UAV until they are released. All the sensors are also periodically monitored during the flight to ensure they function as intended. In certain applications, a video relay between the UAV and the ground station helps the pilot know the UAV's orientation in First person view (FPV) and control it when it is not in the pilot's line of sight (LOS). 

\noindent \textbf{Return to Land:} 
Most UAVs are built to be reusable, similar to human-crewed aircraft. After a UAV completes its mission, it either returns to the ground station or lands in a safe, designated area. Depending on the UAV type and its capabilities, this descent process could be either manual or autonomous. In an autonomous system, the UAV follows the pre-planned path to land.

\noindent \textbf{Post Flight:} 
After the UAV lands, operators perform post-flight system checks, including visual inspection and software and sensor checks. This ensures each component is in good condition and without degradation. Depending on the application, mission-related data can also be downloaded as part of this phase.

\noindent \textbf{Emergency Situation:}
During any operation phase, if the UAV goes into a state outside of the original flight plan for the mission or its critical components malfunction, the UAV goes into an emergency state that consists of a series of predefined procedures and checklists to prevent an accident. Emergency procedures aim to recover the UAV to continue the mission or attempt to land the UAV in the safest manner possible while considering other physical structures, UAVs, and living beings in the vicinity of the UAV. Depending on the mission, the procedures may also include deleting sensitive flight data.

\begin{figure}[t]
    \centering
    \includegraphics[width=2.0in]{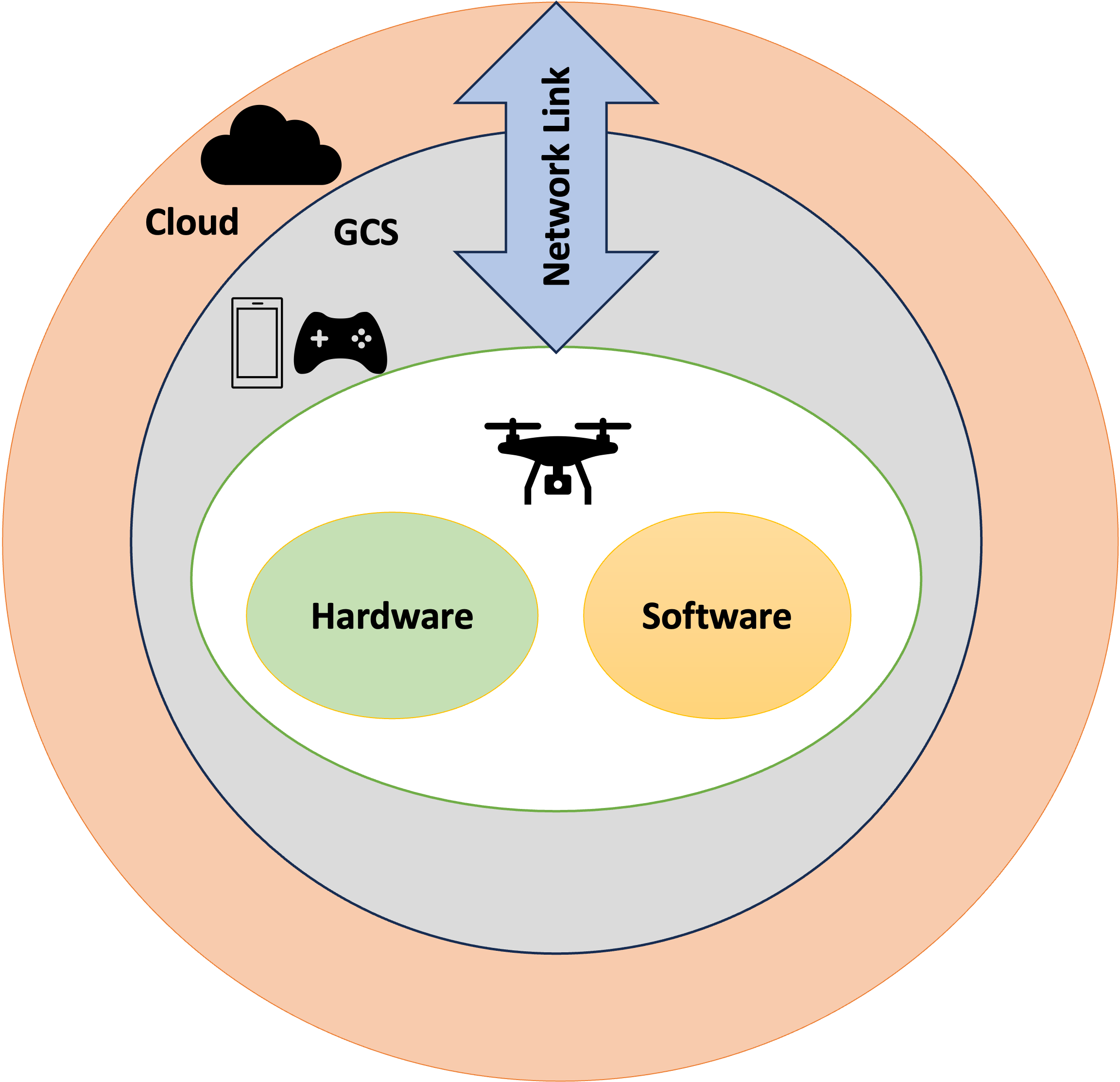}
    \caption{UAV components can be categorized into: UAV hardware, UAV software, Network link, GCS, and Cloud.}
    \label{fig:componentsOfUAV}
\end{figure}

\section{ \textbf{Vulnerabilities in UAS}\label{sec:VulnerabilitiesinUAS}} 

Like most embedded and cyberphysical systems, UAVs are prone to (physical and cyberphysical) attacks. Understanding the vulnerabilities is the first step to addressing them.
Figure \ref{fig:componentsOfUAV} describes the components of a UAV system. Every component of the UAV system is prone to attack. Failure in one component adversely affects other components and might cause an overall system failure. This section describes possible cyberphysical vulnerabilities in UAS in terms of the components: hardware, software, GCS, network link, and server/cloud.

\subsection{Hardware Vulnerabilities}
Hardware includes physical devices in UAVs, such as sensors (e.g., gyroscopes and accelerometers), microprocessors, circuit boards, propellers, and the UAV in general. 
In addition, other physical layer devices, such as the power system, control system, communication module, fans, and rotors~\cite{li2020survey} are also targets of attacks. Attacks on the physical subsystem or the complete UAV system may result in human, economic, and environmental harm. 
Sensors that rely on external agents for sensing are highly vulnerable to attacks. 
For example, adversarial AI can generate subtle perturbations that deceive deep learning-based sensors and navigation systems on UAVs~\cite{tian2021adversarial}. This can cause dangerous spoofing attacks or result in the UAV attempting to avoid collisions with non-existent obstacles, highlighting the critical importance of designing UAV hardware and software with robustness and security in mind. Other onboard sensors, such as GPS, RADAR, LIDAR, and vision-based sensors, are vulnerable to attackers. In what follows, we enumerate the different hardware vulnerabilities.

\noindent{\textbf{Spoofing:}} Adversaries may spoof the identities of certain devices in order to trick an entity into believing false data is coming from a trusted UAV. Adversaries can either generate a copy of actual signals from other sources or generate false signals, tricking the UAV into either failing its mission, losing control leading to damage to the UAV, or hijacking it entirely. The adversary can deceive the receiver into false actions based upon false information about its environment, such as location, time, or other conditional propositions regarding its mission. 
A jamming signal can break the pairing between the controller and UAV. This could be followed up by connecting the UAV to the attacker's controller to take control of the UAV~\cite{trujano2016security}. 
GPS signals used in civilian UAVs are unencrypted. With GPS spoofing, adversaries send false higher-power GPS signals (higher power than true GPS signal). Once it receives these higher-powered false GPS signals, the UAV uses it as if it were a true GPS signal. GPS spoofing is hence easy to perform as it only requires sending a strong GPS signal to a UAV with a slightly higher power than the actual GPS signal it receives~\cite{kerns2014unmanned}.
   
\noindent{\textbf{Jamming:}} The attacker in a jamming attack disrupts the communication channel between components in the UAV system. 
We differentiate this from network level jamming, which typically involves/affect multiple nodes, which is discussed in the Section~\ref{NetworkLinkVulnerabilities}.
Here, adversaries, also known as jammers, generate high amplitude radio frequency (RF) signals corresponding to the frequency used by UAVs~\cite{park2021survey}. This is typically done by blasting electromagnetic noise at the radio frequencies used by the UAS to send and receive information. Adversaries generally target communication channels or GPS signals. Attackers can also flood the communication network with garbage signals that consume bandwidth and computing resources. Jammers can be categorized as constant, deceptive, random, reactive, and brute force~\cite{grover2014jamming}.
  
\noindent{\textbf{Firmware Flashing:}} Firmware is the code operating the embedded devices of UAV hardware. Firmware flashing involves completely removing and rewriting the existing firmware with a new one. Attack vectors for this attack may be from the same local network or over the internet. Firmware flashing has been demonstrated to affect sensors of commercial IoT devices~\cite{gao2019microcontroller}. If an adversary succeeds in rewriting the firmware, they can take complete control of the UAV.

\noindent{\textbf{Supply Chain Attack:}} Supply chain attack involves infiltrating the production and distribution of UAV/UAS components and systems. This can range from tampering with hardware or software during production and distribution and modifying the software during updates. The attacks manifest as harmful defects in the components of UAVs~\cite{belikovetsky2017dr0wned}. These defects are not limited to the physical aspects of these complex systems but extend to the software responsible for their operation. 
These threats not only compromise the functionality of the UAVs but also have broader implications for systems and environments where these UAVs are deployed. Without adequate supply chain attack protections, these compromised components may be difficult to detect once the UAV/UAS is on a mission.

\subsection{Software Vulnerabilities}
In addition to the operating system, the UAV software system includes exploitable firmware on microcontrollers, such as sensors, motors, and communications hardware. Although these software vulnerabilities may share some similarities, we sub-categorize them based on their distinct types or variations. 

\noindent{\textbf{Code Injection:}}
Most UAVs are either automatic or semi-automatic, and their components are programmable. An adversary adding additional or modifying instructions could change the behavior of UAVs~\cite{habibi2015mavr}. For example, code injection can alter the correct sensor reading or alter the control algorithms to compromise its navigation components. Code injection can also be responsible for memory leaks and high CPU use, which may cause the software to crash or rapid battery drainage. 

\noindent{\textbf{Database Injection:}}
Database injection attacks, such as SQL injection, NoSQL injection, Xpath injection, and lightweight directory access protocol (LDAP) injection, are security vulnerabilities for UAV systems that use a database system~\cite{suhina2007exploiting}. These attacks compromise the confidentiality, integrity, and availability of the database. UASs often use databases for storing flight paths, no-fly zones, telemetry data, and other sensitive information. Adversaries can exploit vulnerabilities in database systems to alter, corrupt, or delete database contents/records, steal data, and manipulate data, resulting in serious outcomes, such as altered flight paths and turned off no-fly zones. 
Unauthorized access to the backend database to execute remote commands or take control of the server can occur if adversaries bypass login protocols~\cite{koubaa2019dronemap}.

\noindent{\textbf{Firmware Modification:}}
UAV firmware includes specialized software to control the UAV and is embedded in its hardware. UAVs, whether autonomous or not, will need this software to operate. 
An adversary can acquire a sample of the UAV's official firmware update. This firmware can be analyzed, disassembled, modified, reassembled, and reinstalled. The update, if installed, will enable the adversary to take complete control of the UAV~\cite{leccadito2018survey, blasch2019cyber}.

\noindent{\textbf{Unauthorized Access:}}
Unauthorized software access involves gaining entry to the software system of a UAV without the permission of the user. This involves bypassing the policies and mechanisms designed to manage UAV resource management. The confidential and time-sensitive data collected by UAVs require immediate transmission to GCS. Any unauthorized access can lead to mission failure~\cite{dahiya2020unmanned}. Additionally, software vulnerabilities can offer backdoors for attackers to exploit. The nature of signal transmission for UAVs, which often takes place over insecure atmospheric mediums, further increases the risk of such attacks. 
Robust software access control mechanisms require stringent policies and mechanisms that verify and permit only authorized entities to interact with or command the UAS.

\noindent{\textbf{Buffer Overflow:}}
UAVs are embedded systems with multiple hardware components operating at various speeds. These hardware components have different priorities. To enable these devices to work together, shared data buffers are used. 
When transmitting data across devices, buffers mitigate the timing differences between various events. However, buffers are susceptible to overflows that occur if a program tries to put more data than the size of the buffer.
An adversary can misguide the program into wrong assumptions about the data size. This can lead the entire UAV application to crash or to execute arbitrary code outside the scope of the running program by engineering a buffer overflow ~\cite{hooper2016securing}. A buffer overflow attack can be implemented in two ways: heap overflow and stack overflow. 
During a heap overflow attack, the program memory space is flooded, while during a stack overflow attack, the program call stack is filled above capacity.

\noindent{\textbf{Malware infection:}}
Software running on UAVs is prone to malware attacks. An adversary could infiltrate the UAV software system and plant malicious software~\cite{rugo2022security}. 
Malware consists of code that steals, infects, and degrades the application by the attacker. Trojans, worms, viruses, spyware, and ransomware are some of the most popular and common malware attacks. This malicious software either steals data from the UAV system or causes the UAV to act abnormally. 
While code injection can be a method for malware infection, it's a specific type of attack, and malware can be introduced in other ways, such as drive-by downloads, malvertising, exploit kits, social engineering (phishing emails), and network propagation. 
Although they may be related, it is important to note that cybersecurity concerns can be divided into two distinct classes.

\noindent{\textbf{Supply Chain Attack:}}
An adversary can attack software vendors in the supply chain to compromise UAV software even before sending it to the customer. These attacks can be implemented via hijacking updates, undermining code signing, and compromising the open source code~\cite{nistsupplychain}. Software Trojans~\cite{rao2021iot} have been observed to infiltrate the UAV manufacturing processes introducing further vulnerabilities and raising significant concerns about the overall integrity and safety of the UAV systems. 

\subsection{Ground Control Station Vulnerabilities}
GCSs vary in their configuration, spanning from a single operator employing a specialized remote device or smartphone to sophisticated, large-scale operations centers where multiple operators manage a fleet of UAVs.
Attack vectors focus on either the human operator(s) or the control station facility. 
It's important to note that even if a UAS is fully autonomous and does not have a human-in-the-loop, the human operators responsible for planning and managing the ground station are still vulnerable and can be targeted (phishing, etc.). The control station facilities can be visualized as cyberphysical systems, consequently having both hardware and software vulnerabilities that are susceptible to exploitation.

Servers at a ground station could be used to operate UAVs. 
Attackers can remotely exploit the GCS by using its network link to communicate with both the UAV and server.
They can perform processing and analytics for a mission or basic/rudimentary tasks such as data storage. 

Mobile phones can be used as controllers. They are available in the Google Play Store or Apple App Store, and adversaries can download, reverse engineer potential security keys, deconstruct the defaults, and even potentially gain information that can be used to override how the UAVs are guided. 
Attackers with access to remote device can exploit known or zero-day~\cite{sun2019tell} vulnerabilities to target the GCS.
Attackers with high access privileges (e.g., root access) may infiltrate systems, forcefully quit applications, and access stored information to disconnect the connection with the UAV.
Smartphones and other computing devices, particularly targeted by attackers, are at risk if security updates are not applied promptly. This is particularly true given the rapid pace at which new attacks emerge and the corresponding slowness of release of patches given the nascency of this domain.

Human factors have been the main subject of cyberattacks on GCS (e.g., scams, phishing, insecure password choices, etc.). 
Several techniques can be used to exploit human errors, such as password leaks and malware installation.

\noindent{\textbf{Remote access:}} 
UAVs are controlled via GCS, but if an adversary can remotely access the GCS, he/she can take remote control of the UAV. 

\noindent{\textbf{Forced quitting application:}}
An adversary can compromise a GCS via embedded malicious code and malware like a Trojan horse. Such an adversary can crash running applications at a GCS which would cause a loss of control of all the UAVs operating under the GCS. 

\noindent{\textbf{Data exfiltration:}}
Adversaries can steal data from GCS via spyware, Trojan horse, or embedded malicious code. 
Keyloggers are dangerous from a security perspective. While it is true that keyloggers can evade detection by some antivirus software, the efficacy of antivirus programs in detecting keyloggers depends on the sophistication of the malware and the effectiveness of the antivirus software. Additionally, keyloggers can capture sensitive information such as passwords and personal information and can transmit this data to a remote server controlled by an attacker, making them a serious security threat~\cite{manesh2019cyber}. Keyloggers were found in the Nevada operating cockpits of Creech Air Force Base, which was used to send commands to the Predator and Reaper UAVs in Afghanistan, Pakistan, and Libya~\cite{nguyenvirus}. 
Additionally, GCS stores the UAV data and other data, such as the pilot's details and supply chain information, which can be extracted without authorization. 

\noindent{\textbf{Password Breaking:}}
Using weak passwords, insecure encryption algorithms, and recovering passwords in transmission are attack vectors that adversaries may use to crack/steal passwords and get access to the GCS. 

\noindent{\textbf{Reverse Engineering:}}
The GCS applications are vulnerable to software reverse engineering, which can grant unauthorized access to sensitive data/information, including authentication tokens. 
It's essential to implement necessary security measures to protect the application against potential cyber threats.

\noindent{\textbf{Social Engineering:}}
Baiting is an information gathering technique adversaries use to gather information by deliberately keeping a trap. Social engineering uses manipulation to exploit human errors. Phishing is an attack that typically involves sending fraudulent messages that appear to come from a trusted source to trick the recipient into revealing sensitive information or clicking on a malicious link. The most common ways for phishing attacks to be carried out are through email or messaging.

\subsection{Network Link Vulnerabilities}\label{NetworkLinkVulnerabilities}
Network communication is critical as it enables UAVs to perform various tasks in different environments, including military, commercial, and civilian applications.
It enables them to receive commands from a ground station or a remote operator to fly to a specific location or perform a specific task. This communication link can include data on the UAV's location, speed, altitude, and other parameters needed for navigation and control. 
UAVs also relay back telemetry data, such as images, video, and other sensor readings, to the ground station. 
This enables the operator to monitor the UAV's performance, track its progress, and make adjustments as necessary. Communication between UAVs also helps coordinate the actions of multiple UAVs in a swarm or formation.
Communication protocols and systems enable UAVs to work together effectively and avoid collisions.
UAVs also use network (neighborhood) communication to maintain situational awareness and avoid collisions with other aircraft or obstacles in the airspace. UAVs must be able to detect and avoid other objects in the airspace, which requires real-time communication with other UAVs and ground-based systems.
Communication links between a UAV and GCS occur through WiFi or radio signals that are vulnerable to external attackers. We present below a comprehensive sub-categorization of network vulnerabilities.

\noindent{\textbf{Blackhole/Grayhole:}}
The blackhole attack in mobile ad-hoc networks includes an attack where a malicious node advertises itself as a forwarding node from a sender to a receiver but drops the packets so that packets from the sender never reach the receiver. Both blackhole and grayhole attacks can be initiated by a malicious attacker who has gained unauthorized access to a network or by an insider with legitimate access intending to harm the network or its users. In grayhole, the only difference is that the attacker selectively drops packets. In the case of UAVs, these attacks are usually orchestrated by the malicious node exploiting the routing protocol to broadcasting itself as the shortest route to forward packets to a favored destination. As a result, they may cause packet dropping between the UAV and the GCS, leading to a denial of service in the UAS~\cite{sumra2018security}. 

\noindent{\textbf{Wormhole:}}
A wormhole attack occurs when an adversary employs multiple nodes to form a tunneling network to direct traffic from one or more compromised nodes to another malicious node in another part of the network. Primarily noticed in ad-hoc networks, the adversarial nodes form a tunnel that gives targeted good nodes in a network a reason to direct traffic through the adversaries' ``best route.'' Nodes sending packets through the wormhole may experience loss of privacy with the data captured by the attackers~\cite{lazos2005preventing}.

\noindent{\textbf{Sybil:}}
In peer-to-peer networks, the Sybil attack is when an entity operates multiple false identities simultaneously to cheat the authority in reputation systems. By acting as honest nodes in a protocol, the adversaries implementing a Sybil attack may sabotage the UAS mission by injecting faulty position information of the UAVs and/or causing collisions~\cite{de2021uavouch}. The vulnerability of a network against this attack is based on how easy it is for advertisers to create false identities. 

\noindent{\textbf{Sinkhole:}}
Sinkhole attacks are implemented by either hacking a good node or introducing a bad adversarial node in the network; the malicious node promotes itself as the node with access to the shortest path to the base station (e.g. GCS) in the network~\cite{ngai2006intruder}. Upon receiving that status in the network, a sinkhole node may carry out the attack by receiving packets communicated between the nodes in the network to the receiver station, from which it may alter or drop the packets. 

\noindent{\textbf{Radio Frequency (RF)-based Jamming:}}
Frequency jamming of radio signals is implemented by blasting an overpowering signal in the same frequency range as the frequency used by the targeted network. This blocks the reception or transmission of signals, leading to loss of control and communication between the UAV and GCS or even among UAVs~\cite{parlin2018jamming}. It is relatively easy to orchestrate this attack since the devices required may be commercially purchased. 

\noindent{\textbf{Protocol-based Jamming (Message Flooding):}}
A Denial of Service (DoS) attack is a cyber-attack intended to overwhelm a system resource with a flood of internet traffic such that it cannot respond to regular requests.
In a Distributed Denial of Service (DDoS) attack, malicious actors flood a target server with traffic from multiple sources to prevent legitimate users from accessing network services. 
The protocol-based jamming, also known as message flooding or ping flooding, attack is a type of DDoS attack where the adversary sends a massive amount of protocol messages, such as ping messages, to check the status of targeted nodes~\cite{he2018flight}. 
This would overload the target nodes' resources with Internet Control Message Protocol (ICMP) pings that cause the dropping of the legitimate user's network requests. Like RF jamming, exploiting the protocol design weaknesses may cause a loss of control and communication between the UAV-UAV or UAV-GCS. The target node's IP address must be known beforehand to implement this attack. 

\noindent{\textbf{Deauthentication:}}
Adversaries may force devices out of a wireless network by implementing the deauthentication attack which sends a deauthentication frame to the targeted infrastructure node or the UAS/UAV to signal that it has been disconnected from the network. Following the device's perceived disconnection from the wireless network, it then attempts to reconnect to the original wireless network. Attackers can then act as UAV controllers, a device communicating with UAV, or take control of the UAV using replay attack and/or message injection~\cite{javaid2012cyber}.
The attacker can also sniff the Wi-Fi Protected Access (WPA) 4-way handshake to perform a brute-force or dictionary-based WPA password-cracking attack. Additionally, a person-in-the-middle (PiTM) may be implemented to collect passwords from the targeted device by which the attacker can intercept and collect the information as it passes through the communication channel.

\noindent{\textbf{Packet Sniffing/Analysis:}}
Packet sniffing is a crucial privacy concern for network devices that communicate over wireless and wired connections. 
Sniffed packets can be analyzed by software to perform different analyses~\cite{alajmi2017uavs}. Additionally, the packets may be logged and saved for further analysis offline. 

\noindent{\textbf{Password Breaking:}}
Solving passwords by various methods is known as the practice of password breaking. Keylogging could also be used to get login details~\cite{jeler2020analysis}. A common approach is using brute force to try all possible combinations for a given input size until the correct password solution is reached by either matching its cryptographic hash or gaining access to the password-protected device/system. Password breaking is achieved via dictionary attack, brute force attack, utilizing cloud resources, social engineering attack, and rainbow table~\cite{minarik2019cybernetics}. 

\noindent{\textbf{Person-In-The-Middle:}}
The person-in-the-middle attack, also known as a man-in-the-middle attack, is a form of eavesdropping attack where the attacker intercepts and relays the packets between sender and receiver by appearing as the sender to the receiver and vice-versa~\cite{vemi2015vulnerability}. By being able to relay the packets, the attacker may also alter the messages being sent and received by the unwitting target devices. Without strong encryption between UAVs and GCS, attackers can listen to communication messages~\cite{manesh2017analysis}. Messages such as Automatic Dependent Surveillance-Broadcast (ADS-B) help aircraft navigate, and 
Federal Aviation Administration (FAA) requires those messages to be broadcast unencrypted. Adversaries could listen to them and gather information, such as the number of UAVs flying, the flight routes, and jam the communication network. They could also broadcast false ADS-B messages and replace original/true ADS-B messages.
ADS-B spoofing is easy to implement and can be done using cheap and easily available hardware~\cite{costin2012ghost}.
Some countries have laws to prevent unintended recipients from listening to broadcasting messages however these cannot prevent adversaries from intercepting data~\cite{manesh2019cyber}. 

\noindent{\textbf{Command Injection:}}
Applications running on devices may be susceptible to attacks where an adversary executes operating system commands on the device. Command injection is a vulnerability in applications allowing attackers to increase the privilege of a process or spawn a remote reverse shell that allows interaction from the adversary to the device~\cite{fouda2018security}. 

\noindent{\textbf{Masquerading:}}
Attackers who can gain a fake authorized identity may use it to gain unauthorized access to devices and systems~\cite{alladi2020secauthuav}. As a result, such attacks may bring the target(s) under the attacker's control while also giving unprotected access to the targets' data and processing capabilities. 

\noindent{\textbf{Replay Attack:}}
Replay attacks intercept the packets but do not alter them in their attack mission, e.g., the replay attack on Global Navigation Satellite Systems (GNSS)~\cite{lenhart2021relay}, and video replay attack~\cite{raja2021efficient}. Instead, they sniff the packet and use it to trick the receiver into believing it is communicating with a legitimate source. The fact that the attacker can send a legitimate looking message to the receiver is assumed to be sufficient for the receiver to trust the attacker. 

\noindent{\textbf{Relay Attack:}}
Relay attacks are similar to person-in-the-middle and replay attacks, where a signal is sniffed. In the relay attack, the attacker relays the signal of an authorized sender through itself to a target receiver for authentication~\cite{fouda2018security}. Relay attacks are common in car thefts, contactless card attacks, and server message block relay attacks. 

\noindent{\textbf{Fuzzing:}}
Fuzzing is a technique where an attacker creates a universe of fake messages to see if which ones can affect the victim. The attacker gains network access and sends these messages to the target to build a pattern for which messages affect the target~\cite{fouda2018security}. 
Fuzzing attacks can be orchestrated without prior understanding of the protocols or security measures in place.

\subsection{Server Vulnerabilities}
Attackers may target information stored on remote servers or in the Cloud, encompassing flight-related data, such as logs, video recordings, and confidential details about operators. Internet-connected servers are traditional focal points for cyberattacks, and servers associated with UAS are no exception. Server attacks can occur in any UAS operation phase.

\noindent{\textbf{Data leakage:}}
Like most devices today, UAVs store all their collected data in a cloud server. 
An attacker can extract video feeds, live camera streams, or other sensitive data from the cloud or a third-party server.
Heiligenstein et al. in~\cite{heiligenstein_ambjer_vanzanella_kramer_biobot_2022} presented multiple instances of Amazon data breaches. Both private and public cloud networks are susceptible to attacks and could be compromised~\cite{poremba_2020}. 

\noindent{\textbf{Pilot identity leakage:}}
Cloud also stores information regarding operators/pilots and other team members. 
An attacker may expose sensitive personal information related to the UAV pilot's identity.

\noindent{\textbf{Location leakage:}}
An attacker can leak a UAV's current (or past) location(s) if they can access the cloud that is storing UAS data. 


\section{\textbf{Risk on UAV based on Likelihood and Severity}\label{likelihoodseverityrisk}}

To evaluate the risks linked to these vulnerabilities, we assign a risk factor based on likelihood of occurrence and severity of the vulnerabilities.
These are defined in the U.S. FAA Order 8000.369, Safety Management System, (SMS),~\cite{faa8000369c} and FAA 8040.4C, Safety Risk Management, (SRM) Policy,~\cite{Order80404C}. 
Likelihood represents the probability of a successful cyberattack, while severity reflects the potential consequences of such an attack. 
SMS is a comprehensive document that details a proactive approach to managing safety in aviation operations. SRM Policy establishes policies to analyze, assess, mitigate, and accept safety risks in the aerospace system.
This risk, likelihood, and severity relationship that we will define in more detail below, serves as the foundation for systematizing knowledge in this article. 


\begin{table}[h]
\footnotesize
\begin{tabular}{| m{6em} | m{19em}|}
\hline 
Category & Operations: Expected Occurrence Rate \\
\hline
 Frequent (A) & (Probability) $\geq$ 1 per 1000  \\
\hline
Probable (B) & 1 per 1000 $>$ (Probability) $\geq$ 1 per 100,000  \\
\hline
Remote (C) & 1 per 100,000 $>$ (Probability) $\geq$ 1 per 10,000,000  \\
\hline
Extremely Remote (D) & 1 per 10,000,000 $>$ (Probability) $\geq$ 1 per 1,000,000,000  \\
\hline
Extremely Improbable (E) & 1 per 1,000,000,000 $>$ (Probability) $\geq$ 1 per 10$^{14}$  \\
\hline
\end{tabular}
\caption{\label{tab:ProbabilityTable} Likelihood Definitions ~\cite{faaatosmsmanual}}
\end{table}

\noindent {\em Likelihood:}
Likelihood refers to the probability or frequency of a hazard's effect or outcome, expressed in quantitative terms~\cite{faa8000369c}. 
Table~\ref{tab:ProbabilityTable} shows the likelihood of an event; the values were derived from a ten-year aviation data analysis~\cite{faaatosmsmanual}.

\begin{table}[h]
\footnotesize
\begin{tabular}{ | m{5em} | m{20em}|}
\hline
 Category &    Description \\
\hline
 Minimal (5) & Discomfort to those on the ground and negligible safety effects.  \\
 \hline
 Minor (4) & Physical discomfort to person and slight damage to aircraft.  \\
 \hline
Major (3) & Physical distress or injury to person. Substantial damage to aircraft.  \\
 \hline
 Hazardous (2) & Multiple serious injuries; fatal injury to a relatively small number of persons (one or two); or damage beyond repair (hull loss) without fatalities. Proximity of fewer than 500 ft to manned aircraft.   \\
 \hline
 Catastrophic (1) & Collision with human-crewed aircraft or fatal injury to non-operators. Multiple fatalities (or fatality to all on board) usually with the loss of aircraft.  \\
\hline
\end{tabular}
\caption{\label{tab:SeverityTable} Severity Definitions ~\cite{faaatosmsmanual}}
\end{table}

\noindent{\em Severity:}
Severity refers to the degree of loss or harm caused by a hazard's effect or outcome.
A low severity denotes that an attack, even if it is successful, has low negative consequences for the system, 
and a high severity denotes that the impact of the attack is severe in the context of human (e.g., accidents) or societal harm~\cite{faa8000369c}. 
%
Table~\ref{tab:SeverityTable} presents the severity of incidents with respect to human casuality, which includes five different categories as defined in~\cite{faaatosmsmanual}. 
As the use of UAS is becoming widespread, safety hazards should be recognized according to situations for applications that are categorized as medium to high risk. 
Such examples may be related to those that deal with medical equipment, where the severity of risks is determined by the mission's safety and the potential impact on sensitive information or payload as defined by the stakeholders~\cite{faaatosmsmanual}. 

\begin{table}[b]
\footnotesize
\begin{tabular}{|m{3em}||*{5}{p{1cm}|}}\hline
\backslashbox{L}{S}
&\makebox[1em]{5}&\makebox[1em]{4}&\makebox[1em]{3} &\makebox[1em]{2}&\makebox[1em]{1}\\\hline\hline
 A &\cellcolor{green} Low& \cellcolor{yellow} Medium& \cellcolor{red} High& \cellcolor{red} High& \cellcolor{red} High\\\hline
 B &\cellcolor{green} Low& \cellcolor{yellow} Medium& \cellcolor{red} High& \cellcolor{red} High& \cellcolor{red} High\\\hline
 C &\cellcolor{green} Low& \cellcolor{yellow} Medium& \cellcolor{yellow} Medium& \cellcolor{red} High& \cellcolor{red} High\\\hline
 D &\cellcolor{green} Low&\cellcolor{green} Low& \cellcolor{yellow} Medium& \cellcolor{yellow} Medium& \cellcolor{red} High\\\hline
 E &\cellcolor{green} Low&\cellcolor{green} Low&\cellcolor{green} Low& \cellcolor{yellow} Medium& \cellcolor{pink} Medium/\newline High\\\hline
\end{tabular}
\caption{\label{tab:RiskMatraix} Risk Matrix with Likelihood (L) and Severity (S)~\cite{faaatosmsmanual}. 
}
\end{table}

\noindent {\em Risk:}
Risk is defined as the chance of an unwanted event occurring and the potential consequences of that event. 
Risk assessments are undertaken at each phase of UAS deployment to determine and prioritize mitigation actions.
A systematic approach to it is outlined in FAA Order 8040.4C policy~\cite{Order80404C}. Risks are categorized as Low, Medium, Medium/High, and High, based on a risk matrix that considers Severity and Likelihood levels. 
Table~\ref{tab:RiskMatraix} is a probability-impact matrix that helps identify, assess, and prioritize risks.
A risk level quantifies potential harm or adverse effects for managing risks.
Levels of Risk are assessed based on the Severity and Likelihood of attacks, with scores ranging from low to high.

This paper contributes to the state-of-the-art by being the first to thoroughly consider both Likelihood and Severity (Table~\ref{tab:RiskMatraix}) according to the SMS Manual~\cite{faaatosmsmanual} and identifying risk in each phase. 
Medium/high and High risk requires greater vigilance and urgency for operators and researchers. To effectively mitigate risk, it is essential to identify appropriate mitigation strategies from the ones we outline later in this paper and implement comprehensive defenses.

\noindent\textbf{Likelihood, Severity, and Risk in UAV Operation Phases:}

To assess the likelihood and severity of cybersecurity vulnerabilities in UAV operations, four students independently assigned scores for likelihood and severity based on their understanding of the reviewed papers. If the deviation among their rankings exceeded more than one point, discussions were held with cybersecurity experts, including faculty members, to reach a consensus on the final ranking. Post discussions, the average of the individual rankings was used to determine the final scores. This approach ensured a balanced evaluation, reducing bias and enhancing reliability with expert insights. 
The vulnerability score developed in this study was submitted and approved as part of a field survey submitted to the US FAA. The FAA ratified these results. The report can be found on the ASSURE Center website as open access~\cite{ASSURE}.

Table~\ref{tab:likelihoodandseveritycombined} presents the likelihood and severity assessment matrix. Each column corresponds to a specific phase of the UAV operation and a component of the UAV, while each row represents a distinct vulnerability. 
{\em This answers both {\bf RQ1} and {\bf RQ2}.}
Each cell  denotes the likelihood and severity score, expressed in an alphanumeric-numeric format. The first part (A–E) represents the likelihood score, while the second part (1–5) indicates the severity score. 
The risk for each UAV phase and vulnerability type is derived using the likelihood and severity assessments from Table~\ref{tab:likelihoodandseveritycombined}, in conjunction with the criteria outlined in Table~\ref{tab:RiskMatraix}. The resulting risk values are presented in Table~\ref{tab:Risk123}.

\clearpage


\begin{landscape}
\begin{table}[tbp]
\fontsize{7}{7}\selectfont
\begin{tabular}{|l|l|l|l|l|l|l|l|l|l|l|l|l|l|l|l|l|l|l|l|l|l|l|}
\hline
\multicolumn{2}{|l|}{\multirow{2}{*}{}} &
  \multicolumn{21}{c|}{\begin{tabular}{c} UAS Phases of Operation \end{tabular}} \\ \cline{3-23} 
\multicolumn{2}{|l|}{} &
  \multicolumn{2}{l|}{ \begin{tabular}{c}Pre-Flight\\ /Mission \\ Planning \end{tabular} }  &
  
  \multicolumn{6}{l|}{ \begin{tabular}{c} Preparation   /System Checks \\ (applicable at almost all \\ phases of mission/flight) \end{tabular} }  &
  \multicolumn{5}{c|}{Launch} &
  \multicolumn{3}{l|}{ \begin{tabular}{c} Mission/ \\ Application \\ /Flight   \\ (Communication) \end{tabular} }  &
  \multicolumn{2}{l|}{ \begin{tabular}{c} Return \\ to Land \end{tabular} }  &
  \multicolumn{2}{l|}{ \begin{tabular}{c} Post- \\ Flight \end{tabular} }  &

  Others \\ \hline
\multicolumn{1}{|l|}{\begin{tabular}{l} Attack \\ Reference \\ Number  \\ \\ \\  \end{tabular}} &
\multicolumn{1}{|l|}{\begin{tabular}{l} \\ Severity Legend: \\ 5: Minimal\\ 4: Minor\\ 3: Major \\ 2: Hazardous\\ 1: Catastrophic  \\ \\ Likelihood Legend: \\ A: Frequent\\ B: Probable\\ C: Remote \\ D: Extremely Remote\\ E: Extremely Improbable \\ \\ \\  \end{tabular}} &

  \rotatebox{270}{\parbox{3cm}{Flight Planning (both \\ for manual and autonomous)}}  & 
  \rotatebox{270}{\parbox{3cm}{Programming \\  flight (autonomous only)}}  & 
  
  \rotatebox{270}{\parbox{3cm}{Ground station}}  &
  \rotatebox{270}{\parbox{3cm}{Flight controls}}  &
  \rotatebox{270}{\parbox{3cm}{Data links}}  &
  \rotatebox{270}{\parbox{3cm}{GPS}}  &
  \rotatebox{270}{\parbox{3cm}{Sensors}}  &
  \rotatebox{270}{\parbox{3cm}{Power-battery/fuel}}  & 
  \rotatebox{270}{\parbox{3cm}{System checks   \\ (similar to those \\ noted above)}}  &
  \rotatebox{270}{\parbox{3cm}{Altimeter   verification}}  &
  \rotatebox{270}{\parbox{3cm}{Flight}}  &
  \rotatebox{270}{\parbox{3cm}{Manual}}  &
  \rotatebox{270}{\parbox{3cm}{Autonomous- \\ Flight plan verification}}  &
  \rotatebox{270}{\parbox{3cm}{Data Relay-  \\ Telemetry}}  &
  \rotatebox{270}{\parbox{3cm}{Payload data-  \\ Video relay}}  &
  \rotatebox{270}{\parbox{3cm}{Payload data  - \\ Sensor Information}}  &
  \rotatebox{270}{\parbox{3cm}{Manual}}  &
  \rotatebox{270}{\parbox{3cm}{Autonomous}}  &
  \rotatebox{270}{\parbox{3cm}{Ground Station}}  &
  \rotatebox{270}{\parbox{3cm}{Data Download}}  &
  \rotatebox{270}{\parbox{3cm}{Emergency \\ Procedures }}  
 \\ \hline
  \rowcolor{green}
\multicolumn{1}{|l|}{HW-ID} &
  UAV Hardware Attack &
  \multicolumn{21}{l|}{} \\ \hline
   \rowcolor{lime}
\multicolumn{1}{|l|}{HW-S/GPS} &
  Spoofing-GPS &
  \multicolumn{1}{l|}{E-5} &
  \multicolumn{1}{l|}{E-5} &
  \multicolumn{1}{l|}{D-5} &
  \multicolumn{1}{l|}{A-5} &
  \multicolumn{1}{l|}{A-5} &
  \multicolumn{1}{l|}{A-5} &
  \multicolumn{1}{l|}{A-5} &
  \multicolumn{1}{l|}{C-5} &
  \multicolumn{1}{l|}{D-2} &
  \multicolumn{1}{l|}{A-1} &
  \multicolumn{1}{l|}{A-1} &
  \multicolumn{1}{l|}{B-1} &
  \multicolumn{1}{l|}{B-1} &
  \multicolumn{1}{l|}{A-5} &
  \multicolumn{1}{l|}{A-5} &
  \multicolumn{1}{l|}{A-5} &
  \multicolumn{1}{l|}{A-1} &
  \multicolumn{1}{l|}{A-1} &
  \multicolumn{1}{l|}{D-5} &
  \multicolumn{1}{l|}{D-5} &
 A-1 \\ \hline
  \rowcolor{lime}
\multicolumn{1}{|l|}{HW-S/OS} &
  Spoofing-Other Sensors &
  \multicolumn{1}{l|}{E-5} &
  \multicolumn{1}{l|}{E-5} &
  \multicolumn{1}{l|}{D-5} &
  \multicolumn{1}{l|}{A-5} &
  \multicolumn{1}{l|}{A-5} &
  \multicolumn{1}{l|}{A-5} &
  \multicolumn{1}{l|}{A-5} &
  \multicolumn{1}{l|}{C-5} &
  \multicolumn{1}{l|}{D-2} &
  \multicolumn{1}{l|}{A-1} &
  \multicolumn{1}{l|}{A-1} &
  \multicolumn{1}{l|}{B-1} &
  \multicolumn{1}{l|}{B-1} &
  \multicolumn{1}{l|}{A-5} &
  \multicolumn{1}{l|}{A-5} &
  \multicolumn{1}{l|}{A-5} &
  \multicolumn{1}{l|}{A-1} &
  \multicolumn{1}{l|}{A-1} &
  \multicolumn{1}{l|}{D-5} &
  \multicolumn{1}{l|}{D-5} &
 A-1 \\ \hline
   \rowcolor{lime}
\multicolumn{1}{|l|}{HW-S/ADSB-ID} &
  Spoofing-ADS-B,   Remote ID &
  \multicolumn{1}{l|}{D-5} &
  \multicolumn{1}{l|}{D-5} &
  \multicolumn{1}{l|}{C-5} &
  \multicolumn{1}{l|}{C-5} &
  \multicolumn{1}{l|}{C-5} &
  \multicolumn{1}{l|}{C-5} &
  \multicolumn{1}{l|}{C-5} &
  \multicolumn{1}{l|}{C-5} &
  \multicolumn{1}{l|}{A-4} &
  \multicolumn{1}{l|}{A-4} &
  \multicolumn{1}{l|}{A-4} &
  \multicolumn{1}{l|}{A-4} &
  \multicolumn{1}{l|}{A-4} &
  \multicolumn{1}{l|}{A-5} &
  \multicolumn{1}{l|}{A-5} &
  \multicolumn{1}{l|}{A-5} &
  \multicolumn{1}{l|}{B-4} &
  \multicolumn{1}{l|}{B-4} &
  \multicolumn{1}{l|}{C-4} &
  \multicolumn{1}{l|}{C-4} &
 A-4 \\ \hline
  \rowcolor{lime}
\multicolumn{1}{|l|}{HW-S/A} &
  Spoofing-Actuator &
  \multicolumn{1}{l|}{E-5} &
  \multicolumn{1}{l|}{E-5} &
  \multicolumn{1}{l|}{D-5} &
  \multicolumn{1}{l|}{A-5} &
  \multicolumn{1}{l|}{A-5} &
  \multicolumn{1}{l|}{A-5} &
  \multicolumn{1}{l|}{A-5} &
  \multicolumn{1}{l|}{C-5} &
  \multicolumn{1}{l|}{D-2} &
  \multicolumn{1}{l|}{A-1} &
  \multicolumn{1}{l|}{A-1} &
  \multicolumn{1}{l|}{B-1} &
  \multicolumn{1}{l|}{B-1} &
  \multicolumn{1}{l|}{A-5} &
  \multicolumn{1}{l|}{A-5} &
  \multicolumn{1}{l|}{A-5} &
  \multicolumn{1}{l|}{A-1} &
  \multicolumn{1}{l|}{A-1} &
  \multicolumn{1}{l|}{D-5} &
  \multicolumn{1}{l|}{D-5} &
 A-1 \\ \hline
  \rowcolor{lime}
\multicolumn{1}{|l|}{HW-J/GPS} &
  Jamming-GPS &
  \multicolumn{1}{l|}{E-5} &
  \multicolumn{1}{l|}{E-5} &
  \multicolumn{1}{l|}{E-5} &
  \multicolumn{1}{l|}{A-5} &
  \multicolumn{1}{l|}{A-5} &
  \multicolumn{1}{l|}{A-5} &
  \multicolumn{1}{l|}{A-5} &
  \multicolumn{1}{l|}{C-5} &
  \multicolumn{1}{l|}{D-1} &
  \multicolumn{1}{l|}{A-1} &
  \multicolumn{1}{l|}{A-1} &
  \multicolumn{1}{l|}{A-1} &
  \multicolumn{1}{l|}{B-1} &
  \multicolumn{1}{l|}{A-5} &
  \multicolumn{1}{l|}{A-5} &
  \multicolumn{1}{l|}{A-5} &
  \multicolumn{1}{l|}{A-1} &
  \multicolumn{1}{l|}{A-1} &
  \multicolumn{1}{l|}{E-5} &
  \multicolumn{1}{l|}{E-5} &
 A-1 \\ \hline
  \rowcolor{lime}
\multicolumn{1}{|l|}{HW-J/OS} &
  Jamming-Other   Sensors &
  \multicolumn{1}{l|}{E-5} &
  \multicolumn{1}{l|}{E-5} &
  \multicolumn{1}{l|}{E-5} &
  \multicolumn{1}{l|}{A-5} &
  \multicolumn{1}{l|}{A-5} &
  \multicolumn{1}{l|}{A-5} &
  \multicolumn{1}{l|}{A-5} &
  \multicolumn{1}{l|}{C-5} &
  \multicolumn{1}{l|}{D-1} &
  \multicolumn{1}{l|}{A-1} &
  \multicolumn{1}{l|}{A-1} &
  \multicolumn{1}{l|}{A-1} &
  \multicolumn{1}{l|}{B-1} &
  \multicolumn{1}{l|}{A-5} &
  \multicolumn{1}{l|}{A-5} &
  \multicolumn{1}{l|}{A-5} &
  \multicolumn{1}{l|}{A-1} &
  \multicolumn{1}{l|}{A-1} &
  \multicolumn{1}{l|}{E-5} &
  \multicolumn{1}{l|}{E-5} &
 A-1 \\ \hline
  \rowcolor{lime}
\multicolumn{1}{|l|}{HW-J/ADSB-ID} &
  Jamming-ADS-B,   Remote ID &
  \multicolumn{1}{l|}{D-5} &
  \multicolumn{1}{l|}{D-5} &
  \multicolumn{1}{l|}{C-5} &
  \multicolumn{1}{l|}{C-5} &
  \multicolumn{1}{l|}{C-5} &
  \multicolumn{1}{l|}{C-5} &
  \multicolumn{1}{l|}{C-5} &
  \multicolumn{1}{l|}{C-5} &
  \multicolumn{1}{l|}{A-4} &
  \multicolumn{1}{l|}{A-4} &
  \multicolumn{1}{l|}{A-4} &
  \multicolumn{1}{l|}{A-4} &
  \multicolumn{1}{l|}{A-4} &
  \multicolumn{1}{l|}{A-5} &
  \multicolumn{1}{l|}{A-5} &
  \multicolumn{1}{l|}{A-5} &
  \multicolumn{1}{l|}{B-4} &
  \multicolumn{1}{l|}{B-4} &
  \multicolumn{1}{l|}{C-4} &
  \multicolumn{1}{l|}{C-4} &
 A-4 \\ \hline
  \rowcolor{lime}
\multicolumn{1}{|l|}{HW-J/A} &
  Jamming-Actuator &
  \multicolumn{1}{l|}{E-5} &
  \multicolumn{1}{l|}{E-5} &
  \multicolumn{1}{l|}{E-5} &
  \multicolumn{1}{l|}{A-5} &
  \multicolumn{1}{l|}{A-5} &
  \multicolumn{1}{l|}{A-5} &
  \multicolumn{1}{l|}{A-5} &
  \multicolumn{1}{l|}{C-5} &
  \multicolumn{1}{l|}{D-1} &
  \multicolumn{1}{l|}{A-1} &
  \multicolumn{1}{l|}{A-1} &
  \multicolumn{1}{l|}{A-1} &
  \multicolumn{1}{l|}{B-1} &
  \multicolumn{1}{l|}{A-5} &
  \multicolumn{1}{l|}{A-5} &
  \multicolumn{1}{l|}{A-5} &
  \multicolumn{1}{l|}{A-1} &
  \multicolumn{1}{l|}{A-1} &
  \multicolumn{1}{l|}{E-5} &
  \multicolumn{1}{l|}{E-5} &
 A-1 \\ \hline
  \rowcolor{lime}
\multicolumn{1}{|l|}{HW-FF} &
  Firmware Flashing &
  \multicolumn{1}{l|}{B-4} &
  \multicolumn{1}{l|}{A-4} &
  \multicolumn{1}{l|}{B-4} &
  \multicolumn{1}{l|}{B-4} &
  \multicolumn{1}{l|}{B-4} &
  \multicolumn{1}{l|}{B-4} &
  \multicolumn{1}{l|}{B-3} &
  \multicolumn{1}{l|}{B-2} &
  \multicolumn{1}{l|}{C-2} &
  \multicolumn{1}{l|}{C-3} &
  \multicolumn{1}{l|}{C-3} &
  \multicolumn{1}{l|}{C-3} &
  \multicolumn{1}{l|}{C-3} &
  \multicolumn{1}{l|}{C-4} &
  \multicolumn{1}{l|}{C-4} &
  \multicolumn{1}{l|}{C-4} &
  \multicolumn{1}{l|}{C-1} &
  \multicolumn{1}{l|}{C-1} &
  \multicolumn{1}{l|}{C-3} &
  \multicolumn{1}{l|}{C-4} &
  C-1 \\ \hline
  \rowcolor{lime}
\multicolumn{1}{|l|}{HW-SCA} &
  Supply Chain Attack &
  \multicolumn{1}{l|}{E-5} &
  \multicolumn{1}{l|}{E-5} &
  \multicolumn{1}{l|}{A-5} &
  \multicolumn{1}{l|}{B-5} &
  \multicolumn{1}{l|}{C-5} &
  \multicolumn{1}{l|}{A-5} &
  \multicolumn{1}{l|}{A-5} &
  \multicolumn{1}{l|}{A-5} &
  \multicolumn{1}{l|}{B-5} &
  \multicolumn{1}{l|}{A-5} &
  \multicolumn{1}{l|}{A-1} &
  \multicolumn{1}{l|}{A-1} &
  \multicolumn{1}{l|}{A-1} &
  \multicolumn{1}{l|}{C-5} &
  \multicolumn{1}{l|}{C-5} &
  \multicolumn{1}{l|}{D-5} &
  \multicolumn{1}{l|}{A-1} &
  \multicolumn{1}{l|}{A-1} &
  \multicolumn{1}{l|}{C-5} &
  \multicolumn{1}{l|}{D-5} &
 A-2 \\ \hline
 \rowcolor{magenta}
\multicolumn{1}{|l|}{SW-ID} &
  UAV Software Attack &
  \multicolumn{1}{l|}{}  &
  \multicolumn{1}{l|}{}  &
  \multicolumn{1}{l|}{}  &
  \multicolumn{1}{l|}{}  &
  \multicolumn{1}{l|}{}  &
  \multicolumn{1}{l|}{}  &
  \multicolumn{1}{l|}{}  &
  \multicolumn{1}{l|}{}  &
  \multicolumn{1}{l|}{}  &
  \multicolumn{1}{l|}{}  &
  \multicolumn{1}{l|}{}  &
  \multicolumn{1}{l|}{}  &
  \multicolumn{1}{l|}{}  &
  \multicolumn{1}{l|}{}  &
  \multicolumn{1}{l|}{}  &
  \multicolumn{1}{l|}{}  &
  \multicolumn{1}{l|}{}  &
  \multicolumn{1}{l|}{}  &
  \multicolumn{1}{l|}{}  &
  \multicolumn{1}{l|}{}  &
   \\ \hline
    \rowcolor{pink}
\multicolumn{1}{|l|}{SW-CI} &
  Code Injection &
  \multicolumn{1}{l|}{A-5} &
  \multicolumn{1}{l|}{A-5} &
  \multicolumn{1}{l|}{A-5} &
  \multicolumn{1}{l|}{A-5} &
  \multicolumn{1}{l|}{A-5} &
  \multicolumn{1}{l|}{A-5} &
  \multicolumn{1}{l|}{A-5} &
  \multicolumn{1}{l|}{A-5} &
  \multicolumn{1}{l|}{D-1} &
  \multicolumn{1}{l|}{E-1} &
  \multicolumn{1}{l|}{E-1} &
  \multicolumn{1}{l|}{E-1} &
  \multicolumn{1}{l|}{E-1} &
  \multicolumn{1}{l|}{E-5} &
  \multicolumn{1}{l|}{E-5} &
  \multicolumn{1}{l|}{E-5} &
  \multicolumn{1}{l|}{E-1} &
  \multicolumn{1}{l|}{E-1} &
  \multicolumn{1}{l|}{E-5} &
  \multicolumn{1}{l|}{E-5} &
  E-1 \\ \hline
   \rowcolor{pink}
\multicolumn{1}{|l|}{SW-DI} &
  Database Injection &
  \multicolumn{1}{l|}{C-5} &
  \multicolumn{1}{l|}{C-5} &
  \multicolumn{1}{l|}{C-5} &
  \multicolumn{1}{l|}{C-5} &
  \multicolumn{1}{l|}{C-5} &
  \multicolumn{1}{l|}{C-5} &
  \multicolumn{1}{l|}{C-5} &
  \multicolumn{1}{l|}{C-5} &
  \multicolumn{1}{l|}{C-1} &
  \multicolumn{1}{l|}{C-1} &
  \multicolumn{1}{l|}{C-1} &
  \multicolumn{1}{l|}{C-1} &
  \multicolumn{1}{l|}{C-1} &
  \multicolumn{1}{l|}{C-5} &
  \multicolumn{1}{l|}{C-5} &
  \multicolumn{1}{l|}{C-5} &
  \multicolumn{1}{l|}{C-1} &
  \multicolumn{1}{l|}{C-1} &
  \multicolumn{1}{l|}{C-5} &
  \multicolumn{1}{l|}{C-5} &
  C-1 \\ \hline
   \rowcolor{pink}
\multicolumn{1}{|l|}{SW-FM} &
  Firmware Modification &
  \multicolumn{1}{l|}{A-5} &
  \multicolumn{1}{l|}{A-5} &
  \multicolumn{1}{l|}{A-5} &
  \multicolumn{1}{l|}{A-5} &
  \multicolumn{1}{l|}{A-5} &
  \multicolumn{1}{l|}{A-5} &
  \multicolumn{1}{l|}{A-5} &
  \multicolumn{1}{l|}{A-5} &
  \multicolumn{1}{l|}{D-2} &
  \multicolumn{1}{l|}{D-3} &
  \multicolumn{1}{l|}{D-1} &
  \multicolumn{1}{l|}{D-1} &
  \multicolumn{1}{l|}{D-1} &
  \multicolumn{1}{l|}{D-5} &
  \multicolumn{1}{l|}{D-5} &
  \multicolumn{1}{l|}{D-5} &
  \multicolumn{1}{l|}{D-1} &
  \multicolumn{1}{l|}{D-1} &
  \multicolumn{1}{l|}{A-5} &
  \multicolumn{1}{l|}{A-5} &
  D-1 \\ \hline
   \rowcolor{pink}
\multicolumn{1}{|l|}{SW-BO} &
  Buffer Overflow &
  \multicolumn{1}{l|}{E-5} &
  \multicolumn{1}{l|}{E-5} &
  \multicolumn{1}{l|}{E-5} &
  \multicolumn{1}{l|}{E-5} &
  \multicolumn{1}{l|}{E-5} &
  \multicolumn{1}{l|}{E-5} &
  \multicolumn{1}{l|}{E-5} &
  \multicolumn{1}{l|}{E-5} &
  \multicolumn{1}{l|}{A-5} &
  \multicolumn{1}{l|}{D-5} &
  \multicolumn{1}{l|}{D-1} &
  \multicolumn{1}{l|}{D-1} &
  \multicolumn{1}{l|}{D-1} &
  \multicolumn{1}{l|}{D-5} &
  \multicolumn{1}{l|}{B-5} &
  \multicolumn{1}{l|}{B-5} &
  \multicolumn{1}{l|}{D-1} &
  \multicolumn{1}{l|}{D-1} &
  \multicolumn{1}{l|}{A-5} &
  \multicolumn{1}{l|}{A-5} &
  E-2 \\ \hline
   \rowcolor{pink}
\multicolumn{1}{|l|}{SW-UA} &
  Unauthorized Access &
  \multicolumn{1}{l|}{A-4} &
  \multicolumn{1}{l|}{A-4} &
  \multicolumn{1}{l|}{A-4} &
  \multicolumn{1}{l|}{A-4} &
  \multicolumn{1}{l|}{A-4} &
  \multicolumn{1}{l|}{A-4} &
  \multicolumn{1}{l|}{A-3} &
  \multicolumn{1}{l|}{B-2} &
  \multicolumn{1}{l|}{B-2} &
  \multicolumn{1}{l|}{A-3} &
  \multicolumn{1}{l|}{A-3} &
  \multicolumn{1}{l|}{A-3} &
  \multicolumn{1}{l|}{A-3} &
  \multicolumn{1}{l|}{A-4} &
  \multicolumn{1}{l|}{A-4} &
  \multicolumn{1}{l|}{B-4} &
  \multicolumn{1}{l|}{A-1} &
  \multicolumn{1}{l|}{A-1} &
  \multicolumn{1}{l|}{A-3} &
  \multicolumn{1}{l|}{A-4} &
 A-1 \\ \hline
 \rowcolor{pink}
\multicolumn{1}{|l|}{SW-MI} &
  Malware infection &
  \multicolumn{1}{l|}{A-4} &
  \multicolumn{1}{l|}{A-4} &
  \multicolumn{1}{l|}{A-4} &
  \multicolumn{1}{l|}{A-4} &
  \multicolumn{1}{l|}{A-4} &
  \multicolumn{1}{l|}{A-4} &
  \multicolumn{1}{l|}{A-3} &
  \multicolumn{1}{l|}{B-2} &
  \multicolumn{1}{l|}{B-2} &
  \multicolumn{1}{l|}{A-3} &
  \multicolumn{1}{l|}{A-3} &
  \multicolumn{1}{l|}{A-3} &
  \multicolumn{1}{l|}{A-3} &
  \multicolumn{1}{l|}{A-4} &
  \multicolumn{1}{l|}{A-4} &
  \multicolumn{1}{l|}{B-4} &
  \multicolumn{1}{l|}{A-1} &
  \multicolumn{1}{l|}{A-1} &
  \multicolumn{1}{l|}{A-3} &
  \multicolumn{1}{l|}{A-4} &
 A-1 \\ \hline
   \rowcolor{pink}
\multicolumn{1}{|l|}{SW-SCA} &
  Supply Chain Attack &
  \multicolumn{1}{l|}{E-5} &
  \multicolumn{1}{l|}{E-5} &
  \multicolumn{1}{l|}{A-5} &
  \multicolumn{1}{l|}{B-5} &
  \multicolumn{1}{l|}{C-5} &
  \multicolumn{1}{l|}{A-5} &
  \multicolumn{1}{l|}{A-5} &
  \multicolumn{1}{l|}{A-5} &
  \multicolumn{1}{l|}{B-5} &
  \multicolumn{1}{l|}{A-5} &
  \multicolumn{1}{l|}{A-1} &
  \multicolumn{1}{l|}{A-1} &
  \multicolumn{1}{l|}{A-1} &
  \multicolumn{1}{l|}{C-5} &
  \multicolumn{1}{l|}{C-5} &
  \multicolumn{1}{l|}{D-5} &
  \multicolumn{1}{l|}{A-1} &
  \multicolumn{1}{l|}{A-1} &
  \multicolumn{1}{l|}{C-5} &
  \multicolumn{1}{l|}{D-5} &
 A-2 \\ \hline
    \rowcolor{gray}
\multicolumn{1}{|l|}{GCS-ID} &
  Ground Control System   (GCS) Attack &
  \multicolumn{21}{l|}{} \\ \hline
   \rowcolor{lightgray}
\multicolumn{1}{|l|}{GCS-RA} &
  Remote access &
  \multicolumn{1}{l|}{C-5} &
  \multicolumn{1}{l|}{C-5} &
  \multicolumn{1}{l|}{D-5} &
  \multicolumn{1}{l|}{C-5} &
  \multicolumn{1}{l|}{B-5} &
  \multicolumn{1}{l|}{C-5} &
  \multicolumn{1}{l|}{C-5} &
  \multicolumn{1}{l|}{D-5} &
  \multicolumn{1}{l|}{C-5} &
  \multicolumn{1}{l|}{C-5} &
  \multicolumn{1}{l|}{C-3} &
  \multicolumn{1}{l|}{C-3} &
  \multicolumn{1}{l|}{C-3} &
  \multicolumn{1}{l|}{B-5} &
  \multicolumn{1}{l|}{B-5} &
  \multicolumn{1}{l|}{B-5} &
  \multicolumn{1}{l|}{A-3} &
  \multicolumn{1}{l|}{A-3} &
  \multicolumn{1}{l|}{D-5} &
  \multicolumn{1}{l|}{D-5} &
 A-3 \\ \hline
   \rowcolor{lightgray}
\multicolumn{1}{|l|}{GCS-FQA} &
  Forced quitting application &
  \multicolumn{1}{l|}{C-5} &
  \multicolumn{1}{l|}{C-5} &
  \multicolumn{1}{l|}{D-2} &
  \multicolumn{1}{l|}{C-2} &
  \multicolumn{1}{l|}{B-2} &
  \multicolumn{1}{l|}{C-3} &
  \multicolumn{1}{l|}{C-2} &
  \multicolumn{1}{l|}{D-2} &
  \multicolumn{1}{l|}{C-1} &
  \multicolumn{1}{l|}{C-1} &
  \multicolumn{1}{l|}{C-1} &
  \multicolumn{1}{l|}{C-1} &
  \multicolumn{1}{l|}{C-1} &
  \multicolumn{1}{l|}{B-5} &
  \multicolumn{1}{l|}{B-5} &
  \multicolumn{1}{l|}{B-3} &
  \multicolumn{1}{l|}{A-1} &
  \multicolumn{1}{l|}{A-1} &
  \multicolumn{1}{l|}{D-5} &
  \multicolumn{1}{l|}{D-5} &
 A-1 \\ \hline
   \rowcolor{lightgray}
\multicolumn{1}{|l|}{GCS-DE} &
  Data exfiltration &
  \multicolumn{1}{l|}{D-5} &
  \multicolumn{1}{l|}{D-5} &
  \multicolumn{1}{l|}{B-5} &
  \multicolumn{1}{l|}{B-5} &
  \multicolumn{1}{l|}{B-5} &
  \multicolumn{1}{l|}{D-5} &
  \multicolumn{1}{l|}{D-5} &
  \multicolumn{1}{l|}{E-5} &
  \multicolumn{1}{l|}{A-5} &
  \multicolumn{1}{l|}{B-5} &
  \multicolumn{1}{l|}{B-5} &
  \multicolumn{1}{l|}{B-5} &
  \multicolumn{1}{l|}{B-5} &
  \multicolumn{1}{l|}{A-5} &
  \multicolumn{1}{l|}{A-5} &
  \multicolumn{1}{l|}{A-5} &
  \multicolumn{1}{l|}{A-5} &
  \multicolumn{1}{l|}{A-5} &
  \multicolumn{1}{l|}{A-5} &
  \multicolumn{1}{l|}{A-5} &
 A-5 \\ \hline
   \rowcolor{lightgray}
\multicolumn{1}{|l|}{GCS-PB} &
  Password Breaking &
  \multicolumn{1}{l|}{D-5} &
  \multicolumn{1}{l|}{D-5} &
  \multicolumn{1}{l|}{A-5} &
  \multicolumn{1}{l|}{D-5} &
  \multicolumn{1}{l|}{D-5} &
  \multicolumn{1}{l|}{D-5} &
  \multicolumn{1}{l|}{D-5} &
  \multicolumn{1}{l|}{D-5} &
  \multicolumn{1}{l|}{D-5} &
  \multicolumn{1}{l|}{D-5} &
  \multicolumn{1}{l|}{D-3} &
  \multicolumn{1}{l|}{D-3} &
  \multicolumn{1}{l|}{D-3} &
  \multicolumn{1}{l|}{D-5} &
  \multicolumn{1}{l|}{D-5} &
  \multicolumn{1}{l|}{D-5} &
  \multicolumn{1}{l|}{D-3} &
  \multicolumn{1}{l|}{D-3} &
  \multicolumn{1}{l|}{A-5} &
  \multicolumn{1}{l|}{D-5} &
 A-1 \\ \hline
   \rowcolor{lightgray}
\multicolumn{1}{|l|}{GCS-RE} &
  Reverse Engineering &
  \multicolumn{1}{l|}{A-5} &
  \multicolumn{1}{l|}{A-5} &
  \multicolumn{1}{l|}{A-5} &
  \multicolumn{1}{l|}{A-5} &
  \multicolumn{1}{l|}{A-5} &
  \multicolumn{1}{l|}{A-5} &
  \multicolumn{1}{l|}{A-5} &
  \multicolumn{1}{l|}{A-5} &
  \multicolumn{1}{l|}{A-5} &
  \multicolumn{1}{l|}{A-5} &
  \multicolumn{1}{l|}{A-3} &
  \multicolumn{1}{l|}{A-3} &
  \multicolumn{1}{l|}{A-3} &
  \multicolumn{1}{l|}{A-5} &
  \multicolumn{1}{l|}{A-5} &
  \multicolumn{1}{l|}{A-5} &
  \multicolumn{1}{l|}{A-3} &
  \multicolumn{1}{l|}{A-3} &
  \multicolumn{1}{l|}{C-5} &
  \multicolumn{1}{l|}{C-5} &
  D-1 \\ \hline
   \rowcolor{lightgray}
\multicolumn{1}{|l|}{GCS-SE} &
  Social Engineering &
  \multicolumn{1}{l|}{A-5} &
  \multicolumn{1}{l|}{A-5} &
  \multicolumn{1}{l|}{A-5} &
  \multicolumn{1}{l|}{A-5} &
  \multicolumn{1}{l|}{A-5} &
  \multicolumn{1}{l|}{A-5} &
  \multicolumn{1}{l|}{A-5} &
  \multicolumn{1}{l|}{A-5} &
  \multicolumn{1}{l|}{D-5} &
  \multicolumn{1}{l|}{D-5} &
  \multicolumn{1}{l|}{D-3} &
  \multicolumn{1}{l|}{A-3} &
  \multicolumn{1}{l|}{B-3} &
  \multicolumn{1}{l|}{A-5} &
  \multicolumn{1}{l|}{A-5} &
  \multicolumn{1}{l|}{A-5} &
  \multicolumn{1}{l|}{D-3} &
  \multicolumn{1}{l|}{D-3} &
  \multicolumn{1}{l|}{E-5} &
  \multicolumn{1}{l|}{E-5} &
 B-1 \\ \hline
  \rowcolor{orange}
\multicolumn{1}{|l|}{NL-ID} &
  Network Link Attack &
  \multicolumn{21}{l|}{} \\ \hline
   \rowcolor{brown}
\multicolumn{1}{|l|}{NL-BH/GH} &
  Black Hole/Gray Hole &
  \multicolumn{1}{l|}{E-5} &
  \multicolumn{1}{l|}{E-5} &
  \multicolumn{1}{l|}{E-5} &
  \multicolumn{1}{l|}{E-5} &
  \multicolumn{1}{l|}{E-5} &
  \multicolumn{1}{l|}{E-5} &
  \multicolumn{1}{l|}{E-5} &
  \multicolumn{1}{l|}{E-5} &
  \multicolumn{1}{l|}{B-5} &
  \multicolumn{1}{l|}{C-4} &
  \multicolumn{1}{l|}{C-1} &
  \multicolumn{1}{l|}{C-1} &
  \multicolumn{1}{l|}{C-1} &
  \multicolumn{1}{l|}{C-5} &
  \multicolumn{1}{l|}{C-5} &
  \multicolumn{1}{l|}{C-5} &
  \multicolumn{1}{l|}{C-1} &
  \multicolumn{1}{l|}{C-1} &
  \multicolumn{1}{l|}{E-5} &
  \multicolumn{1}{l|}{E-5} &
 B-1 \\ \hline
   \rowcolor{brown}
\multicolumn{1}{|l|}{NL-W} &
  Wormhole &
  \multicolumn{1}{l|}{E-5} &
  \multicolumn{1}{l|}{E-5} &
  \multicolumn{1}{l|}{E-5} &
  \multicolumn{1}{l|}{E-5} &
  \multicolumn{1}{l|}{E-5} &
  \multicolumn{1}{l|}{E-5} &
  \multicolumn{1}{l|}{E-5} &
  \multicolumn{1}{l|}{E-5} &
  \multicolumn{1}{l|}{D-5} &
  \multicolumn{1}{l|}{D-3} &
  \multicolumn{1}{l|}{D-1} &
  \multicolumn{1}{l|}{D-1} &
  \multicolumn{1}{l|}{D-1} &
  \multicolumn{1}{l|}{C-5} &
  \multicolumn{1}{l|}{C-5} &
  \multicolumn{1}{l|}{C-5} &
  \multicolumn{1}{l|}{C-1} &
  \multicolumn{1}{l|}{C-1} &
  \multicolumn{1}{l|}{E-5} &
  \multicolumn{1}{l|}{E-5} &
 {B-1} \\ \hline
   \rowcolor{brown}
\multicolumn{1}{|l|}{NL-Syb} &
  Sybil &
  \multicolumn{1}{l|}{E-5} &
  \multicolumn{1}{l|}{E-5} &
  \multicolumn{1}{l|}{E-5} &
  \multicolumn{1}{l|}{E-5} &
  \multicolumn{1}{l|}{E-5} &
  \multicolumn{1}{l|}{E-5} &
  \multicolumn{1}{l|}{E-5} &
  \multicolumn{1}{l|}{E-5} &
  \multicolumn{1}{l|}{C-5} &
  \multicolumn{1}{l|}{D-3} &
  \multicolumn{1}{l|}{C-3} &
  \multicolumn{1}{l|}{D-3} &
  \multicolumn{1}{l|}{D-3} &
  \multicolumn{1}{l|}{D-5} &
  \multicolumn{1}{l|}{D-5} &
  \multicolumn{1}{l|}{D-5} &
  \multicolumn{1}{l|}{D-2} &
  \multicolumn{1}{l|}{D-2} &
  \multicolumn{1}{l|}{E-5} &
  \multicolumn{1}{l|}{E-5} &
  D-2 \\ \hline
   \rowcolor{brown}
\multicolumn{1}{|l|}{NL-Sink} &
  Sinkhole &
  \multicolumn{1}{l|}{E-5} &
  \multicolumn{1}{l|}{E-5} &
  \multicolumn{1}{l|}{E-5} &
  \multicolumn{1}{l|}{E-5} &
  \multicolumn{1}{l|}{E-5} &
  \multicolumn{1}{l|}{E-5} &
  \multicolumn{1}{l|}{E-5} &
  \multicolumn{1}{l|}{E-5} &
  \multicolumn{1}{l|}{E-2} &
  \multicolumn{1}{l|}{E-2} &
  \multicolumn{1}{l|}{E-1} &
  \multicolumn{1}{l|}{E-1} &
  \multicolumn{1}{l|}{E-1} &
  \multicolumn{1}{l|}{E-4} &
  \multicolumn{1}{l|}{E-3} &
  \multicolumn{1}{l|}{E-3} &
  \multicolumn{1}{l|}{E-1} &
  \multicolumn{1}{l|}{E-1} &
  \multicolumn{1}{l|}{E-3} &
  \multicolumn{1}{l|}{E-3} &
  E-1 \\ \hline
   \rowcolor{brown}
\multicolumn{1}{|l|}{NL-RFJam} &
  Radio Frequency (RF)-based Jamming &
  \multicolumn{1}{l|}{E-5} &
  \multicolumn{1}{l|}{E-5} &
  \multicolumn{1}{l|}{E-5} &
  \multicolumn{1}{l|}{E-5} &
  \multicolumn{1}{l|}{E-5} &
  \multicolumn{1}{l|}{E-5} &
  \multicolumn{1}{l|}{E-5} &
  \multicolumn{1}{l|}{E-5} &
  \multicolumn{1}{l|}{D-5} &
  \multicolumn{1}{l|}{D-5} &
  \multicolumn{1}{l|}{D-1} &
  \multicolumn{1}{l|}{D-1} &
  \multicolumn{1}{l|}{D-1} &
  \multicolumn{1}{l|}{C-5} &
  \multicolumn{1}{l|}{C-5} &
  \multicolumn{1}{l|}{C-5} &
  \multicolumn{1}{l|}{A-1} &
  \multicolumn{1}{l|}{A-1} &
  \multicolumn{1}{l|}{E-5} &
  \multicolumn{1}{l|}{E-5} &
 A-1 \\ \hline
   \rowcolor{brown}
\multicolumn{1}{|l|}{NL-PBJam} &
  Protocol-based Jamming &
  \multicolumn{1}{l|}{E-5} &
  \multicolumn{1}{l|}{E-5} &
  \multicolumn{1}{l|}{E-5} &
  \multicolumn{1}{l|}{E-5} &
  \multicolumn{1}{l|}{E-5} &
  \multicolumn{1}{l|}{E-5} &
  \multicolumn{1}{l|}{E-5} &
  \multicolumn{1}{l|}{E-5} &
  \multicolumn{1}{l|}{C-5} &
  \multicolumn{1}{l|}{C-5} &
  \multicolumn{1}{l|}{C-1} &
  \multicolumn{1}{l|}{C-1} &
  \multicolumn{1}{l|}{C-1} &
  \multicolumn{1}{l|}{A-5} &
  \multicolumn{1}{l|}{A-5} &
  \multicolumn{1}{l|}{A-5} &
  \multicolumn{1}{l|}{C-1} &
  \multicolumn{1}{l|}{C-1} &
  \multicolumn{1}{l|}{E-5} &
  \multicolumn{1}{l|}{E-5} &
 A-1 \\ \hline
   \rowcolor{brown}
\multicolumn{1}{|l|}{NL-D} &
  Deauthentication &
  \multicolumn{1}{l|}{E-5} &
  \multicolumn{1}{l|}{E-5} &
  \multicolumn{1}{l|}{E-4} &
  \multicolumn{1}{l|}{E-4} &
  \multicolumn{1}{l|}{E-4} &
  \multicolumn{1}{l|}{E-4} &
  \multicolumn{1}{l|}{E-4} &
  \multicolumn{1}{l|}{E-5} &
  \multicolumn{1}{l|}{A-4} &
  \multicolumn{1}{l|}{A-4} &
  \multicolumn{1}{l|}{A-3} &
  \multicolumn{1}{l|}{B-3} &
  \multicolumn{1}{l|}{A-3} &
  \multicolumn{1}{l|}{A-5} &
  \multicolumn{1}{l|}{A-5} &
  \multicolumn{1}{l|}{A-5} &
  \multicolumn{1}{l|}{B-1} &
  \multicolumn{1}{l|}{A-1} &
  \multicolumn{1}{l|}{A-5} &
  \multicolumn{1}{l|}{A-5} &
 B-1 \\ \hline
   \rowcolor{brown}
\multicolumn{1}{|l|}{NL-PS/A} &
  Packet Sniffing/Analysis &
  \multicolumn{1}{l|}{E-5} &
  \multicolumn{1}{l|}{E-5} &
  \multicolumn{1}{l|}{A-5} &
  \multicolumn{1}{l|}{A-5} &
  \multicolumn{1}{l|}{A-5} &
  \multicolumn{1}{l|}{A-5} &
  \multicolumn{1}{l|}{A-5} &
  \multicolumn{1}{l|}{A-5} &
  \multicolumn{1}{l|}{A-5} &
  \multicolumn{1}{l|}{A-5} &
  \multicolumn{1}{l|}{A-5} &
  \multicolumn{1}{l|}{A-5} &
  \multicolumn{1}{l|}{A-5} &
  \multicolumn{1}{l|}{A-5} &
  \multicolumn{1}{l|}{A-5} &
  \multicolumn{1}{l|}{A-5} &
  \multicolumn{1}{l|}{A-5} &
  \multicolumn{1}{l|}{A-5} &
  \multicolumn{1}{l|}{A-5} &
  \multicolumn{1}{l|}{A-5} &
 A-5 \\ \hline
   \rowcolor{brown}
\multicolumn{1}{|l|}{NL-PB} &
  Password Breaking &
  \multicolumn{1}{l|}{A-5} &
  \multicolumn{1}{l|}{A-5} &
  \multicolumn{1}{l|}{A-1} &
  \multicolumn{1}{l|}{A-5} &
  \multicolumn{1}{l|}{A-5} &
  \multicolumn{1}{l|}{A-5} &
  \multicolumn{1}{l|}{A-5} &
  \multicolumn{1}{l|}{A-5} &
  \multicolumn{1}{l|}{A-3} &
  \multicolumn{1}{l|}{A-5} &
  \multicolumn{1}{l|}{A-1} &
  \multicolumn{1}{l|}{A-1} &
  \multicolumn{1}{l|}{A-1} &
  \multicolumn{1}{l|}{A-5} &
  \multicolumn{1}{l|}{A-5} &
  \multicolumn{1}{l|}{A-5} &
  \multicolumn{1}{l|}{A-1} &
  \multicolumn{1}{l|}{A-1} &
  \multicolumn{1}{l|}{A-5} &
  \multicolumn{1}{l|}{A-5} &
 A-1 \\ \hline
   \rowcolor{brown}
\multicolumn{1}{|l|}{NL-PitM} &
  Person-In-The-Middle &
  \multicolumn{1}{l|}{E-5} &
  \multicolumn{1}{l|}{E-5} &
  \multicolumn{1}{l|}{E-5} &
  \multicolumn{1}{l|}{E-5} &
  \multicolumn{1}{l|}{E-5} &
  \multicolumn{1}{l|}{E-5} &
  \multicolumn{1}{l|}{E-5} &
  \multicolumn{1}{l|}{E-5} &
  \multicolumn{1}{l|}{E-5} &
  \multicolumn{1}{l|}{E-5} &
  \multicolumn{1}{l|}{C-1} &
  \multicolumn{1}{l|}{C-1} &
  \multicolumn{1}{l|}{D-1} &
  \multicolumn{1}{l|}{A-5} &
  \multicolumn{1}{l|}{A-5} &
  \multicolumn{1}{l|}{A-5} &
  \multicolumn{1}{l|}{A-1} &
  \multicolumn{1}{l|}{A-1} &
  \multicolumn{1}{l|}{E-5} &
  \multicolumn{1}{l|}{E-5} &
 A-5 \\ \hline
   \rowcolor{brown}
\multicolumn{1}{|l|}{NL-CJ} &
  Command Injection &
  \multicolumn{1}{l|}{E-5} &
  \multicolumn{1}{l|}{C-5} &
  \multicolumn{1}{l|}{C-5} &
  \multicolumn{1}{l|}{C-5} &
  \multicolumn{1}{l|}{C-5} &
  \multicolumn{1}{l|}{C-5} &
  \multicolumn{1}{l|}{C-5} &
  \multicolumn{1}{l|}{C-5} &
  \multicolumn{1}{l|}{C-3} &
  \multicolumn{1}{l|}{C-5} &
  \multicolumn{1}{l|}{C-1} &
  \multicolumn{1}{l|}{C-1} &
  \multicolumn{1}{l|}{C-1} &
  \multicolumn{1}{l|}{C-5} &
  \multicolumn{1}{l|}{C-5} &
  \multicolumn{1}{l|}{C-5} &
  \multicolumn{1}{l|}{C-1} &
  \multicolumn{1}{l|}{C-1} &
  \multicolumn{1}{l|}{C-5} &
  \multicolumn{1}{l|}{A-5} &
 A-1 \\ \hline
   \rowcolor{brown}
\multicolumn{1}{|l|}{NL-M} &
  Masquerading &
  \multicolumn{1}{l|}{E-5} &
  \multicolumn{1}{l|}{E-5} &
  \multicolumn{1}{l|}{E-4} &
  \multicolumn{1}{l|}{E-4} &
  \multicolumn{1}{l|}{E-4} &
  \multicolumn{1}{l|}{E-4} &
  \multicolumn{1}{l|}{E-4} &
  \multicolumn{1}{l|}{E-5} &
  \multicolumn{1}{l|}{E-4} &
  \multicolumn{1}{l|}{E-4} &
  \multicolumn{1}{l|}{E-3} &
  \multicolumn{1}{l|}{E-3} &
  \multicolumn{1}{l|}{E-3} &
  \multicolumn{1}{l|}{B-5} &
  \multicolumn{1}{l|}{B-5} &
  \multicolumn{1}{l|}{B-5} &
  \multicolumn{1}{l|}{C-1} &
  \multicolumn{1}{l|}{C-1} &
  \multicolumn{1}{l|}{B-5} &
  \multicolumn{1}{l|}{B-5} &
 B-2 \\ \hline
   \rowcolor{brown}
\multicolumn{1}{|l|}{NL-ReplayA} &
  Replay Attack &
  \multicolumn{1}{l|}{E-5} &
  \multicolumn{1}{l|}{E-5} &
  \multicolumn{1}{l|}{E-5} &
  \multicolumn{1}{l|}{E-5} &
  \multicolumn{1}{l|}{E-5} &
  \multicolumn{1}{l|}{E-5} &
  \multicolumn{1}{l|}{E-5} &
  \multicolumn{1}{l|}{E-5} &
  \multicolumn{1}{l|}{A-5} &
  \multicolumn{1}{l|}{A-5} &
  \multicolumn{1}{l|}{A-1} &
  \multicolumn{1}{l|}{B-1} &
  \multicolumn{1}{l|}{A-1} &
  \multicolumn{1}{l|}{A-5} &
  \multicolumn{1}{l|}{A-5} &
  \multicolumn{1}{l|}{A-5} &
  \multicolumn{1}{l|}{E-1} &
  \multicolumn{1}{l|}{E-1} &
  \multicolumn{1}{l|}{E-5} &
  \multicolumn{1}{l|}{E-5} &
 A-1 \\ \hline
   \rowcolor{brown}
\multicolumn{1}{|l|}{NL-RelayA} &
  Relay Attack &
  \multicolumn{1}{l|}{E-5} &
  \multicolumn{1}{l|}{E-5} &
  \multicolumn{1}{l|}{E-5} &
  \multicolumn{1}{l|}{E-5} &
  \multicolumn{1}{l|}{E-5} &
  \multicolumn{1}{l|}{E-5} &
  \multicolumn{1}{l|}{E-5} &
  \multicolumn{1}{l|}{E-5} &
  \multicolumn{1}{l|}{E-5} &
  \multicolumn{1}{l|}{E-5} &
  \multicolumn{1}{l|}{E-1} &
  \multicolumn{1}{l|}{E-1} &
  \multicolumn{1}{l|}{E-1} &
  \multicolumn{1}{l|}{E-5} &
  \multicolumn{1}{l|}{E-5} &
  \multicolumn{1}{l|}{E-5} &
  \multicolumn{1}{l|}{E-1} &
  \multicolumn{1}{l|}{E-1} &
  \multicolumn{1}{l|}{E-5} &
  \multicolumn{1}{l|}{E-5} &
  E-1 \\ \hline
   \rowcolor{brown}
\multicolumn{1}{|l|}{NL-F} &
  Fuzzing &
  \multicolumn{1}{l|}{E-5} &
  \multicolumn{1}{l|}{E-5} &
  \multicolumn{1}{l|}{E-5} &
  \multicolumn{1}{l|}{E-5} &
  \multicolumn{1}{l|}{E-5} &
  \multicolumn{1}{l|}{E-5} &
  \multicolumn{1}{l|}{E-5} &
  \multicolumn{1}{l|}{E-5} &
  \multicolumn{1}{l|}{A-5} &
  \multicolumn{1}{l|}{A-5} &
  \multicolumn{1}{l|}{A-1} &
  \multicolumn{1}{l|}{B-1} &
  \multicolumn{1}{l|}{A-1} &
  \multicolumn{1}{l|}{A-5} &
  \multicolumn{1}{l|}{A-5} &
  \multicolumn{1}{l|}{A-5} &
  \multicolumn{1}{l|}{B-1} &
  \multicolumn{1}{l|}{A-1} &
  \multicolumn{1}{l|}{A-5} &
  \multicolumn{1}{l|}{A-5} &
 A-1 \\ \hline
  \rowcolor{teal}
\multicolumn{1}{|l|}{SRV-ID} &
  Cloud Attack &
  \multicolumn{21}{l|}{} \\ \hline
   \rowcolor{cyan}
\multicolumn{1}{|l|}{SRV-DL} &
  Data leakage &
  \multicolumn{1}{l|}{C-5} &
  \multicolumn{1}{l|}{B-5} &
  \multicolumn{1}{l|}{A-5} &
  \multicolumn{1}{l|}{A-5} &
  \multicolumn{1}{l|}{A-5} &
  \multicolumn{1}{l|}{C-5} &
  \multicolumn{1}{l|}{C-5} &
  \multicolumn{1}{l|}{D-5} &
  \multicolumn{1}{l|}{B-5} &
  \multicolumn{1}{l|}{B-5} &
  \multicolumn{1}{l|}{B-5} &
  \multicolumn{1}{l|}{B-5} &
  \multicolumn{1}{l|}{B-5} &
  \multicolumn{1}{l|}{A-5} &
  \multicolumn{1}{l|}{A-5} &
  \multicolumn{1}{l|}{A-5} &
  \multicolumn{1}{l|}{A-5} &
  \multicolumn{1}{l|}{A-5} &
  \multicolumn{1}{l|}{A-5} &
  \multicolumn{1}{l|}{A-5} &
 A-5 \\ \hline
   \rowcolor{cyan}
\multicolumn{1}{|l|}{SRV-PIL} &
  Pilot identity leakage &
  \multicolumn{1}{l|}{A-5} &
  \multicolumn{1}{l|}{A-5} &
  \multicolumn{1}{l|}{A-5} &
  \multicolumn{1}{l|}{E-5} &
  \multicolumn{1}{l|}{E-5} &
  \multicolumn{1}{l|}{E-5} &
  \multicolumn{1}{l|}{E-5} &
  \multicolumn{1}{l|}{E-5} &
  \multicolumn{1}{l|}{E-5} &
  \multicolumn{1}{l|}{E-5} &
  \multicolumn{1}{l|}{C-5} &
  \multicolumn{1}{l|}{E-5} &
  \multicolumn{1}{l|}{E-5} &
  \multicolumn{1}{l|}{E-5} &
  \multicolumn{1}{l|}{E-5} &
  \multicolumn{1}{l|}{E-5} &
  \multicolumn{1}{l|}{E-5} &
  \multicolumn{1}{l|}{E-5} &
  \multicolumn{1}{l|}{A-5} &
  \multicolumn{1}{l|}{B-5} &
  D-5 \\ \hline
   \rowcolor{cyan}
\multicolumn{1}{|l|}{SRV-LL} &
  Location leakage &
  \multicolumn{1}{l|}{A-5} &
  \multicolumn{1}{l|}{A-5} &
  \multicolumn{1}{l|}{A-5} &
  \multicolumn{1}{l|}{E-5} &
  \multicolumn{1}{l|}{E-5} &
  \multicolumn{1}{l|}{E-3} &
  \multicolumn{1}{l|}{E-3} &
  \multicolumn{1}{l|}{E-4} &
  \multicolumn{1}{l|}{D-3} &
  \multicolumn{1}{l|}{E-3} &
  \multicolumn{1}{l|}{B-3} &
  \multicolumn{1}{l|}{D-3} &
  \multicolumn{1}{l|}{B-3} &
  \multicolumn{1}{l|}{E-5} &
  \multicolumn{1}{l|}{E-5} &
  \multicolumn{1}{l|}{E-5} &
  \multicolumn{1}{l|}{A-3} &
  \multicolumn{1}{l|}{A-3} &
  \multicolumn{1}{l|}{A-5} &
  \multicolumn{1}{l|}{C-5} &
  C-3 \\ \hline
  \end{tabular}

\caption{\label{tab:likelihoodandseveritycombined} UAV Likelihood and Severity Assessment Matrix.}
\end{table}
\end{landscape}


\clearpage

\begin{landscape}
\begin{table}[tbp]
\fontsize{7}{7}\selectfont
\begin{tabular}{|l|l|l|l|l|l|l|l|l|l|l|l|l|l|l|l|l|l|l|l|l|l|l|}
\hline
\multicolumn{2}{|l|}{\multirow{2}{*}{}} &
  \multicolumn{21}{c|}{\begin{tabular}{c} UAS Phases of Operation \end{tabular}} \\ \cline{3-23} 
\multicolumn{2}{|l|}{} &
  \multicolumn{2}{l|}{ \begin{tabular}{c}Pre-Flight\\ /Mission \\ Planning \end{tabular} }  &
  
  \multicolumn{6}{l|}{ \begin{tabular}{c} Preparation   /System Checks \\ (applicable at almost all \\ phases of mission/flight) \end{tabular} }  &
  \multicolumn{5}{c|}{Launch} &
  \multicolumn{3}{l|}{ \begin{tabular}{c} Mission/ \\ Application \\ /Flight   \\ (Communication) \end{tabular} }  &
  \multicolumn{2}{l|}{ \begin{tabular}{c} Return \\ to Land \end{tabular} }  &
  \multicolumn{2}{l|}{ \begin{tabular}{c} Post- \\ Flight \end{tabular} }  &

  Others \\ \hline
\multicolumn{1}{|l|}{\begin{tabular}{l} Attack \\ Reference \\ Number  \\ \\ \\ \end{tabular}} &
\multicolumn{1}{|l|}{\begin{tabular}{l} \\ Risk Legend: \\ L: Low\\ M: Medium\\ H/M: High/Medium \\ H: High  \\ \\ \\ \end{tabular}} &
  \rotatebox{90}{\parbox{3cm}{Flight Planning (both \\ for manual and autonomous)}}  & 
  \rotatebox{90}{\parbox{3cm}{Programming \\  flight (autonomous only)}}  & 
  \rotatebox{90}{\parbox{3cm}{Ground station}}  &
  \rotatebox{90}{\parbox{3cm}{Flight controls}}  &
  \rotatebox{90}{\parbox{3cm}{Data links}}  &
  \rotatebox{90}{\parbox{3cm}{GPS}}  &
  \rotatebox{90}{\parbox{3cm}{Sensors}}  &
  \rotatebox{90}{\parbox{3cm}{Power - battery/fuel}}  & 
  \rotatebox{90}{\parbox{3cm}{System checks   \\ (similar to those \\ noted above)}}  &
  \rotatebox{90}{\parbox{3cm}{Altimeter   verification}}  &
  \rotatebox{90}{\parbox{3cm}{Flight}}  &
  \rotatebox{90}{\parbox{3cm}{Manual}}  &
  \rotatebox{90}{\parbox{3cm}{Autonomous -  \\ Flight plan verification}}  &
  \rotatebox{90}{\parbox{3cm}{Data Relay -   \\ Telemetry}}  &
  \rotatebox{90}{\parbox{3cm}{Payload data -   \\ Video relay}}  &
  \rotatebox{90}{\parbox{3cm}{Payload data   -  \\ Sensor Information}}  &
  \rotatebox{90}{\parbox{3cm}{Manual}}  &
  \rotatebox{90}{\parbox{3cm}{Autonomous}}  &
  \rotatebox{90}{\parbox{3cm}{Ground Station}}  &
  \rotatebox{90}{\parbox{3cm}{Data Download}}  &
  \rotatebox{90}{\parbox{3cm}{Emergency \\ Procedures }}  
  \\ \hline
  \rowcolor{green}
\multicolumn{1}{|l|}{HW-ID} &
  UAV Hardware Attack &
  \multicolumn{21}{l|}{} \\ \hline
   \rowcolor{lime}
\multicolumn{1}{|l|}{HW-S/GPS} &
  Spoofing - GPS &
\multicolumn{1}{l|}{L} &
\multicolumn{1}{l|}{L} &
\multicolumn{1}{l|}{L} &
\multicolumn{1}{l|}{L} &
\multicolumn{1}{l|}{L} &
\multicolumn{1}{l|}{L} &
\multicolumn{1}{l|}{L} &
\multicolumn{1}{l|}{L} &
\multicolumn{1}{l|}{M} &
\multicolumn{1}{l|}{\underline H} &
\multicolumn{1}{l|}{\underline H} &
\multicolumn{1}{l|}{\underline H} &
\multicolumn{1}{l|}{\underline H} &
\multicolumn{1}{l|}{L} &
\multicolumn{1}{l|}{L} &
\multicolumn{1}{l|}{L} &
\multicolumn{1}{l|}{\underline H} &
\multicolumn{1}{l|}{\underline H} &
\multicolumn{1}{l|}{L} &
\multicolumn{1}{l|}{L} &
{\underline H} \\ \hline
  \rowcolor{lime}

\multicolumn{1}{|l|}{HW-S/OS} &
  Spoofing - Other Sensors &
\multicolumn{1}{l|}{L} &
\multicolumn{1}{l|}{L} &
\multicolumn{1}{l|}{L} &
\multicolumn{1}{l|}{L} &
\multicolumn{1}{l|}{L} &
\multicolumn{1}{l|}{L} &
\multicolumn{1}{l|}{L} &
\multicolumn{1}{l|}{L} &
\multicolumn{1}{l|}{M} &
\multicolumn{1}{l|}{\underline H} &
\multicolumn{1}{l|}{\underline H} &
\multicolumn{1}{l|}{\underline H} &
\multicolumn{1}{l|}{\underline H} &
\multicolumn{1}{l|}{L} &
\multicolumn{1}{l|}{L} &
\multicolumn{1}{l|}{L} &
\multicolumn{1}{l|}{\underline H} &
\multicolumn{1}{l|}{\underline H} &
\multicolumn{1}{l|}{L} &
\multicolumn{1}{l|}{L} &
{\underline H} \\ \hline
  \rowcolor{lime}
\multicolumn{1}{|l|}{HW-S/ADSB-ID} &
  Spoofing - ADS-B,   Remote ID &
\multicolumn{1}{l|}{L} &
\multicolumn{1}{l|}{L} &
\multicolumn{1}{l|}{L} &
\multicolumn{1}{l|}{L} &
\multicolumn{1}{l|}{L} &
\multicolumn{1}{l|}{L} &
\multicolumn{1}{l|}{L} &
\multicolumn{1}{l|}{L} &
\multicolumn{1}{l|}{M} &
\multicolumn{1}{l|}{M} &
\multicolumn{1}{l|}{M} &
\multicolumn{1}{l|}{M} &
\multicolumn{1}{l|}{M} &
\multicolumn{1}{l|}{L} &
\multicolumn{1}{l|}{L} &
\multicolumn{1}{l|}{L} &
\multicolumn{1}{l|}{M} &
\multicolumn{1}{l|}{M} &
\multicolumn{1}{l|}{M} &
\multicolumn{1}{l|}{M} &
M \\ \hline
  \rowcolor{lime}
\multicolumn{1}{|l|}{HW-S/A} &
  Spoofing - Actuator &
\multicolumn{1}{l|}{L} &
\multicolumn{1}{l|}{L} &
\multicolumn{1}{l|}{L} &
\multicolumn{1}{l|}{L} &
\multicolumn{1}{l|}{L} &
\multicolumn{1}{l|}{L} &
\multicolumn{1}{l|}{L} &
\multicolumn{1}{l|}{L} &
\multicolumn{1}{l|}{M} &
\multicolumn{1}{l|}{\underline H} &
\multicolumn{1}{l|}{\underline H} &
\multicolumn{1}{l|}{\underline H} &
\multicolumn{1}{l|}{\underline H} &
\multicolumn{1}{l|}{L} &
\multicolumn{1}{l|}{L} &
\multicolumn{1}{l|}{L} &
\multicolumn{1}{l|}{\underline H} &
\multicolumn{1}{l|}{\underline H} &
\multicolumn{1}{l|}{L} &
\multicolumn{1}{l|}{L} &
{\underline H} \\ \hline
  \rowcolor{lime}
\multicolumn{1}{|l|}{HW-J/GPS} &
  Jamming - GPS &
\multicolumn{1}{l|}{L} &
\multicolumn{1}{l|}{L} &
\multicolumn{1}{l|}{L} &
\multicolumn{1}{l|}{L} &
\multicolumn{1}{l|}{L} &
\multicolumn{1}{l|}{L} &
\multicolumn{1}{l|}{L} &
\multicolumn{1}{l|}{L} &
\multicolumn{1}{l|}{\underline H} &
\multicolumn{1}{l|}{\underline H} &
\multicolumn{1}{l|}{\underline H} &
\multicolumn{1}{l|}{\underline H} &
\multicolumn{1}{l|}{\underline H} &
\multicolumn{1}{l|}{L} &
\multicolumn{1}{l|}{L} &
\multicolumn{1}{l|}{L} &
\multicolumn{1}{l|}{\underline H} &
\multicolumn{1}{l|}{\underline H} &
\multicolumn{1}{l|}{L} &
\multicolumn{1}{l|}{L} &
{\underline H} \\ \hline
  \rowcolor{lime}
\multicolumn{1}{|l|}{HW-J/OS} &
  Jamming - Other   Sensors &
\multicolumn{1}{l|}{L} &
\multicolumn{1}{l|}{L} &
\multicolumn{1}{l|}{L} &
\multicolumn{1}{l|}{L} &
\multicolumn{1}{l|}{L} &
\multicolumn{1}{l|}{L} &
\multicolumn{1}{l|}{L} &
\multicolumn{1}{l|}{L} &
\multicolumn{1}{l|}{\underline H} &
\multicolumn{1}{l|}{\underline H} &
\multicolumn{1}{l|}{\underline H} &
\multicolumn{1}{l|}{\underline H} &
\multicolumn{1}{l|}{\underline H} &
\multicolumn{1}{l|}{L} &
\multicolumn{1}{l|}{L} &
\multicolumn{1}{l|}{L} &
\multicolumn{1}{l|}{\underline H} &
\multicolumn{1}{l|}{\underline H} &
\multicolumn{1}{l|}{L} &
\multicolumn{1}{l|}{L} &
{\underline H} \\ \hline
  \rowcolor{lime}
\multicolumn{1}{|l|}{HW-J/ADSB-ID} &
  Jamming - ADS-B,   Remote ID &
\multicolumn{1}{l|}{L} &
\multicolumn{1}{l|}{L} &
\multicolumn{1}{l|}{L} &
\multicolumn{1}{l|}{L} &
\multicolumn{1}{l|}{L} &
\multicolumn{1}{l|}{L} &
\multicolumn{1}{l|}{L} &
\multicolumn{1}{l|}{L} &
\multicolumn{1}{l|}{M} &
\multicolumn{1}{l|}{M} &
\multicolumn{1}{l|}{M} &
\multicolumn{1}{l|}{M} &
\multicolumn{1}{l|}{M} &
\multicolumn{1}{l|}{L} &
\multicolumn{1}{l|}{L} &
\multicolumn{1}{l|}{L} &
\multicolumn{1}{l|}{M} &
\multicolumn{1}{l|}{M} &
\multicolumn{1}{l|}{M} &
\multicolumn{1}{l|}{M} &
M \\ \hline
  \rowcolor{lime}
\multicolumn{1}{|l|}{HW-J/A} &
  Jamming - Actuator &
\multicolumn{1}{l|}{L} &
\multicolumn{1}{l|}{L} &
\multicolumn{1}{l|}{L} &
\multicolumn{1}{l|}{L} &
\multicolumn{1}{l|}{L} &
\multicolumn{1}{l|}{L} &
\multicolumn{1}{l|}{L} &
\multicolumn{1}{l|}{L} &
\multicolumn{1}{l|}{\underline H} &
\multicolumn{1}{l|}{\underline H} &
\multicolumn{1}{l|}{\underline H} &
\multicolumn{1}{l|}{\underline H} &
\multicolumn{1}{l|}{\underline H} &
\multicolumn{1}{l|}{L} &
\multicolumn{1}{l|}{L} &
\multicolumn{1}{l|}{L} &
\multicolumn{1}{l|}{\underline H} &
\multicolumn{1}{l|}{\underline H} &
\multicolumn{1}{l|}{L} &
\multicolumn{1}{l|}{L} &
{\underline H} \\ \hline
  \rowcolor{lime}
\multicolumn{1}{|l|}{HW-FF} &
  Firmware Flashing &
\multicolumn{1}{l|}{M} &
\multicolumn{1}{l|}{M} &
\multicolumn{1}{l|}{M} &
\multicolumn{1}{l|}{M} &
\multicolumn{1}{l|}{M} &
\multicolumn{1}{l|}{M} &
\multicolumn{1}{l|}{\underline H} &
\multicolumn{1}{l|}{\underline H} &
\multicolumn{1}{l|}{\underline H} &
\multicolumn{1}{l|}{M} &
\multicolumn{1}{l|}{M} &
\multicolumn{1}{l|}{M} &
\multicolumn{1}{l|}{M} &
\multicolumn{1}{l|}{M} &
\multicolumn{1}{l|}{M} &
\multicolumn{1}{l|}{M} &
\multicolumn{1}{l|}{\underline H} &
\multicolumn{1}{l|}{\underline H} &
\multicolumn{1}{l|}{M} &
\multicolumn{1}{l|}{M} &
{\underline H} \\ \hline
  \rowcolor{lime}
\multicolumn{1}{|l|}{HW-SCA} &
  Supply Chain Attack &
\multicolumn{1}{l|}{L} &
\multicolumn{1}{l|}{L} &
\multicolumn{1}{l|}{L} &
\multicolumn{1}{l|}{L} &
\multicolumn{1}{l|}{L} &
\multicolumn{1}{l|}{L} &
\multicolumn{1}{l|}{L} &
\multicolumn{1}{l|}{L} &
\multicolumn{1}{l|}{L} &
\multicolumn{1}{l|}{L} &
\multicolumn{1}{l|}{\underline H} &
\multicolumn{1}{l|}{\underline H} &
\multicolumn{1}{l|}{\underline H} &
\multicolumn{1}{l|}{L} &
\multicolumn{1}{l|}{L} &
\multicolumn{1}{l|}{L} &
\multicolumn{1}{l|}{\underline H} &
\multicolumn{1}{l|}{\underline H} &
\multicolumn{1}{l|}{L} &
\multicolumn{1}{l|}{L} &
{\underline H} \\ \hline
  \rowcolor{magenta}
\multicolumn{1}{|l|}{SW-ID} &
  UAV Software Attack &
  \multicolumn{1}{l|}{} &
  \multicolumn{1}{l|}{} &
  \multicolumn{1}{l|}{} &
  \multicolumn{1}{l|}{} &
  \multicolumn{1}{l|}{} &
  \multicolumn{1}{l|}{} &
  \multicolumn{1}{l|}{} &
  \multicolumn{1}{l|}{} &
  \multicolumn{1}{l|}{} &
  \multicolumn{1}{l|}{} &
  \multicolumn{1}{l|}{} &
  \multicolumn{1}{l|}{} &
  \multicolumn{1}{l|}{} &
  \multicolumn{1}{l|}{} &
  \multicolumn{1}{l|}{} &
  \multicolumn{1}{l|}{} &
  \multicolumn{1}{l|}{} &
  \multicolumn{1}{l|}{} &
  \multicolumn{1}{l|}{} &
  \multicolumn{1}{l|}{} &
   \\ \hline
   \rowcolor{pink}
\multicolumn{1}{|l|}{SW-CI} &
  Code Injection &
\multicolumn{1}{l|}{L} &
\multicolumn{1}{l|}{L} &
\multicolumn{1}{l|}{L} &
\multicolumn{1}{l|}{L} &
\multicolumn{1}{l|}{L} &
\multicolumn{1}{l|}{L} &
\multicolumn{1}{l|}{L} &
\multicolumn{1}{l|}{L} &
\multicolumn{1}{l|}{\underline H} &
\multicolumn{1}{l|}{\underline {H/M}} &
\multicolumn{1}{l|}{\underline {H/M}} &
\multicolumn{1}{l|}{\underline {H/M}} &
\multicolumn{1}{l|}{\underline {H/M}} &
\multicolumn{1}{l|}{L} &
\multicolumn{1}{l|}{L} &
\multicolumn{1}{l|}{L} &
\multicolumn{1}{l|}{\underline {H/M}} &
\multicolumn{1}{l|}{\underline {H/M}} &
\multicolumn{1}{l|}{L} &
\multicolumn{1}{l|}{L} &
{\underline {H/M}} \\ \hline
  \rowcolor{pink}
\multicolumn{1}{|l|}{SW-DI} &
  Database Injection &
\multicolumn{1}{l|}{L} &
\multicolumn{1}{l|}{L} &
\multicolumn{1}{l|}{L} &
\multicolumn{1}{l|}{L} &
\multicolumn{1}{l|}{L} &
\multicolumn{1}{l|}{L} &
\multicolumn{1}{l|}{L} &
\multicolumn{1}{l|}{L} &
\multicolumn{1}{l|}{\underline H} &
\multicolumn{1}{l|}{\underline H} &
\multicolumn{1}{l|}{\underline H} &
\multicolumn{1}{l|}{\underline H} &
\multicolumn{1}{l|}{\underline H} &
\multicolumn{1}{l|}{L} &
\multicolumn{1}{l|}{L} &
\multicolumn{1}{l|}{L} &
\multicolumn{1}{l|}{\underline H} &
\multicolumn{1}{l|}{\underline H} &
\multicolumn{1}{l|}{L} &
\multicolumn{1}{l|}{L} &
{\underline H} \\ \hline
  \rowcolor{pink}
\multicolumn{1}{|l|}{SW-FM} &
  Firmware Modification &
\multicolumn{1}{l|}{L} &
\multicolumn{1}{l|}{L} &
\multicolumn{1}{l|}{L} &
\multicolumn{1}{l|}{L} &
\multicolumn{1}{l|}{L} &
\multicolumn{1}{l|}{L} &
\multicolumn{1}{l|}{L} &
\multicolumn{1}{l|}{L} &
\multicolumn{1}{l|}{M} &
\multicolumn{1}{l|}{M} &
\multicolumn{1}{l|}{\underline H} &
\multicolumn{1}{l|}{\underline H} &
\multicolumn{1}{l|}{\underline H} &
\multicolumn{1}{l|}{L} &
\multicolumn{1}{l|}{L} &
\multicolumn{1}{l|}{L} &
\multicolumn{1}{l|}{\underline H} &
\multicolumn{1}{l|}{\underline H} &
\multicolumn{1}{l|}{L} &
\multicolumn{1}{l|}{L} &
{\underline H} \\ \hline
  \rowcolor{pink}
\multicolumn{1}{|l|}{SW-BO} &
  Buffer Overflow &
\multicolumn{1}{l|}{L} &
\multicolumn{1}{l|}{L} &
\multicolumn{1}{l|}{L} &
\multicolumn{1}{l|}{L} &
\multicolumn{1}{l|}{L} &
\multicolumn{1}{l|}{L} &
\multicolumn{1}{l|}{L} &
\multicolumn{1}{l|}{L} &
\multicolumn{1}{l|}{L} &
\multicolumn{1}{l|}{L} &
\multicolumn{1}{l|}{\underline H} &
\multicolumn{1}{l|}{\underline H} &
\multicolumn{1}{l|}{\underline H} &
\multicolumn{1}{l|}{L} &
\multicolumn{1}{l|}{L} &
\multicolumn{1}{l|}{L} &
\multicolumn{1}{l|}{\underline H} &
\multicolumn{1}{l|}{\underline H} &
\multicolumn{1}{l|}{L} &
\multicolumn{1}{l|}{L} &
M \\ \hline
  \rowcolor{pink}
\multicolumn{1}{|l|}{SW-UA} &
  Unauthorized Access &
\multicolumn{1}{l|}{M} &
\multicolumn{1}{l|}{M} &
\multicolumn{1}{l|}{M} &
\multicolumn{1}{l|}{M} &
\multicolumn{1}{l|}{M} &
\multicolumn{1}{l|}{M} &
\multicolumn{1}{l|}{\underline H} &
\multicolumn{1}{l|}{\underline H} &
\multicolumn{1}{l|}{\underline H} &
\multicolumn{1}{l|}{\underline H} &
\multicolumn{1}{l|}{\underline H} &
\multicolumn{1}{l|}{\underline H} &
\multicolumn{1}{l|}{\underline H} &
\multicolumn{1}{l|}{M} &
\multicolumn{1}{l|}{M} &
\multicolumn{1}{l|}{M} &
\multicolumn{1}{l|}{\underline H} &
\multicolumn{1}{l|}{\underline H} &
\multicolumn{1}{l|}{\underline H} &
\multicolumn{1}{l|}{M} &
{\underline H} \\ \hline
\rowcolor{pink}
\multicolumn{1}{|l|}{SW-MI} &
  Malware infection &
\multicolumn{1}{l|}{M} &
\multicolumn{1}{l|}{M} &
\multicolumn{1}{l|}{M} &
\multicolumn{1}{l|}{M} &
\multicolumn{1}{l|}{M} &
\multicolumn{1}{l|}{M} &
\multicolumn{1}{l|}{\underline H} &
\multicolumn{1}{l|}{\underline H} &
\multicolumn{1}{l|}{\underline H} &
\multicolumn{1}{l|}{\underline H} &
\multicolumn{1}{l|}{\underline H} &
\multicolumn{1}{l|}{\underline H} &
\multicolumn{1}{l|}{\underline H} &
\multicolumn{1}{l|}{M} &
\multicolumn{1}{l|}{M} &
\multicolumn{1}{l|}{M} &
\multicolumn{1}{l|}{\underline H} &
\multicolumn{1}{l|}{\underline H} &
\multicolumn{1}{l|}{\underline H} &
\multicolumn{1}{l|}{M} &
{\underline H} \\ \hline
  \rowcolor{pink}
\multicolumn{1}{|l|}{SW-SCA} &
  Supply Chain Attack &
\multicolumn{1}{l|}{L} &
\multicolumn{1}{l|}{L} &
\multicolumn{1}{l|}{L} &
\multicolumn{1}{l|}{L} &
\multicolumn{1}{l|}{L} &
\multicolumn{1}{l|}{L} &
\multicolumn{1}{l|}{L} &
\multicolumn{1}{l|}{L} &
\multicolumn{1}{l|}{L} &
\multicolumn{1}{l|}{L} &
\multicolumn{1}{l|}{\underline H} &
\multicolumn{1}{l|}{\underline H} &
\multicolumn{1}{l|}{\underline H} &
\multicolumn{1}{l|}{L} &
\multicolumn{1}{l|}{L} &
\multicolumn{1}{l|}{L} &
\multicolumn{1}{l|}{\underline H} &
\multicolumn{1}{l|}{\underline H} &
\multicolumn{1}{l|}{L} &
\multicolumn{1}{l|}{L} &
{\underline H} \\ \hline
   \rowcolor{lightgray}
  \rowcolor{gray}
\multicolumn{1}{|l|}{GCS-ID} &
  Ground Control System   (GCS) Attack &
  \multicolumn{21}{l|}{} \\ \hline
   \rowcolor{lightgray}
\multicolumn{1}{|l|}{GCS-RA} &
  Remote access &
\multicolumn{1}{l|}{L} &
\multicolumn{1}{l|}{L} &
\multicolumn{1}{l|}{L} &
\multicolumn{1}{l|}{L} &
\multicolumn{1}{l|}{L} &
\multicolumn{1}{l|}{L} &
\multicolumn{1}{l|}{L} &
\multicolumn{1}{l|}{L} &
\multicolumn{1}{l|}{L} &
\multicolumn{1}{l|}{L} &
\multicolumn{1}{l|}{M} &
\multicolumn{1}{l|}{M} &
\multicolumn{1}{l|}{M} &
\multicolumn{1}{l|}{L} &
\multicolumn{1}{l|}{L} &
\multicolumn{1}{l|}{L} &
\multicolumn{1}{l|}{\underline H} &
\multicolumn{1}{l|}{\underline H} &
\multicolumn{1}{l|}{L} &
\multicolumn{1}{l|}{L} &
{\underline H} \\ \hline
   \rowcolor{lightgray}
\multicolumn{1}{|l|}{GCS-FQA} &
  Forced quitting application &
\multicolumn{1}{l|}{L} &
\multicolumn{1}{l|}{L} &
\multicolumn{1}{l|}{M} &
\multicolumn{1}{l|}{\underline H} &
\multicolumn{1}{l|}{\underline H} &
\multicolumn{1}{l|}{M} &
\multicolumn{1}{l|}{\underline H} &
\multicolumn{1}{l|}{M} &
\multicolumn{1}{l|}{\underline H} &
\multicolumn{1}{l|}{\underline H} &
\multicolumn{1}{l|}{\underline H} &
\multicolumn{1}{l|}{\underline H} &
\multicolumn{1}{l|}{\underline H} &
\multicolumn{1}{l|}{L} &
\multicolumn{1}{l|}{L} &
\multicolumn{1}{l|}{\underline H} &
\multicolumn{1}{l|}{\underline H} &
\multicolumn{1}{l|}{\underline H} &
\multicolumn{1}{l|}{L} &
\multicolumn{1}{l|}{L} &
{\underline H} \\ \hline
   \rowcolor{lightgray}
\multicolumn{1}{|l|}{GCS-DE} &
  Data exfiltration &
\multicolumn{1}{l|}{L} &
\multicolumn{1}{l|}{L} &
\multicolumn{1}{l|}{L} &
\multicolumn{1}{l|}{L} &
\multicolumn{1}{l|}{L} &
\multicolumn{1}{l|}{L} &
\multicolumn{1}{l|}{L} &
\multicolumn{1}{l|}{L} &
\multicolumn{1}{l|}{L} &
\multicolumn{1}{l|}{L} &
\multicolumn{1}{l|}{L} &
\multicolumn{1}{l|}{L} &
\multicolumn{1}{l|}{L} &
\multicolumn{1}{l|}{L} &
\multicolumn{1}{l|}{L} &
\multicolumn{1}{l|}{L} &
\multicolumn{1}{l|}{L} &
\multicolumn{1}{l|}{L} &
\multicolumn{1}{l|}{L} &
\multicolumn{1}{l|}{L} &
L \\ \hline
   \rowcolor{lightgray}
\multicolumn{1}{|l|}{GCS-PB} &
  Password Breaking &
\multicolumn{1}{l|}{L} &
\multicolumn{1}{l|}{L} &
\multicolumn{1}{l|}{L} &
\multicolumn{1}{l|}{L} &
\multicolumn{1}{l|}{L} &
\multicolumn{1}{l|}{L} &
\multicolumn{1}{l|}{L} &
\multicolumn{1}{l|}{L} &
\multicolumn{1}{l|}{L} &
\multicolumn{1}{l|}{L} &
\multicolumn{1}{l|}{M} &
\multicolumn{1}{l|}{M} &
\multicolumn{1}{l|}{M} &
\multicolumn{1}{l|}{L} &
\multicolumn{1}{l|}{L} &
\multicolumn{1}{l|}{L} &
\multicolumn{1}{l|}{M} &
\multicolumn{1}{l|}{M} &
\multicolumn{1}{l|}{L} &
\multicolumn{1}{l|}{L} &
{\underline H} \\ \hline
   \rowcolor{lightgray}
\multicolumn{1}{|l|}{GCS-RE} &
  Reverse Engineering &
\multicolumn{1}{l|}{L} &
\multicolumn{1}{l|}{L} &
\multicolumn{1}{l|}{L} &
\multicolumn{1}{l|}{L} &
\multicolumn{1}{l|}{L} &
\multicolumn{1}{l|}{L} &
\multicolumn{1}{l|}{L} &
\multicolumn{1}{l|}{L} &
\multicolumn{1}{l|}{L} &
\multicolumn{1}{l|}{L} &
\multicolumn{1}{l|}{\underline H} &
\multicolumn{1}{l|}{\underline H} &
\multicolumn{1}{l|}{\underline H} &
\multicolumn{1}{l|}{L} &
\multicolumn{1}{l|}{L} &
\multicolumn{1}{l|}{L} &
\multicolumn{1}{l|}{\underline H} &
\multicolumn{1}{l|}{\underline H} &
\multicolumn{1}{l|}{L} &
\multicolumn{1}{l|}{L} &
{\underline H} \\ \hline
  \rowcolor{lightgray}
\multicolumn{1}{|l|}{GCS-SE} &
  Social Engineering &
\multicolumn{1}{l|}{L} &
\multicolumn{1}{l|}{L} &
\multicolumn{1}{l|}{L} &
\multicolumn{1}{l|}{L} &
\multicolumn{1}{l|}{L} &
\multicolumn{1}{l|}{L} &
\multicolumn{1}{l|}{L} &
\multicolumn{1}{l|}{L} &
\multicolumn{1}{l|}{L} &
\multicolumn{1}{l|}{L} &
\multicolumn{1}{l|}{M} &
\multicolumn{1}{l|}{\underline H} &
\multicolumn{1}{l|}{\underline H} &
\multicolumn{1}{l|}{L} &
\multicolumn{1}{l|}{L} &
\multicolumn{1}{l|}{L} &
\multicolumn{1}{l|}{M} &
\multicolumn{1}{l|}{M} &
\multicolumn{1}{l|}{L} &
\multicolumn{1}{l|}{L} &
{\underline H} \\ \hline
    \rowcolor{brown}
  \rowcolor{orange}
\multicolumn{1}{|l|}{NL-ID} &
  Network Link Attack &
  \multicolumn{21}{l|}{} \\ \hline
    \rowcolor{brown}
\multicolumn{1}{|l|}{NL-BH/GH} &
  Black Hole/Gray Hole &
\multicolumn{1}{l|}{L} &
\multicolumn{1}{l|}{L} &
\multicolumn{1}{l|}{L} &
\multicolumn{1}{l|}{L} &
\multicolumn{1}{l|}{L} &
\multicolumn{1}{l|}{L} &
\multicolumn{1}{l|}{L} &
\multicolumn{1}{l|}{L} &
\multicolumn{1}{l|}{L} &
\multicolumn{1}{l|}{M} &
\multicolumn{1}{l|}{\underline H} &
\multicolumn{1}{l|}{\underline H} &
\multicolumn{1}{l|}{\underline H} &
\multicolumn{1}{l|}{L} &
\multicolumn{1}{l|}{L} &
\multicolumn{1}{l|}{L} &
\multicolumn{1}{l|}{\underline H} &
\multicolumn{1}{l|}{\underline H} &
\multicolumn{1}{l|}{L} &
\multicolumn{1}{l|}{L} &
{\underline H} \\ \hline
    \rowcolor{brown}
\multicolumn{1}{|l|}{NL-W} &
  Wormhole &
\multicolumn{1}{l|}{L} &
\multicolumn{1}{l|}{L} &
\multicolumn{1}{l|}{L} &
\multicolumn{1}{l|}{L} &
\multicolumn{1}{l|}{L} &
\multicolumn{1}{l|}{L} &
\multicolumn{1}{l|}{L} &
\multicolumn{1}{l|}{L} &
\multicolumn{1}{l|}{L} &
\multicolumn{1}{l|}{M} &
\multicolumn{1}{l|}{\underline H} &
\multicolumn{1}{l|}{\underline H} &
\multicolumn{1}{l|}{\underline H} &
\multicolumn{1}{l|}{L} &
\multicolumn{1}{l|}{L} &
\multicolumn{1}{l|}{L} &
\multicolumn{1}{l|}{\underline H} &
\multicolumn{1}{l|}{\underline H} &
\multicolumn{1}{l|}{L} &
\multicolumn{1}{l|}{L} &
{\underline H} \\ \hline
    \rowcolor{brown}
\multicolumn{1}{|l|}{NL-Syb} &
  Sybil &
\multicolumn{1}{l|}{L} &
\multicolumn{1}{l|}{L} &
\multicolumn{1}{l|}{L} &
\multicolumn{1}{l|}{L} &
\multicolumn{1}{l|}{L} &
\multicolumn{1}{l|}{L} &
\multicolumn{1}{l|}{L} &
\multicolumn{1}{l|}{L} &
\multicolumn{1}{l|}{L} &
\multicolumn{1}{l|}{M} &
\multicolumn{1}{l|}{M} &
\multicolumn{1}{l|}{M} &
\multicolumn{1}{l|}{M} &
\multicolumn{1}{l|}{L} &
\multicolumn{1}{l|}{L} &
\multicolumn{1}{l|}{L} &
\multicolumn{1}{l|}{M} &
\multicolumn{1}{l|}{M} &
\multicolumn{1}{l|}{L} &
\multicolumn{1}{l|}{L} &
M \\ \hline
    \rowcolor{brown}
\multicolumn{1}{|l|}{NL-Sink} &
  Sinkhole &
\multicolumn{1}{l|}{L} &
\multicolumn{1}{l|}{L} &
\multicolumn{1}{l|}{L} &
\multicolumn{1}{l|}{L} &
\multicolumn{1}{l|}{L} &
\multicolumn{1}{l|}{L} &
\multicolumn{1}{l|}{L} &
\multicolumn{1}{l|}{L} &
\multicolumn{1}{l|}{M} &
\multicolumn{1}{l|}{M} &
\multicolumn{1}{l|}{\underline {H/M}} &
\multicolumn{1}{l|}{\underline {H/M}} &
\multicolumn{1}{l|}{\underline {H/M}} &
\multicolumn{1}{l|}{L} &
\multicolumn{1}{l|}{L} &
\multicolumn{1}{l|}{L} &
\multicolumn{1}{l|}{\underline {H/M}} &
\multicolumn{1}{l|}{\underline {H/M}} &
\multicolumn{1}{l|}{L} &
\multicolumn{1}{l|}{L} &
{\underline {H/M}} \\ \hline
    \rowcolor{brown}
\multicolumn{1}{|l|}{NL-RFJam} &
  Radio Frequency (RF)-based Jamming &
\multicolumn{1}{l|}{L} &
\multicolumn{1}{l|}{L} &
\multicolumn{1}{l|}{L} &
\multicolumn{1}{l|}{L} &
\multicolumn{1}{l|}{L} &
\multicolumn{1}{l|}{L} &
\multicolumn{1}{l|}{L} &
\multicolumn{1}{l|}{L} &
\multicolumn{1}{l|}{L} &
\multicolumn{1}{l|}{L} &
\multicolumn{1}{l|}{\underline H} &
\multicolumn{1}{l|}{\underline H} &
\multicolumn{1}{l|}{\underline H} &
\multicolumn{1}{l|}{L} &
\multicolumn{1}{l|}{L} &
\multicolumn{1}{l|}{L} &
\multicolumn{1}{l|}{\underline H} &
\multicolumn{1}{l|}{\underline H} &
\multicolumn{1}{l|}{L} &
\multicolumn{1}{l|}{L} &
{\underline H} \\ \hline
    \rowcolor{brown}
\multicolumn{1}{|l|}{NL-PBJam} &
  Protocol-based Jamming &
\multicolumn{1}{l|}{L} &
\multicolumn{1}{l|}{L} &
\multicolumn{1}{l|}{L} &
\multicolumn{1}{l|}{L} &
\multicolumn{1}{l|}{L} &
\multicolumn{1}{l|}{L} &
\multicolumn{1}{l|}{L} &
\multicolumn{1}{l|}{L} &
\multicolumn{1}{l|}{L} &
\multicolumn{1}{l|}{L} &
\multicolumn{1}{l|}{\underline H} &
\multicolumn{1}{l|}{\underline H} &
\multicolumn{1}{l|}{\underline H} &
\multicolumn{1}{l|}{L} &
\multicolumn{1}{l|}{L} &
\multicolumn{1}{l|}{L} &
\multicolumn{1}{l|}{\underline H} &
\multicolumn{1}{l|}{\underline H} &
\multicolumn{1}{l|}{L} &
\multicolumn{1}{l|}{L} &
{\underline H} \\ \hline
    \rowcolor{brown}
\multicolumn{1}{|l|}{NL-D} &
  Deauthentication &
\multicolumn{1}{l|}{L} &
\multicolumn{1}{l|}{L} &
\multicolumn{1}{l|}{L} &
\multicolumn{1}{l|}{L} &
\multicolumn{1}{l|}{L} &
\multicolumn{1}{l|}{L} &
\multicolumn{1}{l|}{L} &
\multicolumn{1}{l|}{L} &
\multicolumn{1}{l|}{M} &
\multicolumn{1}{l|}{M} &
\multicolumn{1}{l|}{\underline H} &
\multicolumn{1}{l|}{\underline H} &
\multicolumn{1}{l|}{\underline H} &
\multicolumn{1}{l|}{L} &
\multicolumn{1}{l|}{L} &
\multicolumn{1}{l|}{L} &
\multicolumn{1}{l|}{\underline H} &
\multicolumn{1}{l|}{\underline H} &
\multicolumn{1}{l|}{L} &
\multicolumn{1}{l|}{L} &
{\underline H} \\ \hline
    \rowcolor{brown}
\multicolumn{1}{|l|}{NL-PS/A} &
  Packet Sniffing/Analysis &
\multicolumn{1}{l|}{L} &
\multicolumn{1}{l|}{L} &
\multicolumn{1}{l|}{L} &
\multicolumn{1}{l|}{L} &
\multicolumn{1}{l|}{L} &
\multicolumn{1}{l|}{L} &
\multicolumn{1}{l|}{L} &
\multicolumn{1}{l|}{L} &
\multicolumn{1}{l|}{L} &
\multicolumn{1}{l|}{L} &
\multicolumn{1}{l|}{L} &
\multicolumn{1}{l|}{L} &
\multicolumn{1}{l|}{L} &
\multicolumn{1}{l|}{L} &
\multicolumn{1}{l|}{L} &
\multicolumn{1}{l|}{L} &
\multicolumn{1}{l|}{L} &
\multicolumn{1}{l|}{L} &
\multicolumn{1}{l|}{L} &
\multicolumn{1}{l|}{L} &
L \\ \hline
    \rowcolor{brown}
\multicolumn{1}{|l|}{NL-PB} &
  Password Breaking &
\multicolumn{1}{l|}{L} &
\multicolumn{1}{l|}{L} &
\multicolumn{1}{l|}{\underline H} &
\multicolumn{1}{l|}{L} &
\multicolumn{1}{l|}{L} &
\multicolumn{1}{l|}{L} &
\multicolumn{1}{l|}{L} &
\multicolumn{1}{l|}{L} &
\multicolumn{1}{l|}{\underline H} &
\multicolumn{1}{l|}{L} &
\multicolumn{1}{l|}{\underline H} &
\multicolumn{1}{l|}{\underline H} &
\multicolumn{1}{l|}{\underline H} &
\multicolumn{1}{l|}{L} &
\multicolumn{1}{l|}{L} &
\multicolumn{1}{l|}{L} &
\multicolumn{1}{l|}{\underline H} &
\multicolumn{1}{l|}{\underline H} &
\multicolumn{1}{l|}{L} &
\multicolumn{1}{l|}{L} &
{\underline H} \\ \hline
    \rowcolor{brown}
\multicolumn{1}{|l|}{NL-PitM} &
  Person-In-The-Middle &
\multicolumn{1}{l|}{L} &
\multicolumn{1}{l|}{L} &
\multicolumn{1}{l|}{L} &
\multicolumn{1}{l|}{L} &
\multicolumn{1}{l|}{L} &
\multicolumn{1}{l|}{L} &
\multicolumn{1}{l|}{L} &
\multicolumn{1}{l|}{L} &
\multicolumn{1}{l|}{L} &
\multicolumn{1}{l|}{L} &
\multicolumn{1}{l|}{\underline H} &
\multicolumn{1}{l|}{\underline H} &
\multicolumn{1}{l|}{\underline H} &
\multicolumn{1}{l|}{L} &
\multicolumn{1}{l|}{L} &
\multicolumn{1}{l|}{L} &
\multicolumn{1}{l|}{\underline H} &
\multicolumn{1}{l|}{\underline H} &
\multicolumn{1}{l|}{L} &
\multicolumn{1}{l|}{L} &
L   \\ \hline
     \rowcolor{brown}
\multicolumn{1}{|l|}{NL-CJ} &
  Command Injection &
 \multicolumn{1}{l|}{L} &
\multicolumn{1}{l|}{L} &
\multicolumn{1}{l|}{L} &
\multicolumn{1}{l|}{L} &
\multicolumn{1}{l|}{L} &
\multicolumn{1}{l|}{L} &
\multicolumn{1}{l|}{L} &
\multicolumn{1}{l|}{L} &
\multicolumn{1}{l|}{M} &
\multicolumn{1}{l|}{L} &
\multicolumn{1}{l|}{\underline H} &
\multicolumn{1}{l|}{\underline H} &
\multicolumn{1}{l|}{\underline H} &
\multicolumn{1}{l|}{L} &
\multicolumn{1}{l|}{L} &
\multicolumn{1}{l|}{L} &
\multicolumn{1}{l|}{\underline H} &
\multicolumn{1}{l|}{\underline H} &
\multicolumn{1}{l|}{L} &
\multicolumn{1}{l|}{L} &
{\underline H} \\ \hline
    \rowcolor{brown}
\multicolumn{1}{|l|}{NL-M} &
  Masquerading &
  \multicolumn{1}{l|}{L} &
\multicolumn{1}{l|}{L} &
\multicolumn{1}{l|}{L} &
\multicolumn{1}{l|}{L} &
\multicolumn{1}{l|}{L} &
\multicolumn{1}{l|}{L} &
\multicolumn{1}{l|}{L} &
\multicolumn{1}{l|}{L} &
\multicolumn{1}{l|}{L} &
\multicolumn{1}{l|}{L} &
\multicolumn{1}{l|}{L} &
\multicolumn{1}{l|}{L} &
\multicolumn{1}{l|}{L} &
\multicolumn{1}{l|}{L} &
\multicolumn{1}{l|}{L} &
\multicolumn{1}{l|}{L} &
\multicolumn{1}{l|}{\underline H} &
\multicolumn{1}{l|}{\underline H} &
\multicolumn{1}{l|}{L} &
\multicolumn{1}{l|}{L} &
{\underline H} \\ \hline
    \rowcolor{brown}
\multicolumn{1}{|l|}{NL-ReplayA} &
  Replay Attack &
  \multicolumn{1}{l|}{L} &
\multicolumn{1}{l|}{L} &
\multicolumn{1}{l|}{L} &
\multicolumn{1}{l|}{L} &
\multicolumn{1}{l|}{L} &
\multicolumn{1}{l|}{L} &
\multicolumn{1}{l|}{L} &
\multicolumn{1}{l|}{L} &
\multicolumn{1}{l|}{L} &
\multicolumn{1}{l|}{L} &
\multicolumn{1}{l|}{\underline H} &
\multicolumn{1}{l|}{\underline H} &
\multicolumn{1}{l|}{\underline H} &
\multicolumn{1}{l|}{L} &
\multicolumn{1}{l|}{L} &
\multicolumn{1}{l|}{L} &
\multicolumn{1}{l|}{\underline {H/M}} &
\multicolumn{1}{l|}{\underline {H/M}} &
\multicolumn{1}{l|}{L} &
\multicolumn{1}{l|}{L} &
{\underline H} \\ \hline
    \rowcolor{brown}
\multicolumn{1}{|l|}{NL-RelayA} &
  Relay Attack &
  \multicolumn{1}{l|}{L} &
\multicolumn{1}{l|}{L} &
\multicolumn{1}{l|}{L} &
\multicolumn{1}{l|}{L} &
\multicolumn{1}{l|}{L} &
\multicolumn{1}{l|}{L} &
\multicolumn{1}{l|}{L} &
\multicolumn{1}{l|}{L} &
\multicolumn{1}{l|}{L} &
\multicolumn{1}{l|}{L} &
\multicolumn{1}{l|}{\underline {H/M}} &
\multicolumn{1}{l|}{\underline {H/M}} &
\multicolumn{1}{l|}{\underline {H/M}} &
\multicolumn{1}{l|}{L} &
\multicolumn{1}{l|}{L} &
\multicolumn{1}{l|}{L} &
\multicolumn{1}{l|}{\underline {H/M}} &
\multicolumn{1}{l|}{\underline {H/M}} &
\multicolumn{1}{l|}{L} &
\multicolumn{1}{l|}{L} &
{\underline {H/M}} \\ \hline
    \rowcolor{brown}
\multicolumn{1}{|l|}{NL-F} &
  Fuzzing &
 \multicolumn{1}{l|}{L} &
\multicolumn{1}{l|}{L} &
\multicolumn{1}{l|}{L} &
\multicolumn{1}{l|}{L} &
\multicolumn{1}{l|}{L} &
\multicolumn{1}{l|}{L} &
\multicolumn{1}{l|}{L} &
\multicolumn{1}{l|}{L} &
\multicolumn{1}{l|}{L} &
\multicolumn{1}{l|}{L} &
\multicolumn{1}{l|}{\underline H} &
\multicolumn{1}{l|}{\underline H} &
\multicolumn{1}{l|}{\underline H} &
\multicolumn{1}{l|}{L} &
\multicolumn{1}{l|}{L} &
\multicolumn{1}{l|}{L} &
\multicolumn{1}{l|}{\underline H} &
\multicolumn{1}{l|}{\underline H} &
\multicolumn{1}{l|}{L} &
\multicolumn{1}{l|}{L} &
{\underline H} \\ \hline
  \rowcolor{teal}
\multicolumn{1}{|l|}{SRV-ID} &
  Server Attack &
  \multicolumn{21}{l|}{} \\ \hline
   \rowcolor{cyan}
\multicolumn{1}{|l|}{SRV-DL} &
  Data leakage &
  \multicolumn{1}{l|}{L} &
\multicolumn{1}{l|}{L} &
\multicolumn{1}{l|}{L} &
\multicolumn{1}{l|}{L} &
\multicolumn{1}{l|}{L} &
\multicolumn{1}{l|}{L} &
\multicolumn{1}{l|}{L} &
\multicolumn{1}{l|}{L} &
\multicolumn{1}{l|}{L} &
\multicolumn{1}{l|}{L} &
\multicolumn{1}{l|}{L} &
\multicolumn{1}{l|}{L} &
\multicolumn{1}{l|}{L} &
\multicolumn{1}{l|}{L} &
\multicolumn{1}{l|}{L} &
\multicolumn{1}{l|}{L} &
\multicolumn{1}{l|}{L} &
\multicolumn{1}{l|}{L} &
\multicolumn{1}{l|}{L} &
\multicolumn{1}{l|}{L} &
L \\ \hline
   \rowcolor{cyan}
\multicolumn{1}{|l|}{SRV-PIL} &
  Pilot identity leakage &
  \multicolumn{1}{l|}{L} &
\multicolumn{1}{l|}{L} &
\multicolumn{1}{l|}{L} &
\multicolumn{1}{l|}{L} &
\multicolumn{1}{l|}{L} &
\multicolumn{1}{l|}{L} &
\multicolumn{1}{l|}{L} &
\multicolumn{1}{l|}{L} &
\multicolumn{1}{l|}{L} &
\multicolumn{1}{l|}{L} &
\multicolumn{1}{l|}{L} &
\multicolumn{1}{l|}{L} &
\multicolumn{1}{l|}{L} &
\multicolumn{1}{l|}{L} &
\multicolumn{1}{l|}{L} &
\multicolumn{1}{l|}{L} &
\multicolumn{1}{l|}{L} &
\multicolumn{1}{l|}{L} &
\multicolumn{1}{l|}{L} &
\multicolumn{1}{l|}{L} &
L \\ \hline
   \rowcolor{cyan}
\multicolumn{1}{|l|}{SRV-LL} &
  Location leakage &
  \multicolumn{1}{l|}{L} &
\multicolumn{1}{l|}{L} &
\multicolumn{1}{l|}{L} &
\multicolumn{1}{l|}{L} &
\multicolumn{1}{l|}{L} &
\multicolumn{1}{l|}{L} &
\multicolumn{1}{l|}{L} &
\multicolumn{1}{l|}{L} &
\multicolumn{1}{l|}{M} &
\multicolumn{1}{l|}{L} &
\multicolumn{1}{l|}{\underline H} &
\multicolumn{1}{l|}{M} &
\multicolumn{1}{l|}{\underline H} &
\multicolumn{1}{l|}{L} &
\multicolumn{1}{l|}{L} &
\multicolumn{1}{l|}{L} &
\multicolumn{1}{l|}{\underline H} &
\multicolumn{1}{l|}{\underline H} &
\multicolumn{1}{l|}{L} &
\multicolumn{1}{l|}{L} &
M \\ \hline
\end{tabular}

\caption{\label{tab:Risk123} UAV Risk Assessment Matrix.}

\end{table}

\end{landscape}

\clearpage
\section{ \textbf{Proposed Mitigation Strategies}\label{ProposedMitigationStrategies}}

\begin{figure*}[t]
    \includegraphics[height=7cm, keepaspectratio]{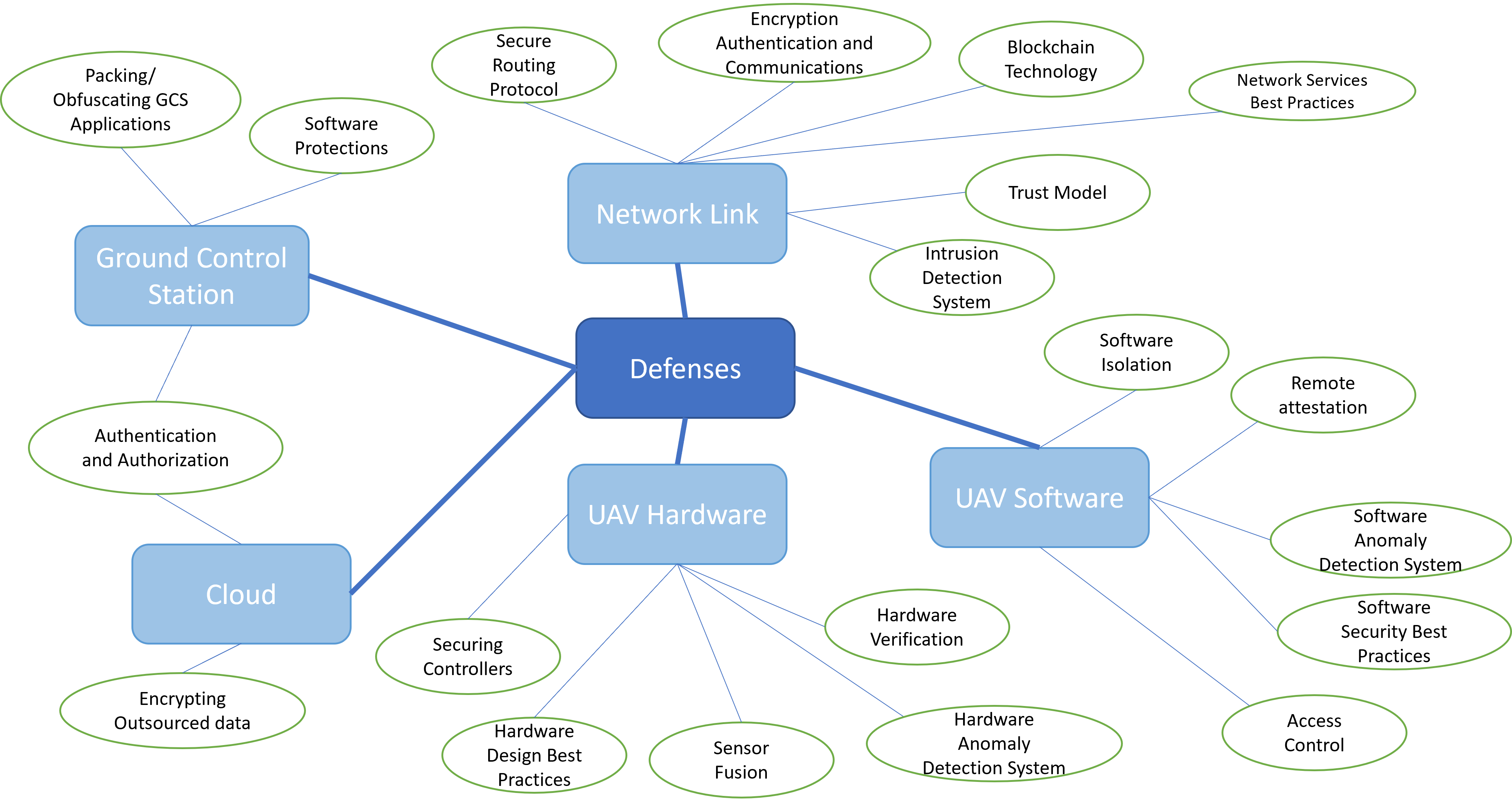}
    \centering
    \caption{Mitigation techniques for vulnerabilities across UAS components: Hardware, Software, Network Link, GCS, and Cloud.}
    \label{fig:mitigation}
\end{figure*} 

Following the approach used for UAS attacks earlier, we classify the proposed countermeasures by the UAS component they are designed to protect. Fig.~\ref{fig:mitigation} provides an overview of defense strategies, categorized by  UAS component.
 UAV mitigation are classified into five categories:  Hardware, Software, Network Link, Ground Control Station, and Cloud. 
It is important to acknowledge that certain measures may fall under multiple categories. For instance, an intrusion detection system (IDS) is a software designed to detect malicious activity and raise alerts when they occur. An IDS may identify hardware, software, network attacks, or a combination of these three based on scope and focus.
In this paper, we refer to hardware IDS as ``hardware anomaly detection system'' and software IDS as ``software anomaly detection system.''

To enhance the structure and clarify the mitigation strategies, we've included comprehensive three-tier trees to represent them. In the first level, we outline the overarching mitigation strategies. The second level lists the methods of implementing these strategies. Finally, the third level identifies the potential vulnerabilities that each strategy is designed to address providing a targeted approach to mitigating risks.

While the implementation of mitigation strategies can often reduce the likelihood of an attack, it is important to note that the impact of the attack may not always remain constant and can be influenced by various factors, such as the nature of the attack, the effectiveness of the mitigation measures, and the specific context in which the attack occurs. 
It is crucial to focus on reducing the likelihood of attacks and consider the potential impact in the specific context in which the UAV is being used. Risks for each threat should be analyzed, and if the risks are medium, medium/high, or high, then the corresponding countermeasures should be prioritized. 
In this section, we aim to address \textbf{RQ3}.

\subsection{Hardware Attack Mitigation \label{sec:hardwarevulnerabilitymitigation}}


UAV hardware mitigation entails protecting physical components, such as sensors, cameras, engines, and control systems from tampering. 
Hardware countermeasures are aimed at safeguarding the hardware components of a UAS, which include cameras, RF transceivers, and sensors for collecting environmental data, such as temperature, humidity, and air quality.  These hardware components can be categorized as controllers, actuators, and sensors. 
An UAV can be defined as a cyber-physical system (CPS). A CPS integrates software and hardware, with software controlling physical interaction with the environment.
In the CPS plant model described by Giraldo et al.~\cite{giraldo2018survey}, these components are important in the functioning of a feedback control system, which is relevant to UAVs in the following context. 
A controller receives input from one or more sensors and sends instructions to one or more actuators. The actuators are responsible for executing physical actions or processes, such as reducing rotor power. Sensors monitor the environment and provide feedback to the controller.

The behavior-based detection approaches for hardware anomaly detection systems (HADS) do not handle unknown events well, while the machine learning (ML)-based approaches, most of them being deep-neural network based, are expensive (both computation and storage). These need to be tuned for scale in UAV systems. A dynamic anomaly detection database, similar to the software common vulnerability enumeration (CVE) database, can enhance effectiveness.

Several robust approaches have been proposed in the literature to secure the controllers including control invariance based assessments~\cite{choi2018detecting}, reach-avoid mechanisms~\cite{huang2016controller} that restrict the controller's operations to well-known paths and detect deviations immediately. However, essential techniques such as the Hardware Trojan attack mitigation or the use of watermarking (prevalent in CPS systems) have not been studied in the UAV space. On the other hand, the approaches to sensor fusion attacks have been rudimentary, including interval based approaches or simple filtering approaches, such as Kalman filters (only two papers in this area). UAVs are being fitted with an increasing number of sensors, which widens the attack surface, requiring greater attention.

Verification of UAV hardware encompasses a wider scope than just verification of computing hardware. A single initiative focused on the remote attestation of UAV hardware to identify firmware updates~\cite{kohnhauser2017scapi}. Other initiatives concentrated on testing hardware verification related to structures, power, propulsion, sensors, and communication systems, yet they primarily addressed these aspects from a software/firmware perspective without including the hardware standpoint. 
This highlights a critical gap, particularly in addressing supply chain and hardware-targeted vulnerabilities.
Current best practices for hardware design, such as obfuscation and sandboxing, enhance hardware's security, but substantial work remains in this domain.

\subsubsection{\textbf{Hardware anomaly detection systems (HADS)} \label{sec:hardwareADS}}

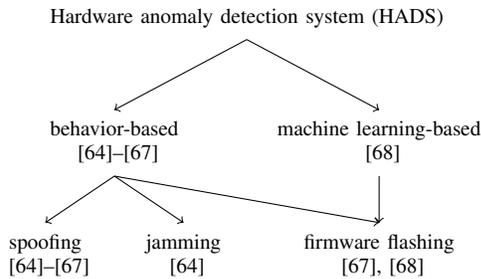
\begin{figure}[ht]
\centering
\begin{forest}
for tree={
    font=\footnotesize,
    parent anchor=south,
    child anchor=north,
    fit=band,
    s sep=5mm, l=15mm,
    align=center,
    edge={->},
    edge label/.append style={font=\scriptsize},
    if n children=0{tier=word}{},
}
[Hardware anomaly detection system (HADS)
    [behavior-based \\ \cite{guo2018roboads,elnaggar2018irl,mitchell2013adaptive,sharma2019briot},name=S1
     [spoofing \\ \cite{guo2018roboads,elnaggar2018irl,mitchell2013adaptive,sharma2019briot}]
        [jamming \\ \cite{guo2018roboads}]
    ]
    [machine learning-based \\ \cite{manesh2019detection},name=S2
        [firmware flashing \\ \cite{sharma2019briot,manesh2019detection},name=ff]
    ]
]
 \draw[->] (S1.south) -- (ff.north);
\end{forest}
  \caption{Hardware anomaly detection systems (HADS) can be implemented via behavior-based and machine learning-based techniques and are effective in dealing with spoofing, jamming, and firmware flashing based vulnerabilities in UAV hardware.}
  \label{fig:HADS12}
\end{figure}

Hardware anomaly detection systems (HADS) are used to detect anomalous behavior in hardware systems, which can indicate a security breach or a hardware failure. 
Based on our literature analysis, 
no knowledge-based detection mechanisms have been suggested. This is probably because of the limitations associated with such an approach, including the need to maintain an updated dictionary of attack patterns and the difficulty in recognizing new forms of attacks, like zero-day vulnerabilities.
Figure \ref{fig:HADS12} comprises three hierarchical levels. The first level, labeled as ``HADS'', can be subdivided into two approach classes:  behavior-based~\cite{guo2018roboads,elnaggar2018irl,mitchell2013adaptive,sharma2019briot} and machine learning-based~\cite{manesh2019detection}, as depicted in Level Two. 
Level Three of the figure highlights the efficacy of these approaches when it comes to addressing significant attacks, such as spoofing~\cite{guo2018roboads,elnaggar2018irl,mitchell2013adaptive,sharma2019briot}, jamming~\cite{guo2018roboads}, and firmware flashing~\cite{sharma2019briot,manesh2019detection}.

In the behavior-based class, HADS relies on a set of pre-defined rules or signatures to identify known anomalies. These rules can be based on simple threshold-based detection, statistical anomaly detection, or rule-based anomaly detection. The advantage of this approach is that it is easy to implement and the rule/signature-based approach tend to have low computational overhead, making it suitable for resource-constrained hardware devices. However, the behavior-based approach may struggle to detect unknown or novel anomalies that do not fit the pre-defined rules or signatures. Machine learning-based approaches use data-driven algorithms to learn the normal behavior of the hardware system and identify deviations from it. These approaches can detect both known and unknown anomalies and have the potential to adapt to changing environments. However, the machine learning approaches require significant computational resources, large amounts of training data, and have a higher risk of false positives or false negatives on account of limited data storage capabilities in the UAVs.

In their study, Guo et al.~\cite{guo2018roboads} showed how behavior-based intrusion detection can detect misbehavior in mobile robots' hardware components.
The authors introduced an anomaly detection system for identifying misbehavior in sensors and actuators. The primary objective was to pinpoint any deviations in sensor data received by the control units and any anomalies in the commands carried out by the robot actuators. 
To accomplish this, the authors engineered a state estimation algorithm and structured their anomaly detection algorithm to detect differences between estimated states and the inputs from actuators and sensors, labeling these disparities as misbehavior.
When applied to two mobile robots, this system successfully identified issues such as signal interference, sensor spoofing, logic bombs, and physical jamming attacks, with minimal detection delay.

Elnaggar and Bezzo~\cite{elnaggar2018irl} performed behavior-based detection using the Bayesian Inverse Reinforcement Learning technique for detecting sensor spoofing attacks.
Their technique utilized previous sensor readings and control inputs on the CPS to predict the goal of sensor spoofing attacks. Consequently, their method identified the compromised sensors that needed attention to reinstate the system's functionality.

Mitchell and Chen~\cite{mitchell2013adaptive} proposed a behavior rule-based intrusion detection system (BRUIDS) for securing sensors and actuators within a UAV. This system is based on behavior specifications for detecting intrusions. 
The authors created a set of rules that the UAV must follow. These rules are then transformed into a state machine that comprises both safe and unsafe states (i.e., attack states). To ensure compliance with the state machine, a neighboring UAV or remote node monitors the UAV to ensure compliance. During the testing phase, the authors adjusted variables to balance false positives and detection rates in the event of an attack. This approach was successful in enhancing the security of UAVs.
Sharma et al.~\cite{sharma2019briot} proposed a tool called `Behavior rule specification-based misbehavior detection for IoT embedded cyber-physical systems' (BRIoT), which uses behavior-specification-based detection to identify misbehavior in IoT devices. 
Users can define an operational profile for the device and generate a set of security requirements and behavioral rules. These rules are then converted into a state machine, which detects any misbehavior.

Manesh et al.~\cite{manesh2019detection} demonstrated that a UAS system is vulnerable to different cybersecurity attacks, such as GPS spoofing, where the attacker misleads a UAV by sending fake messages to the GPS receiver. To address this security challenge, the authors proposed an efficient method based on neural networks to detect GPS spoofing. This approach uses various features, including satellite number, carrier phase, pseudo-range, Doppler shift, and signal-to-noise ratio (SNR), to maximize the accuracy and probability of detection while minimizing false positives.

The absence of suggested knowledge-based hardware anomaly detection mechanisms could stem from their limitations 
in identifying new attack vectors, including zero-day attacks. This highlights the need for additional investigation into alternative techniques to efficiently identify and address emerging attack vectors. While various instances exist of behavior-based and behavior-specification-based detection methods, each has its own set of limitations. This highlights the requirement for more resilient and efficient detection techniques to successfully combat the wide range of cybersecurity threats UAS encounters. 
In summary, the gaps identified in the literature indicate the need for additional research to create anomaly detection mechanisms that effectively tackle the constantly evolving cybersecurity threats that UAS encounters.
Among the promising research areas are the utilization of built-in-self-test (BIST), signal integrity analysis, and power-on-self-test (POST) techniques.

{\bf Non-UAV Specific Solutions}
Minimum Volume Elliptical Principal Component Analysis (MVE-PCA) is a method that can be used to detect anomalies in hardware using multivariant data~\cite{7456266}. MVE-PCA could be used in UASs to monitor the hardware's health and identify if the hardware has been tampered with. 

\subsubsection{\textbf{Securing Controllers}}
\begin{figure}[H]
\vspace{-10pt}
\centering
\begin{forest}
for tree={
    font=\footnotesize,
    parent anchor=south,
    child anchor=north,
    fit=band,
    s sep=5mm, l=15mm,
    align=center,
    edge={->},
    edge label/.append style={font=\scriptsize},
    if n children=0{tier=word}{},
}
    [Securing controllers
    [using control \\ and flow \\ \cite{etigowni2018crystal,choi2018detecting},name=uc2 
        [,phantom]
         [jamming \\ \cite{etigowni2018crystal,demarinis2017toward},name=jm]
          [firmware \\ flashing \\ \cite{etigowni2018crystal}]
    ]
    [redundancy method \\ \cite{demarinis2017toward},name=rm
        [spoofing \\ \cite{etigowni2018crystal,choi2018detecting,demarinis2017toward,huang2016controller,DBLP:journals/corr/abs-1807-11553},name=ss]
    ]
    [reach avoid \\ stages \\ \cite{huang2016controller,DBLP:journals/corr/abs-1807-11553},name=ra
        [,phantom]
    ]
    ]
 \draw[->] (uc2.south) -- (ss.north);
 \draw[->] (ra.south) -- (ss.north);
 \draw[->] (rm.south) -- (jm.north);
\end{forest}
  \caption{Securing controllers effectively safeguards against jamming, spoofing, firmware flashing, and supply chain attacks. Securing controllers can be implemented via control and flow, redundancy, and `reach avoid stages.' Reach avoid stages allow a UAV to be programmed to reach certain target states while simultaneously avoiding specific undesirable conditions. }
  \label{fig:SecuringControllersTree}
  \vspace{-10pt}
\end{figure}
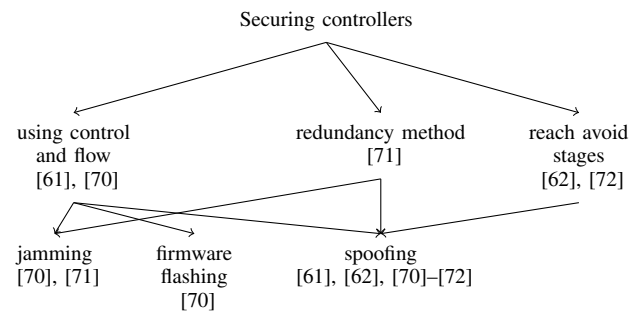

Securing controllers refers to the methods and techniques used to protect the control systems of various machines and devices. 
Research in hardware controller security has received significant attention.
Controllers are often the backbone of UAV systems, providing the necessary input and output controls for the machines to function. However, they can also be vulnerable to attacks that can compromise the system's integrity, safety, and confidentiality. 
Figure \ref{fig:SecuringControllersTree} is composed of a three-level structure. The first level, securing controllers can be further divided into four categories: 
using control and flow ~\cite{etigowni2018crystal,choi2018detecting},
redundancy method~\cite{demarinis2017toward},
reach avoid stages~\cite{huang2016controller,DBLP:journals/corr/abs-1807-11553}
as illustrated in the second level. These methodologies are useful in resolving vulnerabilities, such as 
spoofing~\cite{etigowni2018crystal,choi2018detecting,demarinis2017toward,huang2016controller,DBLP:journals/corr/abs-1807-11553},
jamming~\cite{etigowni2018crystal,demarinis2017toward}, and
firmware flashing~\cite{etigowni2018crystal}, as delineated in the third level of the figure.

Using control and flow~\cite{etigowni2018crystal,choi2018detecting} involves analyzing the control flow of the code to detect potential security vulnerabilities and prevent attacks. It involves verifying the code execution path and ensuring that it adheres to expected patterns. The goal is to prevent attackers from exploiting vulnerabilities in the code and disrupting its intended functionality. Redundant copies of the controller's components are created in the redundancy method~\cite{demarinis2017toward} to ensure that even if one component is compromised, the system can continue functioning properly. This can be achieved through hardware redundancy, such as having multiple processors or sensors, or through software redundancy, such as having multiple copies of the code running in parallel. 
Reach avoid stages~\cite{huang2016controller,DBLP:journals/corr/abs-1807-11553} creates a set of rules that describe the controller's expected behavior and environment. 
It involves actively working towards specific objectives (`reach') while avoiding potential obstacles or pitfalls (`avoid') using a list/rule of `Dos' and `Don'ts.'
The controller's actions are then compared to these rules, and if it deviates from them, the system is flagged as potentially compromised. This approach is particularly useful for detecting UAV attacks involving the controller trying to achieve a goal outside of its expected behavior. 

Etigowni et al.~\cite{etigowni2018crystal} presented a control flow predictor to verify the safety of UAV controllers. Their method involved using a control flow predictor, which monitors the execution state of the flight controller. If the UAV approaches an unsafe state, the control flow predictor deploys pilot-designed countermeasures.
The researchers utilized a data-driven model that employed Kalman Filters to predict future states. They tested their control flow predictor by carrying out a series of controller-based attacks using malware that involved injecting or modifying controller data. In all test cases, the UAV remained safe and alerted the operator of any safety violations.

Choi et al.~\cite{choi2018detecting} introduced Control Invariants (CI) as part of a framework for detecting external physical attacks on robotic vehicles, such as UAVs. Unlike traditional program-based invariants, CI focus on the control and physical dynamics of the vehicle, combining physical characteristics, control algorithms, and physics laws. The CI methodology employs a control system engineering technique known as System Identification (SI), which uses a control invariant template and extensive vehicle measurement data to fine-tune the template's coefficients, ensuring the derived equations closely match the measurement data. These equations predict the vehicle's behavior based on its inputs and states during operation. The CI framework includes a checking code into the vehicle's control program binary, which, during runtime, compares the actual system state against the expected state calculated by the CI equations. An alarm is triggered if any discrepancies are found. Additionally, DeMarinis et al.~\cite{demarinis2017toward} suggested incorporating a redundancy board in the flight controller that allows automatic switching between two firmware versions if the primary controller fails, thus enhancing resilience to attacks in case of a controller compromise.

Reach-avoid problems involve guiding an operating system towards achieving desired configurations while concurrently steering clear of those considered undesirable~\cite{huang2016controller,DBLP:journals/corr/abs-1807-11553}. In their paper, Huang et al.~\cite{huang2016controller} proposed a controller synthesis algorithm capable of solving the ``reach-avoid'' problem in the presence of adversaries. Their approach involved formulating sensor, actuator, and controller attacks to synthesize a secure controller and used a `Satisfiability Modulo Theories' solver~\cite{huang2016controller}.
Attacks that were considered included compromising partial controller software, injecting packets, tampering with actuator signals, and spoofing sensors. 
The developed controller followed a behavioral model to maintain safety in an adverse environment.

{\bf Non-UAV Specific Solutions}
Hardware Trojans~\cite{bellay2021hardware, moein2017hardware} are implemented by adversaries by modifying integrated circuits (ICs) or other semiconductors so that devices show abnormal behavior. These hardware Trojans are hard to detect as they can remain dormant and could be activated under very rare conditions~\cite{pan2022survey}. These attacks generally focus on power and delay side channel signals~\cite{li2008speed}.  
Hardware Trojans in electric circuits can be detected using cell analysis and routing analysis.

Dynamic watermarking can be used to secure CPS control. This technique involves having actuators inject private data into a CPS and observing the response of the sensors connected to the CPS. The CPS can detect threats to sensors and/or actuators by monitoring how private data are handled and read.
Satchidanandan and Kumar~\cite{satchidanandan2016dynamic} explored dynamic watermarking as pattern injection into a medium to detect anomalies in sensors and actuators. The authors proposed using dynamic watermarking to orchestrate secure control over a physical plant by detecting compromised sensors and actuators.
These methods can be adapted for use in UAV systems despite being originally employed for systems other than UAVs.

\subsubsection{\textbf{Sensor Fusion}}
\begin{figure}[ht]
\vspace{-10pt}
\centering
\begin{forest}
for tree={
    font=\footnotesize,
    parent anchor=south,
    child anchor=north,
    fit=band,
    s sep=5mm, l=15mm,
    align=center,
    edge={->},
    edge label/.append style={font=\scriptsize},
    if n children=0{tier=word}{},
}
    [Sensor fusion
    [introducing intervals \\ \cite{ivanov2016attack},name=ii
        [spoofing \\ \cite{ivanov2016attack}]
        [jamming \\ \cite{ivanov2016attack}]
        [supply chain attack \\ \cite{ivanov2016attack}]
    ]
    [filtering approach \\ \cite{nashimoto2018sensor},name=if
        [firmware flashing \\ \cite{nashimoto2018sensor}]
    ]
    ]
\end{forest} 
  \caption{Sensor fusion technique can be deployed by implementing intervals to mitigate vulnerabilities such as spoofing, jamming, and supply chain attacks, or by filtering approach to counteract UAV firmware flashing.}
  \label{fig:SensorFusion12}
  \vspace{-10pt}
\end{figure}
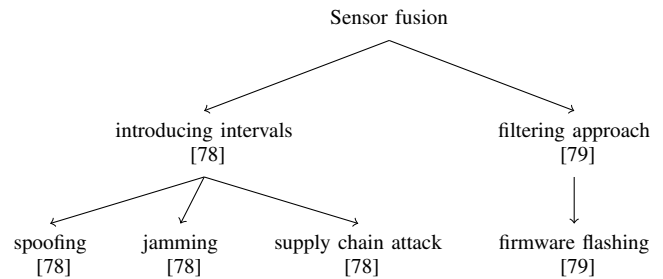

\FloatBarrier

Sensor fusion is a hardware-based countermeasure that uses data from multiple sensors to validate each other and detect attacks.
Figure~\ref{fig:SensorFusion12} illustrates a tree consisting of three levels. The first level, sensor fusion, can be divided into two categories: 
introducing intervals~\cite{ivanov2016attack} and filtering approach~\cite{nashimoto2018sensor}
as shown in the second level. These two techniques are adept at mitigating 
spoofing~\cite{ivanov2016attack},
jamming~\cite{ivanov2016attack},
supply chain attack~\cite{ivanov2016attack}, and
firmware flashing~\cite{nashimoto2018sensor}, respectively,
as outlined in the third level of the figure.

Introducing intervals involves setting up an interval timer that triggers the sensors to take measurements at regular intervals. The measurements are then combined to provide a more accurate picture of the environment being sensed. The filtering approach involves using algorithms to combine and filter sensor data to remove noise, errors, and inconsistencies. Filters, such as a Kalman Filter, can help smooth out the data and produce more accurate results. 

Ivanov et al.~\cite{ivanov2016attack} proposed an algorithm that uses sensor fusion to enhance the resilience of safety-critical cyber-physical systems against attacks.
The study developed a model to show how altering sensor data can impact the accuracy of sensor fusion algorithms. It demonstrated that introducing delays/intervals between sensor communications affects an attacker's ability to disrupt these systems.
The authors improved the accuracy of their sensor fusion algorithm by integrating communication schedules for sensors and utilizing historical sensor readings for improving accuracy. By implementing this approach onto a ground robot, they demonstrated that the impact of faulty sensors was reduced significantly.

Furthermore, Nashimoto et al.~\cite{nashimoto2018sensor} outlined attacks capable of tricking sensor fusion algorithms that employ Kalman Filters by manipulating specific sensors to influence the output of the sensor fusion algorithm. Through experimental analysis, the researchers demonstrated how these attacks could allow attackers to gain partial or full control over the sensor fusion algorithm results.
Surrounding the micro-electro-mechanical system (MEMS) with sound isolation material, the author presented countermeasures for attacks. 
Gravity and geomagnetic sensors' measurement errors can be leveraged to detect attacks.

{\bf Non-UAV Specific Solutions}
Yeong et al.~\cite{yeong2021sensor} presented a complete overview of the perception block in autonomous driving (AD) systems to understand and perceive the surrounding environment. These methodologies highlight the importance of sensor calibration, covering intrinsic, extrinsic, and temporal calibration, and highlight existing open-source calibration packages along with the appropriate use of cameras, LiDAR, and radar. 
The implementation of sensor fusion techniques in non-UAV devices can also be applied to UAVs to mitigate hardware attack impact.

Using multiple sensors provides significant advantages over using a single sensor~\cite{mitchell2007multi} as it provides rich semantics, higher resolution, and improved accuracy. Generally, different algorithms are utilized for different levels of fusion. Some of the algorithms that are commonly used in non-UAV domains that can also be implemented in UAV domain are~\cite{luo2011multisensor}: Kalman Filtering, support vector machine (SVM), Bayesian inference technique, sequential Monte Carlo methods (Particle filter), Dempster-Shafer theory of evidence, K-means clustering, artificial neural networks (ANN), and fuzzy logic. These could be explored more not only for hardware attack mitigation, but beyond. 

\subsubsection{\textbf{Hardware Verification}}
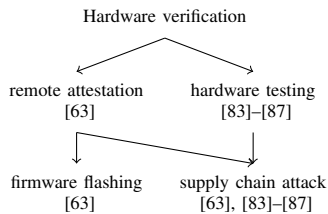
\begin{figure}[ht]
\vspace{-10pt}
\centering
\begin{forest}for tree={
    font=\scriptsize,
    parent anchor=south,
    child anchor=north,
    fit=band,
     align=center,
    edge={->}
    }
    [Hardware verification
        [remote attestation \\ \cite{kohnhauser2017scapi}, name =ra
             [firmware flashing \\ \cite{kohnhauser2017scapi},name=ff]
        ]
        [hardware testing \\ \cite{sullivan2006structural,dantsker2019propulsion,yang2022multi,lombardi2010automated,prabowo2021design},name=ht
            [supply chain attack \\ \cite{kohnhauser2017scapi,sullivan2006structural,dantsker2019propulsion,yang2022multi,lombardi2010automated,prabowo2021design},name=sca]
        ]
    ]
     \draw[->] (ra.south) -- (sca.north);
\end{forest} 
  \caption{Remote attestation and hardware testing are two ways hardware verification can be implemented. These methods can be utilized to assess UAV hardware integrity and are effective to protect against firmware flashing and supply chain attack.} 
  \label{fig:HardVerification123}
  \vspace{-10pt}
\end{figure}

Hardware verification refers to the process of verifying that all hardware components are configured correctly according to pre-defined baselines.
Hardware verification technique as shown in Figure~\ref{fig:HardVerification123} can be divided into
remote attestation \cite{kohnhauser2017scapi} and
hardware testing \cite{sullivan2006structural,dantsker2019propulsion,yang2022multi,lombardi2010automated,prabowo2021design} and can address vulnerabilities such as 
firmware flashing \cite{kohnhauser2017scapi} and supply chain attack \cite{kohnhauser2017scapi,sullivan2006structural,dantsker2019propulsion,yang2022multi,lombardi2010automated,prabowo2021design} as shown in the third level.

Hardware verification is a technique used to detect and recover compromised hardware components. Remote attestation enables UAV operators to remotely verify a device's hardware or software configuration, allowing them to detect any unauthorized changes in hardware and software environments.
Kohnhäuser et al.~\cite{kohnhauser2017scapi} proposed a protocol for remote attestation to detect compromised hardware and software configurations that is resilient to noisy and dynamic networks.
Some of the common hardware verification tests for UAVs include structural testing \cite{sullivan2006structural}, propulsion system testing \cite{dantsker2019propulsion}, power system testing, sensor testing, and communication system testing \cite{yang2022multi,lombardi2010automated,prabowo2021design}. 

Structural testing is carried out to verify that the airframe (mechanical structure of an aircraft) and other structural components of the UAV can withstand the stresses and loads of flight. Propulsion system testing involves testing the motors, propellers, and other components to ensure they function correctly and provide the required thrust. Power system testing includes testing the battery, fuel system, and other power supply components to ensure that they provide sufficient and reliable power to the UAV. Sensor testing verifies that sensors, such as cameras, GPS, RADAR, LiDAR, and other sensors provide accurate and reliable data. Communication system testing checks all communication systems, such as satellite radio, WiFi, and cellular devices to ensure secure communication between the UAV and GCS. These tests are crucial in identifying and addressing any potential issues before the UAV is put into operation, reducing the risk of accidents and ensuring optimal performance.

{\bf Non-UAV Specific Solutions}
Generally, hardware vulnerabilities are undetected by software countermeasures \cite{electronics6030052}. The European CIPSEC (Enhancing Critical Infrastructure Protection with Innovative SECurity Framework) project offers a security framework that tests for vulnerabilities and recommends key personnel training courses, forensics analysis, standardization, and protection against malware and other adversaries using firewalls, IDS, and other security tools \cite{cipsec}. The framework can be used to assess vulnerability in a UAV aviation system identifying potential weaknesses. This framework can also be leveraged to help in standardization of security protocols to ensure high-level security. These standards can be used in the UAV domain to protect infrastructure need for its deployment.

Memory disturbance attack or the RowHammer attack \cite{electronics6030052, polychronou2021securing, li2019detecting} is a backdoor attack that can occur in Dynamic random-access memory (DRAM) chips. 
Hardware Error Correction Code (ECC) helps detect Rowhammer attacks by identifying and correcting unusual single-bit errors.
These attacks could be mitigated by enforcing an ECC or increasing DRAM's refresh rate \cite{kim2014flipping}. Since ECC cannot detect fault if multiple bits flip, Kim et al. proposed probabilistic adjacency row activation (PARA) mechanism that refreshes the adjacent row of the accessed row~\cite{kim2014flipping} or Ghasempour et al.~\cite{ghasempour2015armor}'s proposed method of using catch buffer to activate rows frequently to help detect RowHammer attack. DDR4 memory, although slightly expensive, includes Target Row Refresh (TTR), which tracks row activation frequency to refresh vulnerable rows, making it harder for adversaries to implement the RowHammer attack \cite{versen2020row}. 
RowHammer can also be detected by observing four events: cache references, cache misses, branch instructions retired, and branch mispredictions. These data points can be trained using machine learning classifiers, such as logistic regression, support vector machine, or artificial neural network to detect such attacks~\cite{li2019detecting}.

Sensitive data can also be leaked via side-channel attacks by analyzing hardware characteristics such as power dissipation, computation time, and electromagnetic emission. An attacker can analyze these characteristics of a UAV and deduce data and cryptographic keys~\cite{electronics6030052, polychronou2021securing}. To prevent adversaries from obtaining side channel data, the time taken for code execution can be made constant or random without considering the actual time taken by the processor \cite{ge2018survey}. Hardware virtualization can be used to create isolation, making it difficult for external interference-related attacks to be executed~\cite{electronics6030052}. 

Hardware probing is a side-channel attack technique where an attacker physically accesses hardware components to gather information about devices such as a microchip. Probing attack includes decapsulation to expose the silicon die followed by reverse engineering of the device to extract sensitive information such as encryption keys or data \cite{cad,bellay2021hardware}. As stated in \cite{ishai2003private}, private circuits could be used to secure a system even if adversaries observed certain internal bits during computation.
These hardware verification techniques employed in non-UAV systems can be effectively adapted for use in UAVs, enhancing their overall security. 

\subsubsection{\textbf{Hardware design best practices}}

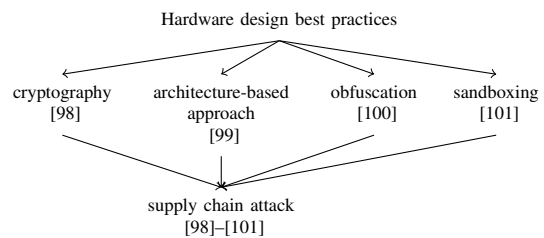
\begin{figure}[H]
\vspace{-10pt}
\centering
\begin{forest}for tree={
font=\scriptsize,
    parent anchor=south,
    child anchor=north,
    fit=band,
    align=center,
    edge={->}
    }
    [Hardware design best practices
    [cryptography \\ \cite{shoufan2015secure},name=cry]
    [architecture-based \\ approach \\ \cite{gomes2017interconnected},name=arc
        [supply chain attack \\ \cite{shoufan2015secure,gomes2017interconnected,quadir2016survey,mead2016defeating},name=sca]
    ]
    [obfuscation \\ \cite{quadir2016survey},name=obf]
    [sandboxing \\ \cite{mead2016defeating},name=sand]
    ]
     \draw[->] (cry.south) -- (sca.north);
     \draw[->] (obf.south) -- (sca.north);
     \draw[->] (sand.south) -- (sca.north);
\end{forest} 
  \caption{Cryptography, architecture-based approach, obfuscation, and sandboxing are essential best practices in UAV hardware design that can help mitigate potential vulnerabilities within the hardware supply chain.}
  \label{fig:HardwareDesignBestPractices1231}
  \vspace{-10pt}
\end{figure}

Hardware design best practices refer to the recommended techniques and methods used to enhance the security of hardware systems. 
These best practices for UAV hardware components design ought to prioritize security measures--security at the cradle.
We have identified four ways best design practices could be followed to reduce the impact of UAV attacks and improve the security of hardware systems as shown in Figure~\ref{fig:HardwareDesignBestPractices1231}, a three-level structure.
These practices can be achieved via four approaches: cryptography~\cite{shoufan2015secure}, architecture-based approach \cite{gomes2017interconnected}, obfuscation \cite{quadir2016survey}, and sandboxing \cite{mead2016defeating} as shown in Level Two of the figure. These can be used to handle the supply chain attack \cite{shoufan2015secure,gomes2017interconnected,quadir2016survey,mead2016defeating}, as indicated in the third level of the figure.
Cryptography secures data through encryption in hardware devices, architecture-based approach ensures secure hardware architecture, obfuscation makes hardware designs complex to counteract tampering, and sandboxing isolates hardware components to isolate potential breaches and deter supply chain hardware attacks. In what follows we discuss each of them in detail.

Cryptography \cite{shoufan2015secure} involves using mathematical algorithms and protocols to secure the communication and data between different hardware components. It can be used for data encryption, authentication, and access control tasks. The architecture-based approach~\cite{gomes2017interconnected} is used to design hardware systems with security in mind from the beginning, rather than adding it as an afterthought. This approach involves implementing security features into the hardware architecture itself to prevent potential attacks. Obfuscation \cite{quadir2016survey} is a technique that makes it difficult for attackers to reverse engineer or understand the hardware design and functions. This can include methods such as hiding data and functions and creating random or complex code structures. Sandboxing \cite{mead2016defeating} is a method used to isolate and protect different hardware components from each other. This can include physically separating the components or using virtualization techniques to create secure zones within the hardware system.

Abdulhadi et al. \cite{shoufan2015secure} proposed a lightweight hardware mechanism to safeguard the privacy and integrity of the data transmitted between the GCS and the UAV.
The security measures for the UAV were implemented using a Field Programmable Gate Array (FPGA) module with the cryptographic engine as a central component of the architecture. This engine is pivotal in the FPGA's design for executing security protocols. It stores the authentication and encryption keys necessary for Advanced Encryption Standard (AES) operations within FPGA's registers when setting up the UAV, ensuring these keys are shielded from potential breaches. The design allows only the cryptographic engine to read the keys, preventing any external read access.

Gomes et al.~\cite{gomes2017interconnected} proposed an interconnected system architecture to maintain UAS integrity. The authors explained that each UAS was composed of eleven interdependent systems within its architecture. These systems encompass communications, sensing, weather reports, power, maintenance, and diagnostics systems, flight management systems, control units, path planning systems, and emergency response. Their architecture was equipped with sensors to detect potential threats and trigger an emergency response system when an alarm is raised.
 
Hardware security also involves protecting devices from reverse-engineering and hardware-specific attacks.
Quadir et al.~\cite{quadir2016survey} explored methods to safeguard chips, boards, and firmware from reverse engineering. 
These include obfuscating/camouflaging designs, using external keys to prevent piracy, utilizing unmarked chips, and creating tamper-proof fittings. 
Mead et al.~\cite{mead2016defeating} used hardware-based sandboxing to secure UAV hardware by isolating non-trusted system-on-chip components. To prevent RF-based jamming attacks, they developed an approach that involved passing input signals through a property checker that verified the legitimacy of the signals received by the RF receiver. By conducting simulated testing, they demonstrated that their design was capable of detecting and thwarting such attacks.

{\bf Non-UAV Specific Solutions}
Devices such as UAVs need to be compact. All the necessary circuits and parts are often designed and built into a system-on-chip (SoC) to get this compactness. 
Incorrect or ambiguous security specifications and flawed design~\cite{dessouky2019hardfails} can make hardware vulnerable. Trusted execution environments (TEEs) like Intel's Software Guard Extension (SGX) could be used in critical applications to prevent potentially dangerous software execution \cite{dessouky2019hardfails, fei2021security} caused by underlying hardware vulnerabilities. 

Secure virtual architecture (SVA) have been used to abstract unnecessary processor details to achieve memory safety, control-flow integrity, and type safety~\cite{dharsee2017software}. If any vulnerabilities are detected, then the manufacturers can create and release a software patch to SVA to mitigate the new bugs \cite{dharsee2017software}. These techniques can be used in the UAV domain to mitigate hardware vulnerabilities via software updates. 

\subsubsection{Research Gaps}
The literature surveyed points to a lack of knowledge-based UAV hardware anomaly detection systems, which is essential to address the rapidly evolving advanced persistent threats (ATPs) landscape. ML is being widely used for anomaly detection, but for UAVs they need to be space and compute efficient. 
For hardware resilience other promising approaches include the creation of built-in-self-test (BIST), signal integrity analysis, and power-on-self-test (POST) techniques, widely used in CPS systems. Further, there is a need for studying hardware Trojans and watermarking in general in UAV hardware, not only in the processors/microcontrollers, but also in the hardware sensors. 
Sensor fusion approaches need to be improved beyond the simple approaches proposed. Approaches to consider include support vector machine, Dempster-Shafer~\cite{shafer1992dempster} theory of evidence, K-means clustering, and neural networks.

These days, software attacks targeting hardware vulnerabilities (SATHVs) have increased significantly, which could be detected using hardware performance counters (HPCs)~\cite{polychronou2021securing}. These registers are used to count special hardware events that could be used to detect attacks on UAV hardware.
%
Trusted Platform Modules (TPMs) and trusted execution environments (TEEs) have been used widely for the preservation of sensitive data, ensuring secure key storage, robust encryption mechanisms, and the facilitation of remote attestation. 
There is a need to consider them in UASes, with an eye on the increased  complexity of UAV design and architecture. The aim is to have seamless integration of TPMs with UAV OS and applications with limited overheads. 
Fault injection and reverse engineering techniques, such as lowering SNR, masking (binding), emission filtering, and shielding can be investigated in UAVs to reduce chip emissions so that it becomes difficult for attackers to extract information \cite{moein2017hardware}. 


Hardware vulnerabilities may occur due to a defect in production or when vulnerabilities are not caught before being released from the supply chain, which makes an unprotected supply chain a contributing hardware vulnerability factor~\cite{bellay2021hardware, pan2022survey}. The potential for counterfeits and Trojans in low-cost replacements are an important challenge to address on the hardware side~\cite{hoque2018hardware}. The uniqueness of the UAS ecosystem with the many players make it a unique challenge that needs to be studied. Further, procedures need to developed to securely decommission and dispose of UAV hardware to ensure that no sensitive information remains accessible. 

\subsection{Software Attack Mitigation  \label{sec:softwarevulnerabilitymitigation}}

UAV Software Mitigation includes the software that controls the UAV's flight and other functions. Defenses in this area include intrusion detection, authentication, and access control to prevent unauthorized access and manipulation.
This section will discuss ways to safeguard the UAS software and firmware, including the applications, processes, and the underlying operating system(s), and identify the research gaps.
\begin{figure*}[!ht]
\vspace{-15pt}
\centering
\begin{forest}for tree={
    font=\scriptsize,
    parent anchor=south,
    child anchor=north,
    fit=band,
    align=center,
    edge={->}
    }
    [Software anomaly detection systems (SADS)
        [behavior-based \\ \cite{stracquodaine2016unmanned}, name=bb[
            [database injection \\ \cite{stracquodaine2016unmanned}]
            [firmware modification \\ \cite{stracquodaine2016unmanned}]
            [code injection \\ \cite{stracquodaine2016unmanned,vuong2015performance,lu2019data}, name=ci]
            ]
            ]
        [data-driven mechanism\\ (machine learning/ data mining)\\ \cite{vuong2015performance,lu2019data}, name=ml[
            [buffer overflow \\ \cite{stracquodaine2016unmanned,vuong2015performance,lu2019data}, name=bo]
            [malware infection \\ \cite{stracquodaine2016unmanned,vuong2015performance,lu2019data}, name = mi]
            ]
    ]
    ]
     \draw[->] (ml.south) -- (ci.north);
     \draw[->] (bb.south) -- (bo);
     \draw[->] (bb.south) -- (mi.north);
\end{forest} 
  \caption{Software anomaly detection system (SADS) can be categorized as a behavior-based and data-driven mechanism. The behavior-based method tackles vulnerabilities such as database injection, firmware modification, code injection, buffer overflow, and malware infection. The data driven approaches are good at addressing code injection, buffer overflow, and malware infection.}
  \label{fig:SADS234}
    \vspace{-15pt}
\end{figure*}
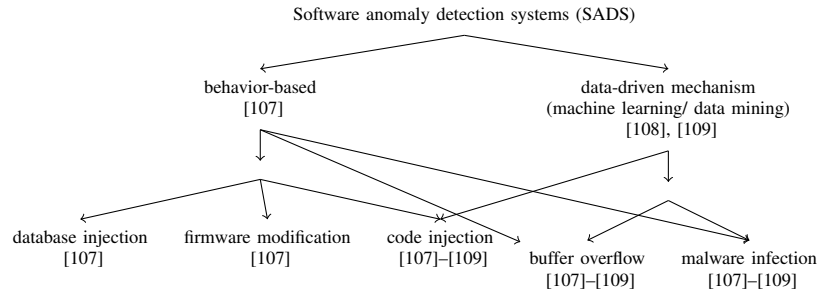

\subsubsection{\textbf{Software Anomaly Detection Systems (SADS)}  \label{sec:softwareADS}}
Software anomaly detection systems (SADS) are designed to detect abnormal behavior in software systems. 
The literature contains fewer proposals for detecting anomalies/intrusions targeting the software applications, operating systems, or firmware of UAS compared to HADSes that protect the hardware components and the UAS's network IDS that protect the network.
More emphasis is placed on the UAS's capture and processing of data, such as RF signals or network packets, rather than the software running on it. 
The software-based ADSes proposed in the literature lack knowledge-based and behavior-specification-based detection mechanisms.
The literature lacks knowledge-based detection due to lack of mechanisms to identify new attack vectors and a regularly updated attack dictionary. Additionally, the main drawback of behavior-based detection mechanisms is that they require significant effort and expert knowledge to create a specification for the entity being detected.

Figure \ref{fig:SADS234} is a three level hierarchical structure that classifies the types of SADS. Proposed SADes may be classified into two categories: behavioural-based and data-driven mechanism (machine learning and data mining) as shown in the second level. Approaches in both categories can be used to mitigate vulnerabilities such as database injection, firmware modification, code injection, buffer overflow, and malware infection as shown in the third level of the figure. 
Similar to that in hardware mitigation, behavior-based \cite{stracquodaine2016unmanned} approaches rely on defining a set of rules or signatures that describe normal behavior in a system. Any deviation from these rules is flagged as an anomaly. 
However, they may not be as effective in detecting previously unseen or complex anomalies. 
Machine learning approaches \cite{vuong2015performance,lu2019data} learn the normal behavior of a system from a large dataset of training examples. The solutions can identify deviations from the learned behavior as anomalies. 
Data mining techniques are used in extracting valuable information and patterns from large datasets using statistics, machine learning, and database systems. 
These approaches can be effective in detecting complex or previously unseen anomalies, but require more computational resources and may require a larger dataset for training, similar to HADS.

Stracquodaine et al.~\cite{stracquodaine2016unmanned} proposed a behavior-based SADS for detecting hardware failure, communication channel corruption, sensor spoofing, and malware on the UAS.
The SADS monitored the control flow of the UAS operating system and autopilot software. The data was then fed into an event processor that compared it to a normal profile (derived offline) to detect anomalies in real-time. 
The authors used simulations to showcase their effectiveness.
    
Vuong et al.~\cite{vuong2015performance} developed a SADS to detect DoS, command injection, and malware threats for   robotic vehicles. Their method involved using the system's existing processes to gather logs detailing both the cyber and physical activities of the device. These logs were analyzed to extract features, which served as training data for a streamlined decision tree algorithm. The lightweight decision tree algorithm enabled their system to locate attacks accurately with minimal latency.

\begin{figure*}[b]
\vspace{-15pt}
\centering
\begin{forest}
for tree={
    font=\scriptsize,
    parent anchor=south,
    child anchor=north,
    fit=band,
    align=center,
    edge={->}
    }
    [Remote attestation
        [cryptography \\ \cite{kohnhauser2017scapi}, name=cry
            [firmware modification \\ \cite{kohnhauser2017scapi}]
            [supply chain attack \\ \cite{kohnhauser2017scapi}]
            [malware infection \\ \cite{kohnhauser2017scapi,ambrosin2017toward, asokan2015seda}, name =mi]
        ]
        [knowledge/model based \\ \cite{ambrosin2017toward, asokan2015seda}, name=kmb
            [code injection \\ \cite{kohnhauser2017scapi,ambrosin2017toward, asokan2015seda}, name=ci]
            [database injection \\ \cite{ambrosin2017toward, asokan2015seda}]
            [buffer overflow \\ \cite{ambrosin2017toward, asokan2015seda}]
        ]
    ]
     \draw[->] (cry.south) -- (ci.north);
     \draw[->] (kmb.south) -- (mi.north);
\end{forest} 
  \caption{Remote attestation can be used to mitigate software vulnerabilities, such as firmware modification, supply chain attack, malware infection, code injection, database injection, and buffer overflow using cryptography and knowledge/model-based techniques. }
  \label{fig:RemoteAttestation1324}
  \vspace{-15pt}
\end{figure*}
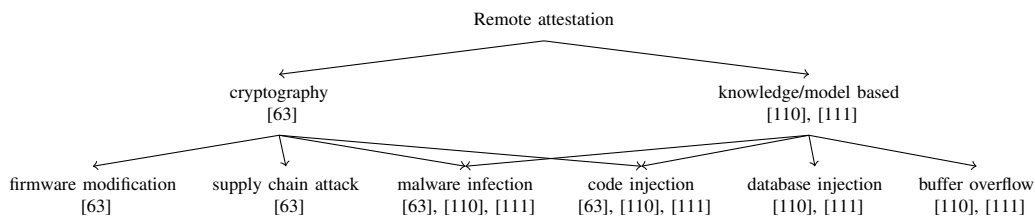
    
Lu et al.~\cite{lu2019data} proposed a ``normal timing method'', which is a timing data driven malware detection approach using several anomaly detection techniques, such as range-based, distance-based, and support vector machine classification. The system behavior model divided the system timing into smaller parts instead of using a single timing model. Furthermore, the data-driven malware detection employed a comprehensive threat model that included a wide range of real-world malware attacks. This approach utilized hardware detectors for efficient and non-intrusive range and distance-based detection, as well as SVM anomaly detection.
The authors discovered that using a sub-component timing model improved detection accuracy compared to the lumped timing model. 
This study is a good guide for designers to evaluate tradeoffs between detection accuracy, hardware area, and power consumption based on specific application needs.

{\bf Non-UAV Specific Solutions}
There are several tools that can help to detect anomalies/errors in programs and identify the cause. Dynamic Invariant Detection $\cup$ Checking Engine (DIDUCE)~\cite{hangal2002tracking} is a tool that can distinguish the behavior between a successful and failing program. In addition to unit testing, DIDUCE can also be used to test code with inputs that lead to unknown outputs. This tool is effective in finding bugs that result from algorithmic errors while handling wrong input or rare combination of values (corner cases). 
There are several other tools that can be used to detect software vulnerabilities, such as: static application security testing (SAST), dynamic application security testing (DAST), interactive application security testing (IAST), and software composition analysis (SCA) \cite{mateo2020combining,morga2023integration}. 
These tools can be used to detect anomalies in UAV software. 

UAS enterprises may incentivize bug bounties~\cite{votipka2018hackers, walshe2020empirical, hata2017understanding}  to encourage researchers to report product vulnerabilities. These programs can be an effective way to identify and fix software vulnerabilities and help build relationships with the security research community. Prior research illustrated that the participation rate increased if the payments were based on the criticality of vulnerability~\cite{finifter2013empirical}. 

\subsubsection{\textbf{Remote Attestation}}
Remote attestation is a security mechanism used to ensure that a remote device or system is running trusted software and has not been compromised by malicious actors. It involves verifying the integrity of a remote system's software components by exchanging cryptographic messages between the remote system and a trusted verifier. 
There are two main approaches to remote attestation: cryptography and knowledge/model-based, as shown in the second level of Figure \ref{fig:RemoteAttestation1324}. The third level of the figure shows the specific vulnerabilities each method can address. 
The cryptography-based \cite{kohnhauser2017scapi} approach uses digital signatures and other cryptographic techniques to verify the integrity of the software components on the remote system. This approach relies on using secure keys and exchanging signed messages between the remote system and the verifier. On the other hand, the knowledge/model-based \cite{ambrosin2017toward, asokan2015seda} approach relies on pre-defined models or knowledge about the system's expected behavior to detect anomalies or deviations from normal behavior. This approach involves comparing the system's current state to the expected state based on the pre-defined model or knowledge to check for discrepancies. 

Malware can be used to launch software-level attacks on a UAS and can acquire higher privilege levels to accomplish malicious objectives. 
Verifying the software configuration of a remote UAS node can help prevent software-level attacks and detect if the node is compromised during operation. Remote attestation is a method that allows a trusted source node to verify the configuration of a remote node.

Kohnhäuser et al.~\cite{kohnhauser2017scapi} proposed a remote attestation protocol to detect software and hardware compromises in embedded devices such as UAV. The protocol uses a challenge-response mechanism to verify the integrity of software and hardware components, ensuring that they have not been tampered with. The protocol is also designed to be flexible, allowing it to be adaptive to different hardware and software configurations. This method can be used to detect DoS attacks.

Ambrosin et al.~\cite{ambrosin2017toward} proposed a secure collective remote attestation protocol tailored for the dynamic UAV swarms to ensure the all UAV configurations are up-to-date. In this protocol a UAV initiates a local attestation to verify if its operational software matches a known good/optimal configuration. Subsequently, the UAV shares its attestation results and expands its network knowledge by engaging in a consensus algorithm. To ensure the authenticity of consensus messages, each UAV includes a Trusted Execution Environment (TEE). This TEE maintains the network's status and authenticates the messages shared among UAVs, adding a timestamp to each. For network insights, a verifier can query any network device. A non-compromised UAV will then provide the consensus state, offering information on each node.

Asokan et al.~\cite{asokan2015seda} introduced Scalable Embedded Device Attestation (SEDA). It is an attestation scheme for remotely verifying the configurations of entire device swarms--e.g., UAVs and remote vehicles--based on their security model assumptions.
They proposed using remote attestation to detect software and physical attacks, addressing the limitations of existing protocols.
SEDA could detect various attacks with low overhead and was effectively scalable on various embedded devices, accommodating swarms of up to one million nodes.

{\bf Non-UAV Specific Solutions}
Several software-based remote attestation techniques exist for large IoT networks. 
For instance, efficient and secure distributed remote attestation (ESDRA)~\cite{kuang2019esdra} is a remote attestation technique for device swarms that uses many-to-one attestation. This eliminates the possibility of a single point of failure in the network.
Practical attestation for highly dynamic swarm topologies (PADS)~\cite{ambrosin2018pads} is a remote attestation protocol that is effective in large networks with dynamic topologies that can be implemented in highly dynamic UAV swarms. 

\subsubsection{\textbf{Access Control}\label{sec:softwaremitigationAccessControl}}

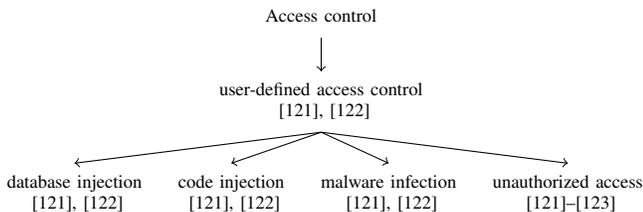
\begin{figure}[H]
\vspace{-10pt}
\centering
\begin{forest}
for tree={font=\scriptsize,align=center,edge={->}}
    [Access control
        [user-defined access control \\ \cite{liu2017protc,yoon2017virtualdrone}
            [database injection \\ \cite{liu2017protc,yoon2017virtualdrone},name=di]
            [code injection \\ \cite{liu2017protc,yoon2017virtualdrone},name=ci]
            [malware infection \\ \cite{liu2017protc,yoon2017virtualdrone},name=mi]
            [unauthorized access \\ \cite{liu2017protc,yoon2017virtualdrone,araghizadeh2016efficient},name=ua]
        ]
    ]
\end{forest} 
  \caption{Effective access control technique can mitigate database injection, code injection, malware infection, and unauthorized access. }
  \label{fig:AccessControl}
  \vspace{-10pt}
\end{figure}

Access control \cite{liu2017protc,yoon2017virtualdrone} is a method of limiting access to resources based on identity and permissions. 
Although access control generally refers to application or operating system (OS) level access here, we club medium (wireless) access control into this too. 
Figure~\ref{fig:AccessControl} is a three level tree that represents access control in UAV software. Access control on UAV software, represented in Level One of the figure, is implemented via user-defined access control, as shown in Level Two of the figure. This method could mitigate vulnerabilities, such as database injection, code injection, malware infection, and unauthorized access as shown in Level Three of the figure. 
User-defined access control is a customizable mechanism that allows for flexible management of access rights by users or administrators, providing a more granular approach than predefined models, such as mandatory access control (MAC) or Role-based access control (RBAC).

Liu and Srivastava proposed a mechanism that accomplished UAV software access control 
by defining trusted and untrusted computing blocks, enforcing user-defined access controls to regulate peripheral access, and establishing secure communication channels between computing blocks using encryption and signature schemes. \cite{liu2017protc}. They developed PROTC (Protecting drone’s peripherals through arm trustzone), a security framework, designed to protect the peripherals of UAVs using ARM TrustZone technology. PROTC ensured the safety of a UAV and the integrity of its data even if the UAV's operating system has been compromised. It also allows for easy installation of third-party applications, which provides high flexibility. The evaluation results showed that PROTC secures the UAV peripherals and incurs low-performance overheads. 
Additionally,~Yoon et al.~\cite{yoon2017virtualdrone} proposed a framework that utilized virtualization to switch to a trusted control state following an unsafe condition.

Muhammed et al.~\cite{araghizadeh2016efficient} proposed an Advanced Prioritize (AP) Medium Access Control (MAC) protocol in wireless sensor network (WSN)-UAV. 
AP-MAC, a channel access method for UAVs, operates in four steps within fixed-length time intervals. Initially, the UAV announces its presence to nearby sensors. Next, unregistered sensors attempt to send registration frames to the UAV. Then the UAV creates and sends an efficient TDMA (Time Division Multiple Access) schedule to the registered sensors. Finally, in the last step, these sensors transmit their data according to the allocated TDMA schedule.

{\bf Non-UAV Specific Solutions}
Qin et al. in \cite{qiu2020survey} surveyed a list of access control methods for IoT devices. There are several types of access control models, such as Bell-LaPadula (BLP), Clark-Wilson, discretionary access control (DAC), role-based access control (RBAC), attribute-based access control (ABAC) model.
These access control mechanisms can be mapped to UAVs as both rely on real-time data collection and transmission. IoT technology improves communication channels between UAVs and GCS, resulting in more efficient management and control of UAV operations.

Tourani et al. in \cite{tourani2018tactic} proposed Tag-based Access Control Framework for the Information-Centric Wireless Edge Networks (TACTIC) which is a lightweight access control for ICN wireless edge that eliminates the need of always-on authentication server. In TACTIC security is achieved combining tag validation and path authentication. 
In~\cite{dougherty2021apecs} Doughetry et al. proposed Distributed Access Control Framework for Pervasive Edge Computing Services (APECS) that allows user and services–pervasive edge computing (PEC) to mutually authenticate and authorize each other via a federated access control model. Similar to OAuth, APECS adds a token-based authorization on the top of OAuth scheme that provides an authentication method to verify tokens. Since APECS is operated on the edge it can also be implemented in UAVs.

\subsubsection{\textbf{Software Isolation}}

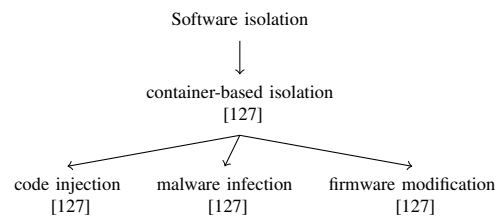
\begin{figure}[ht]
\vspace{-10pt}
\centering
\begin{forest}
for tree={font=\scriptsize,align=center,edge={->}}
    [Software isolation
        [container-based isolation \\ \cite{chen2019container},name=cb
            [code injection \\ \cite{chen2019container},name=ci]
            [malware infection \\ \cite{chen2019container},name=mi]
            [firmware modification \\ \cite{chen2019container},name=fm]
        ]
    ]
\end{forest} 
  \caption{Software isolation technique can be utilized to mitigate code injection, malware infection, and firmware modification. }
  \label{fig:SoftwareIsolation23333}
  \vspace{-10pt}
\end{figure}

Software isolation is the practice of creating secure boundaries between software components to prevent unauthorized access and data breaches. 
If a UAS has potential for compromise, 
software-level attacks can be mitigated by isolating untrusted software so that they can not perform malicious actions on UASs.
Figure~\ref{fig:SoftwareIsolation23333} represents a hierarchical tree that represents software isolation categorization. At the second level, method of software isolation is highlighted: container-based isolation. 
Software virtualization can also be used for isolation. However, it typically exhibits slower responsiveness and encounters various efficiency challenges. Hence, container-based isolation is more common.
It isolates software by enforcing restrictions on what resources or functions it can access.
The third level lists vulnerabilities that this method can address.  
Container-based \cite{chen2019container} isolation involves running software in isolated containers with their own file system, network stack, and memory space. This allows applications to run independently of each other without interfering with each other's data or operations.
Software isolation can be used as a second-level defense mechanism following IDSs and remote attestation.

Jiyang et al.~\cite{chen2019container} proposed a software framework to address DoS attacks 
for real-time UAV systems using containers. Their `Container Drone' framework included defense mechanisms for three critical system resources: CPU, memory, and the communication channel.
To protect CPU resources, they utilized the Linux kernel's control group (c-group) feature and Docker's integrated isolation mechanism.
MemGuard, a Linux kernel module, was used for memory protection.
To protect communication channels, IPtables were used to limit communication and protect the system from DoS attacks.

{\bf Non-UAV Specific Solutions}
Filippos et al. \cite{kolimbianakis2022software} proposed a software-defined technique that supports physical memory isolation and tackles indirect access to off-chip and on-chip memory. Software-defined interconnect (SDI) blocks set specific memory boundaries to prevent illegal address space access by accelerators using a customized direct memory access (DMA) block.
Instead of a job manager-assistant, IMMU, a general packet processing unit (GPPU) controls DMA operation for command and data. The buffers were predefined for allocation and deallocation of memory. 

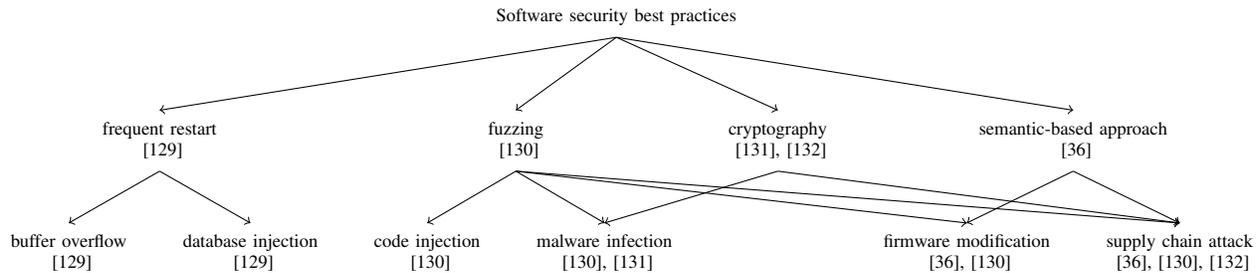
\begin{figure*}[ht]
\vspace{-15pt}
\centering
\begin{forest}for tree={
    font=\scriptsize,
    parent anchor=south,
    child anchor=north,
    fit=band,
    s sep=5mm,l=15mm,
    align=center,
    edge={->}
    }
    [Software security best practices
        [frequent restart \\ \cite{abdi2018guaranteed}, name=fr
            [buffer overflow \\ \cite{abdi2018guaranteed}]
            [database injection \\ \cite{abdi2018guaranteed}]
        ]
        [fuzzing \\ \cite{kim2019rvfuzzer}, name=f
            [code injection \\ \cite{kim2019rvfuzzer}, name = ci]
            [malware infection \\ \cite{kim2019rvfuzzer,calnoor2020secure}, name=mi]
        ]
        [cryptography \\ \cite{cho2020random,calnoor2020secure}, name=c]
        [semantic-based approach \\ \cite{sun2019tell},name=s
            [firmware modification \\ \cite{kim2019rvfuzzer,sun2019tell}, name = fm]
            [supply chain attack \\ \cite{kim2019rvfuzzer,sun2019tell,cho2020random}, name = sca]
        ]
    ]
     \draw[->] (f.south) -- (fm.north);
     \draw[->] (f.south) -- (sca.north);
     \draw[->] (c.south) -- (sca.north);
     \draw[->] (c.south) -- (mi.north);
\end{forest} 
\caption{\label{fig:SoftwareSecurityBestPractices} Software security best practices can used to mitigate UAV software vulnerabilities such as buffer overflow, database injection, code injection, malware infection, firmware modification, and software supply chain attack. These best practices can be implemented using frequent restarts, fuzzing, cryptography, and semantic-based approaches.}
\vspace{-15pt}
\end{figure*}

Victor et al. in \cite{costan2016sanctum} proposed Sanctum that provides software isolation using enclave at the user level that can protect software from other malicious software running in the hardware.
It provides advanced security measures against various software attacks, including catch-timing attacks. 
A cache timing attack is a technique where an adversary can infer sensitive information about a system by observing and analyzing the timing of cache accesses. 
Unlike Intel's SGX, Sanctum does not use encryption for enclave memory. Sanctum has a significant advantage over SGX in its ability to protect against side-channel attacks by isolating the memory access patterns of different processes, an area where SGX's defenses are relatively limited. 
Several other software isolation techniques can be used to isolate software components in UAV, such as virtualization, sandboxing, microkernel architecture, secure enclaves, language-based isolation, microservice and serverless computing, and software fault isolation (SFI), these have been largely unexplored.

\subsubsection{\textbf{Software Security Best Practices}}
Software security best practices are a set of guidelines, principles, and techniques that aim to enhance the security and reliability of software systems. These practices cover different aspects of software development, including design, development, testing, deployment, and maintenance. By following these best practices, software developers and organizations can reduce the risk of security breaches, data loss, and other security-related issues. 
Figure \ref{fig:SoftwareSecurityBestPractices} portrays a hierarchical tree centered on software security best practices. On the second level, four primary software security practices are highlighted: frequent restart, fuzzing, cryptography, and semantic-based approach. The third level, specific vulnerabilities that each practice can address are listed.
Fuzzing~\cite{kim2019rvfuzzer} refers to feeding large amounts of random unexpected data to a software program to detect vulnerabilities and weaknesses in the program's code. Semantic-based approach~\cite{sun2019tell} analyzes the semantics of a software program to identify potential security issues, such as buffer overflows, race conditions, and other vulnerabilities. Frequent restart~\cite{abdi2018guaranteed} of a software program can help prevent the accumulation of vulnerabilities and potential security risks over time. The use of encryption and other cryptographic~\cite{cho2020random,goyal06,waters11,calnoor2020secure} techniques can help protect data and prevent unauthorized access to sensitive information. 
Code obfuscation, anti-tampering, anti-debugging, and digital rights management (DRM) are other techniques that can be used to protect UAV software. 

When building UAS software, integrating security best practices has the potential to mitigate vulnerabilities that may arise during UAS deployment. 
Although there exists literature that applies directly to UAS software best practices for software security, it is a wide-ranging topic that spans far beyond UAS-specific software development, similar to hardware design best practices. 
Specifically, the literature we found included various tools for analyzing software binaries to detect vulnerabilities and enforce real-time software security best practices.

\textbf{Fuzzing:} 
An effective method for detecting software vulnerabilities is fuzzing. This automated testing process involves assessing test cases that include randomly generated data.
Kim et al.~\cite{kim2019rvfuzzer} worked on research on fuzzing techniques to identify vulnerabilities in robotic vehicles. They developed a policy-guided fuzzer that ensures the vehicle adheres to safety and functional policies using user commands, configuration parameters, and physical states.
This was called RVFuzzer, which is a tool designed to find input validation bugs in robotic vehicles through control-guided testing. The tool analyzes the control flow of the vehicle and generates test inputs that can trigger potential bugs. 
The authors evaluated this fuzzer on three robotic vehicle controller programs, including the ArduPilot UAS controller software, and discovered over 150 previously unknown software bugs.

\textbf{Sematic-based:} 
Embedded software binaries in UAVs can be reverse-engineered and searched for vulnerabilities. 
Sun et al.\cite{sun2019tell} created a tool that enhances reverse engineering software by extracting semantic data from the executable, which is then utilized for vulnerabilities and binary patching. This tool extracts the control flow graph from a designated function and creates a symbolic expression through symbolic execution. This expression is compared with the abstract syntax tree of the algorithm executed through the binary. 
It was able to discover a zero-day vulnerability in a Linux kernel controller. This tool operates as a plugin for the Interactive Disassembler (IDA) Pro decompiler showcasing the feasibility of conducting vulnerability assessments on UAS firmware binaries without needing access to the source code.

\textbf{Frequent restarts:} 
Abdi et al.~\cite{abdi2018guaranteed} presented a method to ensure real-time software security by proposing frequent restarts and diversifications for embedded controllers to increase the difficulty of launching attacks.
Their recommended approach for software restoration involved rebooting the system and reloading a secure, uncorrupted version of the controller software. This strategy is preferred over solely relying on intrusion detection, as no detection system can guarantee complete security. 
Regularly restarting and varying the software on embedded controllers can make it harder for attackers to conduct successful attacks.
Initiating system restart or restart of components while in operation can counteract software aging by clearing resource leaks and eliminating temporary files that may have built up. This practice is particularly effective in scenarios where the restart process is significantly quicker than the dynamics of the physical system.

\textbf{Cryptography:} 
Cho et al.~\cite{cho2020random} proposed a novel random number generator specifically crafted for UAV applications called DroneRNG. Traditional random number generators typically rely on sources developed on desktop environments rather than UAV-specific contexts. Generally, UAVs employ open-source cryptographic libraries such as  OpenSSL or the standard C random function for random number generation, drawing on random sources from user resources found on PCs, such as mouse or keyboard inputs, interrupt request times, and disk operation times, to produce high-quality Pseudo-Random Numbers (PRNG). However, these conventional methods fall short for UAV use due to the absence of such peripherals and sometimes no UAV operating system. DroneRNG addresses this gap by leveraging the sensor data available in UAVs. It utilizes signals from accelerometers, gyroscopes, and barometers—sensors that UAVs are equipped with during operation and stationary phases to generate high-quality random numbers tailored for UAV security needs.

Ciphertext Policy Attribute-Based Encryption (CP-ABE) is a well-known cryptographic technique~\cite{goyal06,waters11}. Rajasekar et al. \cite{calnoor2020secure} proposed using CP-ABE in the context of UAS security, by using an access tree to describe the access policy of encrypted messages. Attributes that a UAV has is used to decrypt ciphertext messages. An authentication server, also called Attribute Authority (AA), generates a public key (PK) which was shared among all other UAVs. This AA also generated a secret key (SK) per UAV based on attributes UAVs have.

{\bf Non-UAV Specific Solutions}
Several machine learning techniques~\cite{jie2016survey}, rule-based analysis~\cite{engler2001bugs}, symbolic execution~\cite{cadar2008klee}, and fuzzy testing~\cite{sutton2007fuzzing} are used to detect software vulnerabilites.  
A rule-based analysis identifies vulnerabilities in software by assessing based on a set of rules or guidelines and then examining the software to see if it follows those rules. 
Symbolic execution detects vulnerabilities in software by analyzing the code at a symbolic level rather than at a concrete level. Variables in the code are treated as symbolic expressions rather than as concrete values. However, symbolic execution can be computationally intensive and may not be practical for analyzing very large codebases. It may also be less precise than other vulnerability detection methods, as it relies on approximations of the code's behavior rather than an exact analysis.
Fuzzy testing \cite{liu2012software} provides the software with invalid or unexpected input and observed how it responds. 
However, fuzz testing has some limitations. It may not be able to identify all vulnerabilities in the software, as it relies on the software being provided with invalid input. It may also generate a large number of false positives, where the software appears to be vulnerable when it is actually functioning correctly.
These techniques while traditionally explored in non-UAV contexts, should be studied and applied within the UAV domain.

Machine learning techniques can use a large amount of data to model and learn about software vulnerabilities. Machine learning models can use program analysis, feature extraction, and vulnerability knowledge for vulnerability analysis. Lexical, syntactic, and semantic analyses are used to analyze any software to identify potential vulnerabilities, such as security vulnerabilities, performance issues, and reliability problems. 
There were several other machine learning methods proper for other domain, that can be implemented in UAV domain to detect software vulnerabilities. These include principal component analysis (PCA), K-means algorithm, logistic regression, and data mining techniques~\cite{jie2016survey}. 

These methods and tools that can be used to detect, explored, and patched software vulnerabilities automatically could be implemented in UAS. These methods can be divided into three categories: static analysis, dynamic analysis, and mixed analysis \cite{ji2018coming}. Static analysis analyzes the program without running it, dynamic analysis analyzes program behavior during execution, and mixed analysis combined the above two. Static vulnerability detection is suitable in the early stages of software development life cycle (SDLC) and uses pattern matching, lexical analysis, parsing, data flow analysis, and taint analysis. Dynamic vulnerability detection is highly accurate and has fewer false positives since it is performed after the execution of a program. The dynamic analysis can be implemented via fault injection and fuzz testing. The aforementioned techniques can also be applied in the field of UAVs.

\subsubsection{Research Gaps}
Most commercial UAVs use commercial-off-the-shelf (COTS) software, which might contain known and unknown software vulnerabilities. While the Software Anomaly Detection Systems (SADS) in Figures~\ref{fig:SADS234} provide a foundation for vulnerability detection, they may not comprehensively cover all software weaknesses. The first step to mitigate these vulnerabilities is to identify them~\cite{kannan2004economic}. There are several ways to analyze and assess vulnerabilities. 
The CVE Database, Common Weakness Enumeration (CWE) Database, and National Vulnerability Database (NVD) are, respectively, databases with a list of publicly known cybersecurity vulnerabilities, common software weaknesses, and a comprehensive collection of information gathered by National Institute of Standards and Technology (NIST) about cybersecurity vulnerabilities. According to the National Vulnerability Database (NVD), there are over 100,000 recorded software vulnerabilities \cite{nvd}. NVD also listed vulnerability characteristics for each of these records. However, integrating these databases with UAV-specific detection methods remains an area that requires further exploration to enhance real-time threat intelligence and proactive threat mitigation. 

UAS software security team can prioritize the vulnerabilities based on vulnerability scoring. In~\cite{spanos2018multi}, vulnerability scores were calculated using text analysis and multi-target classification techniques. There are several other ways to classify software vulnerabilities, such as Aslam classification, Knuth classification, Grammar-based classification, and Endres classification~\cite{krsul1998software}. If UAV software vulnerabilities are known and archived into common knowledge through similar database repositories, developing a more resilient UAS becomes easier. 

Several software vulnerability detection tools and methods are available to detect vulnerable software such as AppScan DE by Watchfire, N-Stealth by N-Stalker, NTOSpider by NTObjectives, Spike Proxy by Immunity, and TestMaker by Pushtotes; however, a single tool cannot detect all of the vulnerabilities \cite{amankwah2017evaluation}. The effectiveness of SADS could be improved by incorporating such tools into the software development lifecycle, allowing for more comprehensive vulnerability coverage and improved detection accuracy.


Mitigation of software bugs should be based on severity (high to low). The top fifteen software bugs based on severity \cite{chang2011trend} were Buffer overflow, Integer overflow, Format string, PHP remote file inclusion SQL injection, Authentication, Directory traversal, Denial of Service, Privilege action, Cross-Site Request, Forgery (CSRF), Carriage Return and Line Feed (CRLF) injection, Race condition, Cryptographic error, Information leak/ disclosure, and Cross-Site Scripting (XSS). The severity of these bugs in UAV software may differ from non-UAV systems,requiring further investigation.

Analysis is required to investigate the impact of frequent restarts on UAV software during different phases of UAV operation. Restarting the system before launch is expected to be simpler than restarting it just before payload deployment, as the latter has the potential to significantly affect operational and safety aspects. Frequent system restarts may cause data loss, dependency issues, and configuration errors, as well as introduce new vulnerabilities and state inconsistency. 

In conclusion, while Fig~\ref{fig:SADS234} and Fig.~\ref{fig:RemoteAttestation1324} outlines detection mechanisms, research gaps remain in enhancing software vulnerability detection tools, better ranking vulnerabilities based on their impact and relevance to UAV operations, and using cybersecurity databases like CVE, CWE, and NVD to identify threats. Addressing these gaps by incorporating classification techniques and text analysis into UAV software frameworks will strengthen security and ensure mission-critical resilience.

\subsection{Ground Control Station (GCS) Attack Mitigation \label{sec:GCSVulnerabilitymitigation}}

Ground Control Station Attack Mitigation includes the software and hardware used to control and monitor the UAV's flight. Authentication, access control, and intrusion detection are employed in this area to prevent unauthorized access and manipulation. 
As the GCS and the UAVs are connected through the Command and Control (C2) link, a breach of the GCS opens the door for a variety of future attack scenarios.
Compared to UAV hardware, software, and communication links, attacks and countermeasures for GCS are not widely studied in the literature.
We identified a few approaches that can address GCS attacks such as packaging/obfuscating GCS applications, software protection, and authentication and authorization.

\begin{figure}[h]
\vspace{-10pt}
\centering
\begin{forest}
for tree={font=\scriptsize, align=center,edge={->}}
    [Packing/obfuscating GCS applications
        [code obfuscation \\ \cite{nassi2021sok}
        [reverse engineering \\ \cite{nassi2021sok}]]
    ]
\end{forest} 
\caption{\label{fig:Packing/Obfuscating GCS Applications} Packing/obfuscating GCS applications can mitigate reverse engineering attacks.}
\vspace{-10pt}
\end{figure}
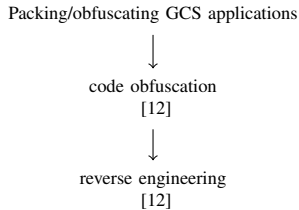

\subsubsection{\textbf{Packing/Obfuscating GCS Applications}}
Nassi et al.~\cite{nassi2021sok} proposed obfuscating code to make orchestration of reverse engineering attacks on GCS applications more difficult for attackers.
Figure~\ref{fig:Packing/Obfuscating GCS Applications} is a three-level tree where Level One represents a mitigation strategy that involves packing/obfuscating GCS applications.
Level Two shows the technique that can be used; and Level Three shows vulnerabilities it can mitigate. 
Code obfuscation is the process of modifying software to make it difficult to understand.
This can be used to prevent reverse engineering attacks and includes ``packers'', 
which can obfuscate a binary program, hide its true functionality, thus making vulnerability analysis more difficult. 
This approach could thwart attacks similar to the one Aaron Luo presented at DEFCON in 2017~\cite{luo2016drones}, where the team reverse-engineered a GCS application to uncover hard-coded authentication tokens. These uncovered tokens were used to obtain unauthorized access to a UAV in a demo.
Code obfuscation techniques discussed in UAV hardware and software attack mitigations are equally relevant and applicable in GCS applications.

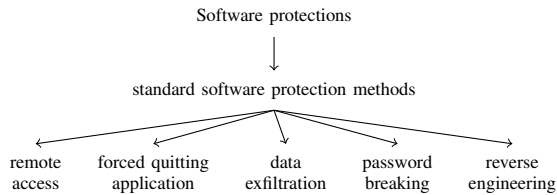
\begin{figure}[ht]
\vspace{-10pt}
\centering
\begin{forest}
for tree={font=\scriptsize,align=center,edge={->}}
    [Software protections
    [standard software protection methods
        [remote\\ access]
        [forced quitting\\ application]
        [data\\ exfiltration]
        [password\\ breaking]
        [reverse\\ engineering]
    ]
    ]
\end{forest} 
\caption{\label{fig:Software Protections} Software protections in GCS can be implemented using standard software protection methods to mitigate attacks, such as remote access, force quitting applications, data exfiltration, password breaking, and reverse engineering. }
\vspace{-10pt}
\end{figure}    

\subsubsection{\textbf{Software Protections}}
Another strategy for safeguarding GCS from attackers involves implementing standard software protection (as shown in Figure~\ref{fig:Software Protections}), such as an IDS, antivirus, firewall, or other security solutions designed to block known attacks or isolate the GCS from an untrusted network. 
Despite their critical role in defending GCS against known attacks, there exists a notable research gap in their exploration. For instance, a firewall can shield a GCS from malicious network traffic initiated by an attacker, while an IDS can identify known attack techniques targeting the GCS and notify the user.
We could not find any paper on GCS software protection in the last ten years (2013-2023). 
The techniques for mitigating UAV software attacks (discussed in Section \ref{sec:softwarevulnerabilitymitigation}) are equally applicable to GCS software.

GCS software vulnerabilities could also be detected using machine learning techniques, such as deep learning \cite{hanif2021rise}. Frequent code reusing without considering the security context is a major reason previously unknown vulnerabilities occur. As such, techniques like Code Clone Verification (CLORIFI) could be used in GCS to detect software vulnerabilities via code clone verification techniques \cite{li2016clorifi}. 

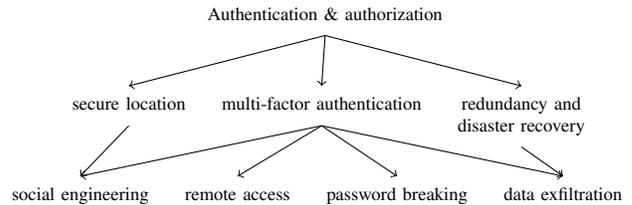
\begin{figure}[ht]
\vspace{-10pt}
\centering
\begin{forest}
for tree={font=\scriptsize,align=center,edge={->}, l=12mm}
    [Authentication \& authorization
        [secure location,name=sl]
        [multi-factor authentication,name=mfa
            [social engineering,name=se]
            [remote access,name=ra]
            [password breaking,name=pb]
            [data exfiltration,name=de]
        ]
        [redundancy and \\ disaster recovery,name=rdr]
    ]
    \draw[->] (sl.south) -- (se.north);
     \draw[->] (rdr.south) -- (de.north);
\end{forest} 
\caption{\label{fig:AuthenticationAuthorization} Authentication \& authorization in GCS can be implemented by securing the location, multi-factor authentication, and redundancy and disaster recovery to mitigate vulnerabilities, such as remote access, data exfiltration, password breaking, and social engineering.}
\vspace{-10pt}
\end{figure}

\subsubsection{\textbf{Authentication \& authorization}}
Authentication,  authorization, and access control of the GCS is important to guaranteeonly authorized users have access to the GCS. Our literature search showed a gap in this area.
There are existing methods for meeting this requirement, such as multi-factor authentication, secure location, and redundancy and disaster recovery as shown in Level Two of Figure~\ref{fig:AuthenticationAuthorization}. Level Three of the figure represents vulnerabilities these solutions can mitigate. 

GCS location should be physically secured using physical barriers, regular maintenance, and proper visitor management procedures so that unauthorized personnel can not enter the restricted area. 
Multi-factor authentication (MFA) is a security procedure where a user uses two or more forms of authentication. The categories for authenticaiton factor are: something you know (like a personal identification number (PIN), password that user enters), something you have (like a mobile device, security token/card), and something you are (like a biometrics, iris scan, face recognization validation, voice validation). Having two of the three categories to authenticate and authorize a user adds an extra layer of security before granting access to GCS and/or GCS application. 

Having multiple copies of data allows the user to recover access data from GCS in case of data failure or physical damage to GCS such as natural disaster. Segregating critical data from less sensitive data and storing it in secure, isolated environments can limit an attacker's ability to exfiltrate sensitive information. 

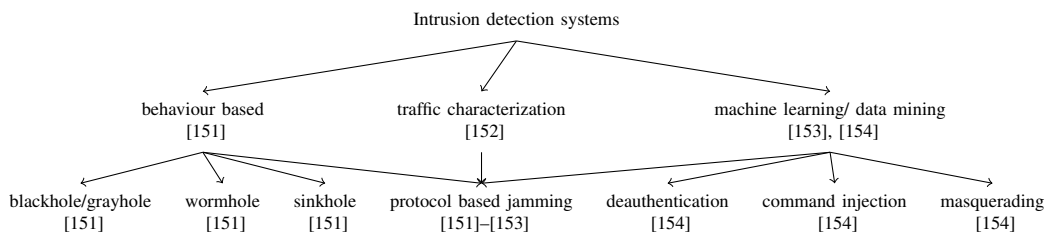
\begin{figure*}[b]
  \vspace{-15pt}
\centering
\begin{forest}for tree={align=center,
    font=\scriptsize,
    parent anchor=south,
    child anchor=north,
    fit=band,
    s sep=2mm,l=12mm,
    edge={->}
    }
    [Intrusion detection systems
        [behaviour based \\ \cite{zhang2018design},name=bb1
            [blackhole/grayhole\\ \cite{zhang2018design}]
            [wormhole\\ \cite{zhang2018design}]
            [sinkhole\\ \cite{zhang2018design}]
        ]
        [traffic characterization \\ \cite{miquel2017design},name=tc1
            [protocol based jamming \\ \cite{zhang2018design,moustafa2020autonomous,miquel2017design},name=pbj1]
        ]
        [machine learning/ data mining \\ \cite{moustafa2020autonomous,ying2019detecting},name=ml1
            [deauthentication\\ \cite{ying2019detecting}]
            [command injection\\ \cite{ying2019detecting}]
            [masquerading\\ \cite{ying2019detecting}]
        ]
    ]
     \draw[->] (ml1.south) -- (pbj1.north);
     \draw[->] (bb1.south) -- (pbj1.north);
\end{forest} 
  \caption{Intrusion detection systems can be implemented using three primary techniques: behavior-based, traffic characterization, and machine learning. These techniques can be used to address threats, such as based jamming, blackhole/grayhole attacks, and deauthentication.}
  \label{fig:Intrusion0DetectionSystems}
  \vspace{-15pt}
\end{figure*}

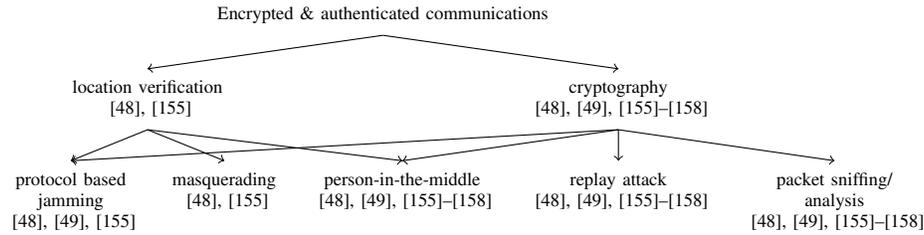
\begin{figure*}[ht]
  \vspace{-15pt}
\centering
\begin{forest}for tree={align=center,
    font=\scriptsize,
    parent anchor=south,
    child anchor=north,
    fit=band,
    edge={->}
    }
    [Encrypted \& authenticated communications
        [location verification \\ \cite{manesh2017analysis,strohmeier2014security},name=lv
            [protocol based \\jamming\\ \cite{manesh2017analysis,strohmeier2014security,costin2012ghost},name=pbj]
            [masquerading\\ \cite{manesh2017analysis,strohmeier2014security},name=mas]
        ]
        [cryptography \\ \cite{allouch2019mavsec,he2016communication,tanveer2020lake,strohmeier2014security,manesh2017analysis,costin2012ghost},name=cryp
            [person-in-the-middle \\ \cite{allouch2019mavsec,he2016communication,tanveer2020lake,strohmeier2014security,manesh2017analysis,costin2012ghost},name=pitm]
            [replay attack \\ \cite{allouch2019mavsec,he2016communication,tanveer2020lake,strohmeier2014security,manesh2017analysis,costin2012ghost},name=repa]
            [packet sniffing/ \\ analysis \\ \cite{allouch2019mavsec,he2016communication,tanveer2020lake,strohmeier2014security,manesh2017analysis,costin2012ghost},name=psa]
        ]
    ]
     \draw[->] (lv.south) -- (pitm.north);
     \draw[->] (cryp.south) -- (pbj.north);
\end{forest} 
  \caption{There are two primary security mechanisms to encrypted \& authenticated communications: location verification and cryptography. These methods can be used to mitigate network attacks, such as protocol-based jamming, masquerading, person-in-the-middle, replay attacks, and packet sniffing/analysis. }
  \label{fig:EncryptedAuthenticatedCommunications}
  \vspace{-15pt}
\end{figure*}

{\bf Non-UAV Specific Solutions}
Similar to how cloud companies and critical infrastructure companies protect their physical infrastructure/building, GCS should be protected as a secure location. The GCS's information technology (IT) team should work with the risk management and physical security teams. Staff working at GCS have different backgrounds and responsibilities, and they should have different levels of clearance based on their responsibility. Access control can be implemented via RBAC and ABAC. Personnel working at a GCS should also be trained in standard practices and incident management policy. 

Such access control can be divided into physical access control and logical access control \cite{camilleri}. Physical access control included protecting physical infrastructure by monitoring facilities and external perimeter using closed circuit television (CCTV) and alarm systems. Alarms are useful for detecting environmental threats such as fire, floods, and electric outages. Rooms in GCS should be locked so that only the authorized user can access them via PIN code and/or bio-metrics access. Any devices that are decommissioned should also be disposed of accordingly to prevent sensitive data leakage. 
Logical access control can be used to manage access to digital resources, such as data, networks, and software, in order to protect the confidentiality, integrity, and availability of those resources.
Logical access should be accomplished using a strong authentication system with multi-factor authentication whenever resources are accessed. Passwords should be updated periodically, like most companies with a policy of changing passwords every 90 days \cite{summers_2022_pwd_90_days}. Logs of all the events should be saved and the system should have a security incident, and event management (SIEM) capability \cite{dave2011security}. In addition to these techniques, firewalls should be implemented to prevent intrusions, and antivirus software should be used to prevent malware, trojans, worms, and viruses. Similar to data center security, a layered approach could be applied to GCS security. The layered approach uses several defense layers that might overlap with each other \cite{el2013proposed}.

\subsubsection{Research Gaps}
GCS are vulnerable to attackers who could target them anytime, regardless of whether UAVs are in operation or not. As a critical part of the UAV systems, data collected from UAVs are relayed back to GCS, where they are stored and processed. Despite GCS being an integral part of any UAV operation, we did not find significant focus on the nuances of their security. 
We believe that GCS security should be held to the same standard as we hold data center security today. This means there has to be an emphasis on both cybersecurity (system and application) as well as physical security; both of these area need research focus.

\subsection{Network Link Attack Mitigation \label{sec:Networkvulnerabilitymitigation}}
Network Link Attack Mitigation covers the communication link between UAVs or UAV and the ground station, and defenses in this category include encryption, frequency hopping, and anti-jamming measures to prevent interception and disruption. 
The network link in this subsection refers to  the communication link used between UAVs, GCS, cloud environments/third-party servers, or other nodes in the UAV operating environment.
The existing literature on network link defense strategies has mainly concentrated on protecting the communication links between UAVs and GCSs that use the IEEE 802.11 suite (WiFi) for Radio Frequency (RF) communication. However, other wireless technologies, such as sub-GHz RF, cellular networks (like 3G, 4G/LTE, 5G), and satellite communications (SATCOM) could get used with greater UAS proliferation. 
We need to examine whether their existing protection mechanisms are suitable in the context of UAVs; if not new protection mechanisms would need to be custom-designed.
Attacks on network links (listed in Subsection~\ref{NetworkLinkVulnerabilities}) can be mitigated through various strategies, including intrusion detection systems, encrypted \& authenticated communications, secure routing protocols, blockchain technologies, trust models, and network service best practices.

\subsubsection{\textbf{Intrusion Detection Systems}}
Network intrusion detection systems (IDS) are designed to monitor and analyze network traffic for signs of malicious activity or security breaches. 
Network IDS can be applied using three primary techniques: behavior-based, traffic characterization, and machine learning, as shown in Level Two of Figure~\ref{fig:Intrusion0DetectionSystems}. These techniques provide a comprehensive defense against various cyber threats, as illustrated in Level Three of the figure. 
Behavior-based~\cite{zhang2018design} approach involves analyzing network traffic patterns and looking for deviations from expected behavior. For example, if a UAV suddenly starts sending a large amount of data from a previously inactive account, it may be flagged as suspicious behavior. 
The machine learning-based \cite{moustafa2020autonomous,ying2019detecting} approach uses machine learning algorithms to analyze network traffic and identify patterns that may indicate malicious activity. 
Machine learning models are trained on large network traffic datasets to learn to distinguish between normal and abnormal behavior. 
Data mining~\cite{moustafa2020autonomous} analyzes large datasets to discover patterns and extract meaningful insights using various computational techniques.
The traffic characterization-based~\cite{miquel2017design} approach involves analyzing network traffic at a more granular level to identify specific types of traffic that are associated with known security threats. For example, certain types of network traffic may be associated with malware or phishing attacks. As in the case of hardware-based ADS~(section~\ref{sec:hardwareADS}) and software-based ADS~(section~\ref{sec:softwareADS}), more behavior-based intrusion detection systems were available in the current literature than knowledge-based and behavior-specification-based detection systems. 
This is because knowledge-based detection necessitates the creation of a refreshed dictionary of attack patterns, and a fundamental limitation of this method is its ineffectiveness in identifying novel attack strategies.
Detecting attacks based on behavior specification is tricky due to the complexity of defining normal behavior, dynamic environments, and the risk of false alarms caused by new attack types.

Zhang et al.~\cite{zhang2018design} proposed a hybrid solution that uses spectral traffic analysis and a resilient controller/observer to detect anomalies in UAV networks.
This approach was based on the functional and dynamic behavior of Transmission Control Protocol (TCP) and User Datagram Protocol (UDP) networking protocols.
The authors suggested analyzing the statistical signature of the network traffic exchanged in this hybrid approach. Anomalies were identified by comparing this signature with a database of known signatures.

Moustafa and Jolfaei~\cite{moustafa2020autonomous} presented an autonomous intrusion detection system to identify cybersecurity threats such as DoS, DDoS, and probing attacks using machine learning.
They simulated attack events on the UAS communication systems to create a dataset.
They used a Kali Linux virtual machine to launch DoS, DDoS, and probing attacks in a testbed environment. They used this dataset to train a machine learning model for intrusion detection system.
They performed classification using decision trees, K-nearest neighbors, multi-layer perceptron, etc., evaluating the models using typical evaluation metrics.

Ying et al. ~\cite{ying2019detecting} proposed the SODA (Spoofing Detector for ADS-B) framework, employing deep neural networks, to detect ADS-B spoofing. This framework consists of an aircraft classifier and a message classifier. The aircraft classifier identifies counterfeit communications by using the phases of incoming messages as input. Meanwhile, the message classifier detects malicious network traffic from ground-based adversaries by analyzing PHY-layer features, including in-phase and quadrature (IQ) samples.

Lyapunov-Krasovskii functional is a mathematical tool used in the stability analysis of dynamical systems, particularly those with time delays. 
This tool can be used to study TCP adaptation to varying network conditions, such as bandwidth and latency changes, to assess a network's performance.
Miquel et al.~\cite{miquel2017design} proposed an IDS for UAV fleets, utilizing a method based on the Lyapunov-Krasovskii functional and dynamic behavior for TCP. A bottleneck link is used to handle multiple links that are shared by $N$ flows. They developed a controller/observer algorithm aimed at identifying traffic irregularities. 
This method analyses traffic characteristics and recognizes DDoS attack patterns.

{\bf Non-UAV Specific Solutions}
Lauf et al.~\cite{lauf2010distributed} proposed a behavior-based IDS against network-link attacks. Their distributed IDS establishes a predefined set of behaviors offline, tailored to the nodes operating in the ad-hoc network, and examines patterns in the application's probability density function to identify the network's standard behavior. The authors demonstrated through evaluation that their IDS effectively identifies spoofing and jamming attacks. This approach utilizes semantic information from the node's application to pinpoint compromised nodes in the network.
UAVs often communicate in an ad-hoc manner, creating a dynamic and decentralized network, much like the nodes in an ad-hoc network. 
This distributed IDS is relevant to UAVs as it addresses specific challenges of ad-hoc, dynamic UAVs by detecting spoofing and jamming attacks, the most common threats to UAV systems. 

\subsubsection{\textbf{Encrypted \& Authenticated Communications}}
Encryption and authentication are two powerful tools that can be utilized to ensure a high level of security. When network traffic is sent to/from a UAV, the traffic must be encrypted to protect privacy. 
The messages should be protected against tampering to ensure their integrity and authenticity, and the communicating parties must be authenticated.
Without these safeguards, an attacker might intercept and analyze traffic, modify the messages, transmit counterfeit messages, or masquerade as a legitimate network node.
In this subsection, we will explore different strategies for safeguarding confidentiality, ensuring message integrity, verifying authenticity, and authenticating users in communication networks.
Additionally, we will also list measures that guarantee protocol-specific security.
Encrypted and authenticated communication can be achieved via cryptography and location verification as shown in Level Two of Figure~\ref{fig:EncryptedAuthenticatedCommunications}. 
The use of strong encryption algorithms~\cite{allouch2019mavsec,he2016communication,tanveer2020lake,strohmeier2014security,manesh2017analysis,costin2012ghost} can prevent attackers from accessing data transmitted over the network. Authentication ensures secure communication between the intended parties and prevents unauthorized users from accessing the communication. Using digital signatures can prevent the modification of the transmitted data. 
Location verification~\cite{manesh2017analysis,strohmeier2014security} can prevent data interception and modification by ensuring communication occurs from an expected location.
This can be achieved by using geolocation technology to verify the location of the parties involved in the communication.

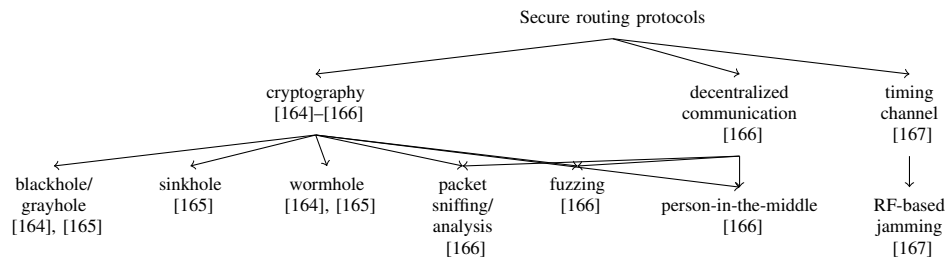
\begin{figure*}[b]
  \vspace{-15pt}
\centering
\begin{forest}
for tree={align=center,
    font=\scriptsize,
    parent anchor=south,
    child anchor=north,
    fit=band,
    s sep=5mm,l=10mm,
    edge={->}
    }
    [Secure routing protocols
        [cryptography \\ \cite{maxa2016extended,lei2019securing,agron2019secure},name=c
            [blackhole/ \\ grayhole \\ \cite{maxa2016extended,lei2019securing},name=bgh]
            [sinkhole \\ \cite{lei2019securing},name=sh]
            [wormhole \\ \cite{maxa2016extended,lei2019securing},name=wh]
            [packet \\ sniffing/ \\ analysis \\ \cite{agron2019secure},name=psa]
            [fuzzing \\ \cite{agron2019secure},name=fz]
        ]
        [decentralized \\ communication \\ \cite{agron2019secure},name=dc
                    [person-in-the-middle \\ \cite{agron2019secure},name=pitm]
        ]
        [timing \\ channel \\ \cite{thiha2019new},name=tc
            [RF-based \\ jamming \\ \cite{thiha2019new},name=rbj]
        ]
    ]
    \draw[->] (dc.south) -- (psa.north);
    \draw[->] (c.south) -- (pitm.north);
    \draw[->] (dc.south) -- (fz.north);
\end{forest} 
  \caption{Secure routing protocols can be implemented via cryptography (mitigating blackhole/grayhole, sinkhole, wormhole, packet sniffing attacks), decentralized communication (targeting fuzzing, person-in-the-middle attacks), and timing channel (countering RF-based jamming). }
  \label{fig:SecureRoutingProtocols}
    \vspace{-15pt}
\end{figure*}

Manesh and Kaabouch~\cite{manesh2017analysis} examined the risks present in ADS-B, identifying potential attacks such as eavesdropping, message deletion, modification, injection, and jamming in their study. They suggested various countermeasures, dividing them into two primary categories: secure broadcast authentication and location verification. The secure broadcast authentication approaches verify the authenticity of ADS-B broadcasts, differentiating between cryptographic and non-cryptographic methods to ensure that messages come from a trusted source.
Cryptographic schemes involved asymmetric encryption and aircraft address message authentication code (AA-MAC). Non-cryptographic schemes used fingerprinting and spread spectrum techniques such as direct sequence spread spectrum (DSSS) and frequency hopping spread spectrum (FHSS).
Secure location verification mechanisms are used to verify a broadcaster's location by employing techniques, such as distance bounding, Kalman Filtering, and data fusion.

Strohmeier et al. \cite{strohmeier2014security} also subdivided ADS-B security measures into secure broadcast authentication and secure location verification.
Secure broadcast authentication used software-based, hardware-based, and location-based fingerprinting techniques and randomized frequency hopping.
Location verification is achieved through multilateration. This process computes an object's location by analyzing the time difference of arrival (TDOA) of a signal emitted from the object at multiple receivers.

Symmetric (or secret-key) and asymmetric (or public-key) key cryptography are used to ensure confidentiality, preventing data from being intercepted and analyzed by eavesdroppers. 
To ensure data confidentiality in the MAVLink protocol, symmetric encryption algorithms can be utilized~\cite{allouch2019mavsec}.
Encryption algorithms, such as ChaCha20 (a stream cipher designed by D. J. Bernstein), AES, and RC4, only added slight CPU/memory overheads. 
Additionally, He et al.~\cite{he2016communication} proposed that 
if the communication medium employs the IEEE 802.11 suite, activating WPA2 and employing large keys can likewise guarantee data confidentiality.

Several techniques have been proposed in the literature to enable secure ADS-B communications.
Costin and Francillon~\cite{costin2012ghost} specified several security vulnerabilities in Automatic Dependent Surveillance-Broadcast (ADS-B), including eavesdropping, jamming, spoofing, and the injection or modification of messages. Using readily available commercial transceivers, they carried out successful replay and impersonation attacks. To mitigate these vulnerabilities, they recommended the implementation of a lightweight public key infrastructure (PKI) suitable for devices with limited resources, the addition of message authentication codes into ADS-B transmissions, and the establishment of key distribution mechanisms via certifying authorities such as the FAA and EUROCONTROL.

In addition to ensuring message authenticity, it is important to authenticate users to prevent unauthorized use of resources (access control).
Tanveer et al.~\cite{tanveer2020lake} proposed a key exchange protocol for secure communications in Internet of Drone (IoD) environments. The protocol provides mutual authentication between mobile users and secure session key derivation. It is composed of multiple phases including user registration, authentication, key exchanges, and updates for passwords and biometrics, effectively safeguards against user impersonation, person-in-the-middle, DoS, and replay attacks.

{\bf Non-UAV Specific Solutions}
Error checking \cite{samaila2018security} can be used to detect and prevent data corruption and transmission errors. For example, checksums can be used to verify the integrity of the data. If the checksum does not match the calculated value, it indicates that the data has been corrupted or modified. 
When transmitting a message through a network, verifying message integrity involves checking that the message remains unchanged. Establishing message authenticity ensures it is sent by a trustworthy source. Using a message authentication code to create a checksum for the message, which is appended before sending, can protect the integrity of the data.
A hashed message authentication code (HMAC) with a symmetric key can provide two-way authentication and protect message integrity~ \cite{samaila2018security}.

\subsubsection{\textbf{Secure Routing Protocols}}
Secure routing protocols are designed to prevent attacks on network routing, which can cause disruption or interception of data flow. Level Two of a hierarchical tree in Figure \ref{fig:SecureRoutingProtocols} indicates four ways to implement a secure routing protocol, and Level Three of the tree shows vulnerabilities these methods can mitigate.
Decentralized communication \cite{agron2019secure,hu2003packet} protocols are designed to enable communication between nodes without relying on a central authority. This approach can be more resilient against attacks that target central authorities, such as denial-of-service attacks. Decentralized communication can also provide greater privacy and anonymity, which are important in UAS communication. Cryptographic protocols~\cite{maxa2016extended,lei2019securing} can be used to secure the routing information exchanged between nodes (UAV-UAV, UAV-GCS). For example, a routing protocol might use digital signatures to ensure that routing information has not been modified in transit. Cryptographic protocols can also be used to protect the privacy of routing information, preventing attackers from intercepting or analyzing routing messages. 
Timing channel~\cite{thiha2019new} offer a method to identify and thwart attacks predicated on timing-based events. An attacker, for instance, may try to intercept or alter routing messages at specific time to evade detection. By incorporating random delays or jitter into the communication process, timing channel protocols enhance security. This approach disrupts the predictability of message transmission times, thereby complicating the attacker's efforts to time their malicious activities effectively.

Malicious manipulations of UAS routing protocols, such as wormhole, blackhole, and Sybil attacks, are among the attacks targeting UAS routing protocols. To maintain the network link's availability and ensure its operational status throughout the UAS mission, safeguards can be implemented against attacks aimed at disrupting message routing among network participants.
Maxa et al.~\cite{maxa2016extended} presented a secure reactive routing protocol (SUAP). The authors utilized public key cryptography, hash chains, and geographical leashes to ensure message authenticity that is successful in detecting and preventing attacks on network link routing, such as wormhole and blackhole attacks.

Implementation of novel network architectures, such as Named Data Networking (NDN) within UAV ad-hoc networks (UAANETs) introduced new challenges, such as content poisoning~\cite{lei2019securing}. The architecture of NDN-based UAANETs, which includes a caching mechanism within the network, can lead to altered router caches and reduced performance when attacked.
The authors in ~\cite{lei2019securing} present a unique framework to mitigate this issue. This comprehensive solution effectively identifies and eliminates poisoned content. It combines the Interest-Key-Content Binding (IKCB) method, a specialized forwarding strategy, and a system for verification on demand. 
They implemented the `permissioned blockchain technology' to establish a decentralized repository for IKCB and detect internal threats. This ensured the secure verification and recording of IKCB regulations that link content name, publisher public key digest (PPKD), and content digest. In addition, they proposed an `Adaptive Delegate Consensus Algorithm' (ADCA) for the blockchain to eliminate the need for traditional mining and provide high performance, scalability, and consistent reliability throughout the network.

A secure routing protocol can be used for a Flying Ad Hoc Network (FANET) between a GCS and UAS. It ensures integrity, confidentiality, and authentication \cite{agron2019secure, hu2003packet}. 
To maintain the integrity of the data, the said authors implemented a nonce hash mechanism. For confidentiality, they utilized the TWINE algorithm, which is lightweight and straightforward. Two key sizes of 80-bit and 128-bit are supported and a 64-bit clock size is used to secure the essential fields in routing messages. To ensure authentication, the authors employed a hybrid authentication procedure.
A combination of symmetric and asymmetric key encryption algorithms and digital signatures was employed to safeguard the packet field containing sensitive information, such as the geographic information of UASs. Additionally, the authors implemented a security measure called a `packet leash mechanism' to thwart wormhole attacks. 

Adversaries can detect the synchronization header (SHR) of a data-frame  and use it to disrupt the communication channel between sender and receiver. Thiha et al.~\cite{thiha2019new} proposed sending a short dummy packet before sending the actual packet to counter these attacks.

{\bf Non-UAV Specific Solutions}
Geographical routing~\cite{samaila2018security} protocols use location information to route traffic between nodes. This approach can be effective against certain attacks, such as spoofing or routing loops since they rely on physical proximity rather than trust-based routing decisions. However, geographical routing may be less efficient in large, complex networks. Multi-path routing~\cite{samaila2018security} protocols use multiple routes to transmit data between nodes. Using multiple routes, multi-path routing can provide redundancy and resistance against network disruptions that are caused by attacks or failures. However, multi-path routing can increase network overhead. Multi-path routing can be applied in UAV communication to reduce the risk of data loss from single point of failure or unstable connection.

\begin{figure}[ht]
\vspace{-10pt}
\centering
\begin{forest}
for tree={align=center,font=\scriptsize,edge={->}}
    [Blockchain technology solutions
        [group key \\ distribution scheme \\ \cite{li2019blockchain}
            [deauthentication \\ \cite{li2019blockchain},name=deauth]
        ]
        [decentralized \\ communication \\ \cite{aggarwal2019new},name=dc
            [masquerading \\ \cite{aggarwal2019new}]
            [replay\\ attack \\ \cite{aggarwal2019new}]
        ]
        [routing with \\ blockchain \\ \cite{liu2019blockchain}
            [blackhole/ \\grayhole\\ \cite{liu2019blockchain}]
            [wormhole \\ \cite{liu2019blockchain}]
            [sinkhole \\ \cite{liu2019blockchain}]
        ]
    ]
     \draw[->] (dc.south) -- (deauth.north);
\end{forest} 
  \caption{Blockchain technology solutions can be categorized as group key distribution scheme, decentralized communication, and routing. These methods can be used to mitigate vulnerabilities, such as authentication, masquerading, wormhole, and sinkhole attacks. }
  \label{fig:BlockchainTechnology}
  \vspace{-10pt}
\end{figure}
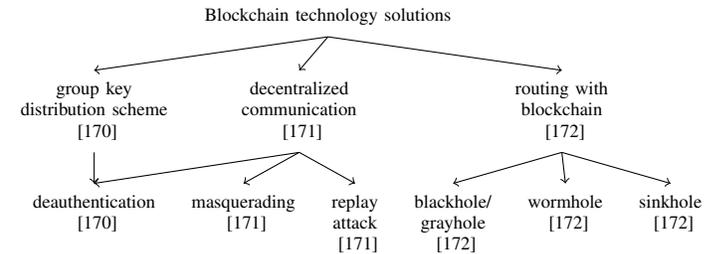

\subsubsection{\textbf{Blockchain Technology Solutions}}
Blockchains have been used for implementing group key distribution schemes, decentralized communication, and routing as shown in Level Two of Figure~\ref{fig:BlockchainTechnology}. These methods help mitigate network vulnerabilities, such as authentication, masquerading, replay attacks, blackhole/grayhole, wormhole, and sinkhole attacks as shown in Level Three of the figure.
In a group key distribution scheme \cite{li2019blockchain}, a blockchain is used to distribute cryptographic keys to authorized parties. This approach can prevent unauthorized access to network communications, as the keys can be distributed securely and transparently. Blockchain can facilitate secure routing \cite{liu2019blockchain} in a network. Distributed ledger can provide a tamper-proof record of routing decisions and enable secure communication between nodes. This approach can effectively prevent routing table poisoning and person-in-the-middle attacks. Blockchain can be used to allow for decentralized communication~\cite{aggarwal2019new} between nodes in a network. This approach can prevent attacks that target central authorities or rely on centralized communication channels. 
Blockchain's tamper-resistant distributed storage properties can also create  
secure protocols, which is essential to protect UASs and their networks against potential threats.

Li et al.~\cite{li2019blockchain} developed a specialized private blockchain for GCSs communications aimed at distributing and preserving group key broadcast messages. This approach enabled them to formulate a secure and efficient group key distribution scheme for UAS networks. The effectiveness of the protocol was evaluated against two adversary models, and the results showcased the scheme's capability to safeguard against multiple types of attacks with minimal impact on time and storage resources.
Additionally, Liu et al.~\cite{liu2019blockchain} presented a routing approach that leverages blockchain technology to shield the network structure from exposure, even in cases where certain network peers are compromised. They employed the consensus mechanism found in blockchain applications to autonomously identify and counteract compromised nodes, thereby safeguarding the routing rules from deliberate alterations or broad disclosure.

Similarly, Aggarwal et al.~\cite{aggarwal2019new} proposed a system architecture designed to secure data distribution within an IoD framework, employing Ethereum blockchain technology. Their model supports secure, decentralized interactions between UAVs and users. Utilizing a blockchain framework to gather data from UASs, their proposed approach ensured the integrity, authentication, and authorization of collected data.

{\bf Non-UAV Specific Solutions}
Mohammed in \cite{mohammed2021hybrid} proposes a distributed security environment using blockchain to enhance security in IoT networks to protect against DoS attacks. This framework used cryptographic techniques, access control methods, and hashing to ensure the security of data transmission in IoT environments.

\begin{figure}[ht]
\vspace{-10pt}
\centering
\begin{forest}
for tree={
    font=\scriptsize,
    edge={->},
    align=center             
}
    [Trust models
        [behavior monitoring \\ \cite{keshavarz2020real,ge2020provenance}
            [protocol-based \\ jamming \\ \cite{keshavarz2020real,ge2020provenance}]
            [PITM \\ \cite{keshavarz2020real,ge2020provenance}]
            [replay \\ attack \\ \cite{keshavarz2020real}]
            [blackhole/ \\ grayhole \\ \cite{ge2020provenance}]
            [wormhole \\ \cite{ge2020provenance}]
            [sybil \\ \cite{ge2020provenance}]
        ]
    ]
\end{forest} 
\caption{Trust models can be implemented using behavior monitoring techniques to mitigate software vulnerabilities, such as protocol-based jamming, PiTM, replay attacks, blackhole, grayhole, wormhole, and sybil attacks.}
\label{fig:Trust Models}
\vspace{-10pt}
\end{figure}
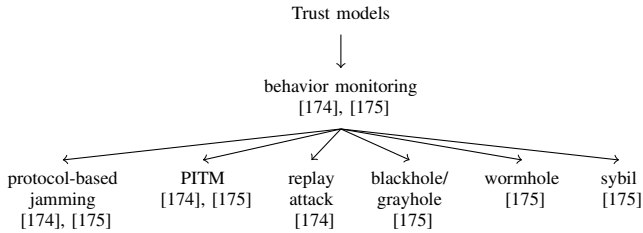

\subsubsection{\textbf{Trust models}}
An alternative approach to safeguarding UAS network links is to use trust models.
Figure~\ref{fig:Trust Models} shows that behavior monitoring are used to build trust models to address network vulnerabilities, such as protocol-based jamming, person-in-the-middle attacks, replay attacks, blackhole attacks, grayhole attacks, wormhole attacks, and sybil attacks.
This technique, increasingly popular among UAV swarms, involves monitoring the behavior of nodes within the UAS network. It calculates a trust score for each node, and those with scores falling below a predetermined threshold are deemed untrustworthy and removed from the network.

Keshavarz et al.~\cite{keshavarz2020real} introduced a method for assessing trust within a UAS, utilizing a central entity such as the GCS to continuously evaluate the UAVs' behaviors, including their travel routes, energy usage, and task completion. This enabled relative trust score calculation for the UAVs, allowing for the real-time identification of any unusual behaviors. The mechanism distinguishes between anomalous behaviors resulting from cyber-physical attacks and those that may arise from challenging environmental conditions, such as turbulence or irregular energy consumption, through the use of an audit unit.
This trust assessment strategy measures the performance of UAVs based on the audit unit's observations, taking into account the potential for uncertainty in these observations. The trust framework can detect and flag malicious UAVs engaged in cyber-security threats such as flooding attacks, person-in-the-middle attacks, and GPS spoofing in real time.
Ge et al.~\cite{ge2020provenance} used a trust-scoring system designed to identify a range of network-based threats, including black/gray hole attacks, Sybil attacks, DDoS attacks, and person-in-the-middle attacks.

{\bf Non-UAV Specific Solutions}
In~\cite{arabsorkhi2016conceptual} Arabsorkhi et al. proposed a decentralized trust model for IoT devices similar to that in human societies. These devices utilize past experiences to inform trust decisions. Without such history, devices collect opinions from their network peers. The nodes will proceed with the service if they receive positive feedback from their peers. Conversely, if the information gathered is inadequate, the node refers to its trust threshold to decide. Post-interaction, the device updates its records with the service provider's trustworthiness for future interactions. At any stage, if the acquired information is lacking or the trust level falls below the acceptable limit, the device will refrain from using the service. This method of gaining trust can also be implemented in UAV swarms.

\subsubsection{\textbf{Network Service Best Practices}}
\begin{sidewaysfigure}
\centering
\begin{forest}
for tree={align=center,font=\scriptsize,
    parent anchor=south,
    child anchor=north,
    fit=band,
    s sep=5mm,l=22mm,
    edge={->}
    }
    [Network service best practices
        [cryptography \\ \cite{ samland2012ar}, name=c
            [password breaking \\ \cite{samland2012ar}]
            [blackhole/ \\ grayhole \\ \cite{he2016communication},name=bgh ]
            [wormhole \\ \cite{he2016communication},name=wh]
            [sybil \\ \cite{he2016communication},name=syb]
            [sinkhole \\ \cite{he2016communication}, name=sin]
            [person-in-the-middle \\ \cite{he2016communication}, name=pitm]
            [masquerading \\ \cite{he2016communication}, name=mas]
            [fuzzing \\ \cite{he2016communication}, name=fuz]
            [protocol-based \\ jamming \\ \cite{hooper2016securing},name=pbj]
        ]
        [device/asset \\ management \\ \cite{he2016communication},name=rma]
        [fuzzy \\ technique \\ \cite{hooper2016securing},name=ft]
    ]
     \draw[->] (rma.south) -- (bgh.north);
     \draw[->] (rma.south) -- (wh.north);
     \draw[->] (rma.south) -- (syb.north);
     \draw[->] (rma.south) -- (sin.north);
     \draw[->] (rma.south) -- (pitm.north);
     \draw[->] (rma.south) -- (mas.north);
     \draw[->] (rma.south) -- (fuz.north);
     \draw[->] (ft.south) -- (pbj.north);
\end{forest} 
  \caption{Network service best practices can be implemented using cryptography, device/asset management, and fuzzy technique. These methods can be used to mitigate a large number of known vulnerabilities, such as password breaking, blackhole/grayhole attacks, wormhole attacks, sybil, person-in-the-middle, masquerading, fuzzing, and protocol-based jamming.}
  \label{fig:NetworkServiceBestPractices}
\end{sidewaysfigure}
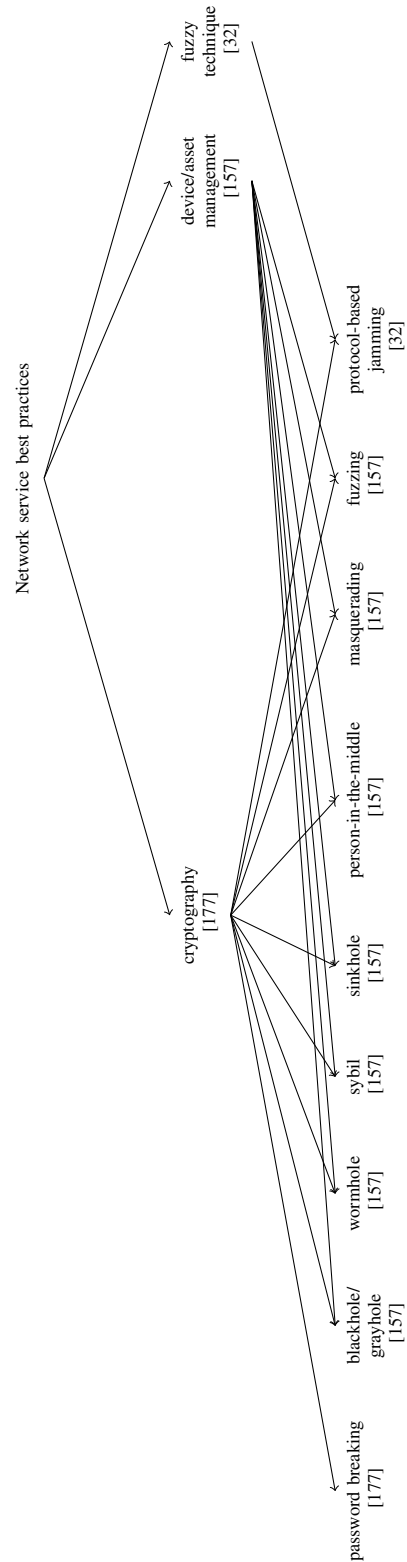

It is essential to defend the UAV network and the associated network services utilized by the UAVs and other network entities from network link attacks.
This emphasis is recurrent across multiple studies in the literature. Most of the recommended network service best practices can be easily implemented within the UAS or the devices that support the UAV network.
Figure \ref{fig:NetworkServiceBestPractices} is a three-level tree listing ways network service best practices can be utilized to secure UAV network. Level Two outlines various methods to implement these practices, including cryptography, fuzzy techniques, and device/asset management. Level Three lists vulnerabilities these methods can mitigate, such as wormhole, Sybil, masquerading, and protocol-based jamming.

Samland et al.~\cite{samland2012ar} explored various approaches and methodologies pertaining to best practices in network service. 
These consist of using WPA2 with strong passwords and replacing the use of Telnet and FTP services with SSH.
He et al.~\cite{he2016communication} investigated ways of increasing WiFi network security, such as disabling SSID broadcasts and restricting network access only to devices with pre-registered MAC addresses.

It is recommended that network services be activated only for crucial UAS tasks and that the operating system and network services are always kept up to date.
Standard AR Discovery and Wi-Fi access points on UAVs are vulnerable to cyberattacks \cite{hooper2016securing,bonilla2018common}. 
Hooper et al.~\cite{hooper2016securing} investigated how to identify vulnerabilities in software services, such as DoS and buffer overflow attacks, that could be exploited via the network. Their research focused on utilizing a fuzzy technique to uncover these vulnerabilities in the Parrot Bebop UAV's software services during AR Discovery. They suggested a security framework tailored for Wi-Fi-based UAS to protect against cyber-attacks. For example, to prevent buffer overflow attacks, they recommended filtering the input the UAV receives over the network, thus ensuring that this input does not compromise the memory integrity of the network service.

{\bf Non-UAV Specific Solutions}
Samaila et al.~\cite{samaila2018security} emphasized the importance of educating end-users on cybersecurity best practices, enforcing strong passwords, and activating and supervising security event logging.
In~\cite{kamoun2005toward} Kamoun listed thirteen functional areas for best practices to maintain communication networks. 
Technicians need continuous training and skills development to maintain networks effectively. 
The author also emphasized predictive and preventive maintenance procedures, post-mortem and root-cause analysis, and efficient contractor management. 
Developing appropriate maintenance policies and organizational structures, ensuring leadership support, and using Computerized Maintenance Management Systems (CMMS) can be implemented for secure communication. 
These methods can also be implemented in the UAS network. 

\subsubsection{Research Gaps}
We identified that the network contained the most vulnerabilities and hence was the largest attack vector for adversaries to steal data or take complete/partial control of the UAV. Accordingly, most of the papers surveyed emphasized mitigating attacks in the network link, such as GPS spoofing and jamming 
Unencrypted ADS-B communications in UAS can raise security concerns. UAVs generally have smaller batteries, which can limit their processing capability. This hinders the deployment of a robust encryption scheme that protects the privacy 
of ADS-B data. 
Developing a low-overhead robust security framework is crucial. 
Such frameworks would consist of a robust IDS at the network layer combined with simple backup methods (such as switching communication channels) in the event of an attack when a GCS detects that an aircraft's ADS-B may be compromised. 

Ongoing efforts are being developed to detect ADS-B spoofing attacks, but well-developed and comprehensive solutions are required. For example, physical-layer features classified as valid messages may provide security only if adversaries are unaware of the parameters, such as IQ (In-phase and Quadrature) samples or phases of specific aircraft messages.
I and Q components describe signal's amplitude and phase, crucial for signal manipulation in communication. 
IQ samples and phases may vary over time, but in some cases, IQ samples or phases are statically adjusted to optimize data transmission or interference effects. Future works implementing statistical or ML models should be developed based on less static signature-based features but based on robust dynamic features of an aircraft that are more difficult to spoof by adversaries. 

Secure routing protocols were investigated in UAS to ensure integrity, confidentiality, availability 
of the data transmitted over the network. 
These protocols help manage the communications between various components of a UAS, such as the flight controller, sensors, and payload, to support the mission of the UAS 
The common theme among all routing protocol was using lightweight and simple algorithms in conjunction with packet leashing~\cite{hu2003packet} to prevent wormhole attacks.
Packet leashing is a technique to restrict the unauthorized forwarding of packets by using temporal or geographical constraints. 

Interestingly, future internet architectures such as Information-Centric Networks (ICN) have been studied for vehicular ad-hoc networks (VANET) for scalable and reliable networks, which are particularly useful in multi-to-multi communications when swarms of UAVs are communicating with each other \cite{safwat2022survey}. 
ICN focuses on content, making it effective for UAVs as it enhances network resilience, supports scalability, reduces latency through caching, and improves security. 
These features make it perfect for the dynamic and mobile nature of UAV operations. 
However, even though content in ICN may be encrypted, they are susceptible to wormhole attacks~\cite{mick2017laser}. 
Hence, future research for lightweight and secure protocols in ICN-based UAV Ad hoc Networks (UAANETs) 
should address this gap for scalable and secure UAV networks.

\subsection{Cloud/Server Attack Mitigation}
Cloud Attack Mitigation involves addressing the attacks on the cloud-based servers used to store and process the data collected by the UAV. It is important to distinguish between data storage and data processing, as processing involves computational tasks that raise concerns about verifying the accuracy of results and detecting potential manipulation. This section primarily focuses on data storage.

\subsubsection{\textbf{Encrypting outsourced data}}

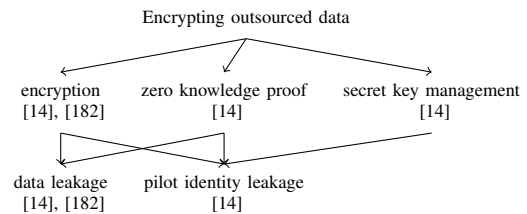
\begin{figure}[ht]
\vspace{-10pt}
\centering
\begin{forest}
for tree={align=center,font=\scriptsize,edge={->}}
    [Encrypting outsourced data
        [encryption  \\ \cite{lin2018security,xu2015secure}, name=enn
            [data leakage \\ \cite{lin2018security,xu2015secure}, name=dl]
        ]
        [zero knowledge proof \\ \cite{lin2018security}, name=zkp
            [pilot identity leakage \\ \cite{lin2018security},name=pil]
        ]
        [secret key management \\ \cite{lin2018security},name=skm
        ]
    ]
     \draw[->] (enn.south) -- (pil.north);
     \draw[->] (skm.south) -- (pil.north);
     \draw[->] (zkp.south) -- (dl.north);
\end{forest} 
  \caption{Outsourced data can be encrypted to mitigate vulnerabilities in cloud/servers, such as data leakage and pilot identity leakage.}
  \label{fig:Encrypting Outsourced Data}
  \vspace{-10pt}
\end{figure}

Data should be encrypted before being transmitted to the Cloud to protect the confidentiality of information being outsourced from UAS networks to the Cloud.
Figure~\ref{fig:Encrypting Outsourced Data} is a three-level tree centered on securing outsourced data and its confidentiality at Level One; Level Two lists three methods to achieve it: encryption, secret key management, and zero-knowledge proof. Level Three of the figure lists the attacks those methods can mitigate.

Lin et al.~\cite{lin2018security} identified two privacy challenges in the Internet of Devices (IoD): location/identity privacy and outsourced data privacy. To address the first challenge, they proposed using lightweight and efficient symmetric key encryption algorithms, key management systems, and zero-knowledge proofs as potential solutions.
Drone location information was encrypted using lightweight and energy-efficient algorithms like Elgamal and Advanced Encryption Standard (AES).
Sensitive information during navigation was protected via zero-knowledge range proof to maintain confidentiality.
For the latter challenge, their solution was a lightweight identity-based encryption scheme, leveraging both asymmetric and symmetric key cryptography to secure data privacy and provide flexible access when necessary.
Xu and Zhu~\cite{xu2015secure} proposed a customized cryptographic tool to protect data outsourced to the cloud from networked control systems and to guarantee the integrity of results computed in the cloud.

{\bf Non-UAV Specific Solutions}
We note that this area has been extensively studied and proposed approaches may be relevant to the UAS domain.
Data-oriented attacks focus on stealing or corrupting data in the cloud/server via malware infections, phishing scams, and SQL injection. These attacks could be protected via data execution prevention (DEP) and address space layout randomization (ASLR)~\cite{morton2018security}. 
DEP monitors programs to make sure that they use system memory safely. ASLR randomizes the memory location of data, making it harder for adversaries to execute malicious code in server/cloud storage. Third-party auditors can also verify the integrity of dynamic data stored in the cloud \cite{meenakshi2014cloud}.  

Attacks that are possible in computer networks are also applicable in the Cloud. The most common threats are person-in-the-middle attacks, phishing, eavesdropping, sniffing, DoS, and DDoS~\cite{dou2013confidence, ahmed2014cloud}. Data stored in the Cloud/server can be secured using multi-factor authentication. Using two secret keys, identity-based encryption (IBE) and public key encryption (PKE), could enhance the security of cloud/server \cite{liu2015two}. These advanced security policies, access management, data protection policy, and security techniques \cite{chou2013security} can enhance UASs using public, private, and hybrid cloud resources or servers.  Other encryption techniques listed in hardware attack mitigation (Section~\ref{sec:hardwarevulnerabilitymitigation}), software attack mitigation (Section~\ref{sec:softwarevulnerabilitymitigation}), GCS attack mitigation (Section~\ref{sec:GCSVulnerabilitymitigation}), and network link attack mitigation (Section~\ref{sec:Networkvulnerabilitymitigation}) are equally applicable in encrypting outsourced data to cloud/server.

\subsubsection{\textbf{Authentication and Authorization}}

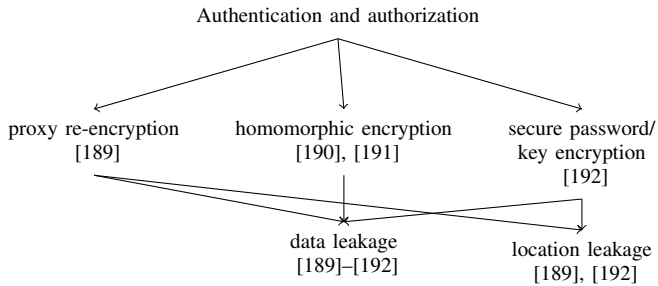
\begin{figure}[ht]
\vspace{-10pt}
\centering
    \begin{forest}for tree={align=center,
    font=\footnotesize,
    parent anchor=south,
    child anchor=north,
    fit=band,
    s sep=5mm,l=15mm,
    edge={->}
    }[Authentication and authorization
        [proxy re-encryption \\ \cite{baboolal2019preserving},name=pre
        ]
        [homomorphic encryption \\ \cite{alzahrani2023protecting,cheon2018toward},name=he
            [data leakage \\ \cite{baboolal2019preserving,wazid2018design,alzahrani2023protecting,cheon2018toward},name=dl]
        ]
        [secure password/ \\ key encryption \\ \cite{wazid2018design},name=spke
            [location leakage \\ \cite{baboolal2019preserving,wazid2018design},name=ll]
        ]
    ]
     \draw[->] (pre.south) -- (ll.north);
     \draw[->] (spke.south) -- (dl.north);
     \draw[->] (pre.south) -- (dl.north);
\end{forest} 
  \caption{Authentication and authorization can be implemented via proxy re-encryption and homomorphic encryption to mitigate data leakage; and secure password and key encryption to mitigate location leakage. }
  \label{fig:AuthenticationandAuthorization}
  \vspace{-10pt}
\end{figure}

Proxy re-encryption, homomorphic encryption, and secure password/ key encryption are three ways authentication and authorization can be implemented to secure the cloud/server attack of a UAS system, as shown in Level Two of Figure~\ref{fig:AuthenticationandAuthorization}. 
Proxy re-encryption (PRE) is a cryptographic system that transforms encrypted data from one encryption key to another without first decrypting the data. 
Homomorphic encryption enables computations on encrypted data without decrypting it first.
These methods can be used to mitigate data leakage, as shown in Level Three of the figure. 
A secure password reduces the risk of unauthorized access to personal or sensitive information, enhancing overall security and privacy, and is designed to make it difficult for adversaries to guess or crack.

Baboolal et al. in \cite{baboolal2019preserving} used PRE to encrypt and decrypt data as it is transmitted between two parties. 
They proposed storing videos in the cloud using a proxy re-encryption technique. The proxy server can be configured only to allow authorized users to access the data, ensuring that the data is secure and protected from unauthorized access. 
The scheme involved generating a one-time pass key for viewing videos, with a trusted control center managing the key. 
Similar to outsourced data encryption, protections are implemented to stop unauthorized entities from accessing the data stored in the cloud.

Alzaharani et al.~\cite{alzahrani2023protecting} proposed a homographic encryption-based technique for secure communication in UAVs. Data that is encrypted is again encrypted with an additional key. Using this dual-key approach, even if an attacker obtains one key, they still need the second key to decrypt the data. Consequently, intercepting a single message will maintain the security of all other messages, making this system more secure than traditional single-key methods.

Cheon et al. in \cite{cheon2018toward} adopted a homorphic cryptography for controller, homomorphic authenticated encryption (LineHAE), that supports linear operation between ciphertexts, with fast encryption, evaluation, and verification procedures. It eliminates the potential risk linked to handling the secret key within the controller by removing the requirement to encrypt and decrypt data for the mathematical operation inside the controller.
LineHAE can detect attacks such as network attack, controller attack, tapping signal attack, and attack on encrypted data. This is effect in protecting against eavesdropping and forgery attacks, unlike homomorphic encryption alone that does not provide means to check whether the received signal at the drone side is authentic or compromised.

Wazid et al.~\cite{wazid2018design} designed an authentication scheme that allowed IoD/UAV users to access data from UAS service providers directly. Either the cloud server or a GCS registered each UAV before deployment. The proposed scheme secured passwords and biometric updates using authentication and key agreement protocols. Through UAV key management, the GCS  established pairwise keys between neighboring UAVs to secure their communication. This approach prevents privileged insider and offline password-guessing attacks, as well as user/server/UAS impersonation attacks and denial-of-service attacks.
Secure password/ key encryption \cite{wazid2018design} is critical for securing user accounts and preventing unauthorized access. Passwords and keys can be encrypted using strong encryption algorithms and stored securely to prevent theft or compromise. Multi-factor authentication can be used to provide an additional layer of security, requiring users to provide multiple forms of authentication (such as a password and a fingerprint) to access cloud resources.

{\bf Non-UAV Specific Solutions}
Blockchain technology has the potential to enhance the security of the cloud/server. One potential application is using blockchain to secure the stored data against tampering. Since each block in a blockchain contains a cryptographic hash of the previous block, it is extremely difficult to alter the data stored in a block without detection \cite{pavan2020server}. 
Smart contracts facilitate automated access control management as self-executing agreements, where the contract terms between the buyer and seller are embedded directly within the code.
They can be used to automatically grant or revoke access to server/cloud-based resources based on predefined conditions \cite{gupta2019cloud}.

\subsubsection{Research Gaps}
A major issue in UAV systems utilizing cloud services is the lack of direct control over the infrastructure.
Most of the cloud resources are controlled by a third party like Google, Amazon AWS, and Microsoft Azure. These companies emphasize the security and privacy of their users, cloud services act as a black box, which may raise trust issues for users. The users can secure login credentials and data being stored to and from the cloud. However, they have limited control over the data once it is stored in the cloud. To address security concerns, customized security measures can be implemented based on user requirements, allowing monitoring and protection of virtual environments. This can be achieved via Service Provider Attack Detection (SPAD) and Tenant Specific Attack Detection (TSAD) \cite{hawedi2018security}. SPAD enforced a security policy to ensure malicious traffic is not included while receiving virtual machine traffic. If the system detects malicious traffic, the VM is isolated, and the user/customer is notified about the incident. TSAD is a system that a cloud provider implements to provide cloud users with additional security measures based on their needs.

\section { \textbf{Discussion}\label{Discussion}}
In this section, we present and analyze some recently proposed enhancements and concepts that could strengthen UAS cybersecurity. We also reflect on lessons learned from existing studies and practical applications in the real world.

\subsection{Regulatory Compliance}
Regulations can be established to mandate the use of Remote ID for UAVs, and these regulations are a part of the broader framework in Integrated Airspace Management.

\subsubsection{Remote ID}
Remote ID, or Remote Identification, is a digital identifier that can be used to identify and broadcast specific information, such as UAV identification number that relevant authorities can access. 
In the US, FAA introduced Remote ID in December 2019~\cite{faa_remote_id_2021}. Other countries and aviation organizations have also developed or implemented their Remote ID standards and regulations~\cite{airsight_remote_id_2023}.
The development of Remote ID standards has also been supported by industry groups~\cite{astm_identification_tracking_2020,astm_f3411_22a}. 

Remote IDs operate similarly to automobile license plates. These digital license plates provide airspace awareness to regulating agencies, government officials, and law enforcement~\cite{faaremoteID} if it is broadcast along with identification info, such as location, registration number, and device details. 
Remote ID can be used to facilitate rules. 
For example, in the US, the FAA has recognized special areas called FAA recognized identification areas (FRIA) where UAVs are not required to broadcast their identification number~\cite{faaremoteID}. 
A Remote ID broadcast could have several elements along with ID numbers such as GCS longitude and latitude, GCS, UAV altitude, UAV longitude, and latitude, the emergency status of the UAV, time mark (UTC when the unmanned aircraft or control station was at a particular set of coordinates), and velocity \cite{faaremoteID}. 
Some of the advantages of using remote ID are:
\begin{enumerate}
\item{Enhanced safety:}
By providing real-time tracking and identification of UAVs, Remote ID can help prevent accidents caused by mid-air collisions or other unsafe behaviors. In addition, emergency responders can use Remote ID information to locate and respond to incidents involving UAVs quickly.
\item{Improved security:}
 Remote ID can help detect and prevent unauthorized or malicious UAV operations in sensitive areas, such as airports and other critical infrastructure sites. By providing authorities with the capabilities of real-time identification and tracking, Remote ID can help prevent criminal activities.
\item{Enhanced public trust:}
Remote ID can help address concerns about privacy and security related to UAV operations by providing a means for the public to identify UAVs in their vicinity. This can help reduce incidents of UAV induced personal privacy violations and enable the public to report suspicious or unsafe UAV activities.
\item{Expanded operational capabilities:}
By enabling the safe integration of UAVs into national airspace systems, Remote ID can open up new applications for UAVs, such as package delivery, inspection services, and emergency response. This can significantly benefit industries, such as logistics, agriculture, and public safety.
\end{enumerate}

The use of Remote ID can significantly improve the security of UAS systems. However, it can also increase the potential attack surface (security and privacy) as there is more information that is transmitted where there was earlier none. This duality warrants further investigation to identify critical gaps in research and understanding of increased attack vectors.

Some of the approaches to make the use of Remote ID more secure and private are: a) encrypt identification and location data, unlike ADS-B, which broadcasts in plaintext. Remote ID can be encrypted in contrast with the the open broadcasting issues of ADS-B, thus enhancing safety and security.  
b) Transmit only necessary, non-sensitive information for compliance, such as the UAV's ID, location, and altitude. 
c) Limited receiver range, or tiered access, which can be used to implement different levels of access. For example, the ATC may have access to comprehensive data, whereas other leisure UAV enthusiasts can only see limited information. 
d) If implemented in compliance with standardized privacy regulations and handled correctly, location privacy can be strengthened. 

\subsubsection{Integrated Airspace Management}
Integrated Airspace Management (IAM) represents a comprehensive methodology for regulating and synchronizing airspace utilization by diverse entities, including UAVs with commercial, military, and private aircraft~\cite{namuduri2023digital}. Within this framework, IAM can be tailored to facilitate a more granular segmentation of airspace, accommodating various classes of UAV operations, such as commercial, recreational, media, and entertainment. 
This necessitates communication and coordination between all airspace operators and management systems, as well as shared flight plans and status updates. Remote ID  is imperative to ensure UAVs operate safely and within the legal and regulatory framework.
Robust IAM systems are critical for safely integrating UASes into airspace alongside traditional manned aircraft.

\subsection{Lessons Learned}
This manuscript provides a comprehensive understanding of the prevailing UAS cybersecurity threats landscape and presents approaches that can be used to improve UAS security and privacy. These insights provide a foundational understanding of the current landscape, notable trends, successful practices, and areas requiring further exploration. To sum up, here we presents some lessons to refine the perspective on the subject matter and present future directions to the community.

Each phase of UAV operation is susceptible to vulnerabilities and requires meticulous attention to address them. Notably, the launch phase (refer Table~\ref{tab:Risk123}) of the UAS operation presents the most significant risks due to a large set of attacks. This requires research on providing the highest levels of security for this phase of the UAS operations. In what follows, we discuss our lessons learned in each component category. 
Also, ML is becoming a major force multiplier for cybersecurity. Although ML techniques can help fingerprint attacks, they are expensive and impractical for resource-constrained UAVs--particularly the new generative models. The challenge is to make ML pervasive in UAV/UAS cyberdefense while accounting for the limited computational power, battery capacity, and payload constraints. 

{\bf Hardware:}
Hardware vulnerabilities are detected by analyzing the behavior of hardware devices, such as sensors and processors and comparing them with established normal operational parameters. For instance, GPS devices are among the most vulnerable, with GPS spoofing being a prevalent and relatively easy method of attack.
Building effective detection mechanisms requires a large dataset of hardware behavior and a comprehensive knowledge base of the same for improved efficiency.
This large dataset may not be obtainable given the diversity of UAVs/UASs, with numerous heterogeneous settings and a range of varied scenarios.
We did not find any work that discusses/addresses these important concerns. 

In our study, we did not find any standards for hardware quality control, supply chain, and device disposal after the end of a UAV life cycle.
ISO 9000~\cite{iso9000} and ISO 9001~\cite{iso9001} are generic standards for quality management and quality assurance. These can be tuned for UASs to improve hardware resilience (design, architecture, manufacturing) by using quality management systems during design, production, and maintenance.  
Use these standards in the supply chain and addressing challenges in their implementation is pertinent.

{\bf Software:}
Most of the state-of-the-art is aimed at securing UAV software centered around preventing tampering. This involves strategies such as fortifying the software through shielding techniques~\cite{moein2017hardware}, employing cryptography to hinder tampering efforts~\cite{cho2020random,calnoor2020secure}, and implementing robust access control measures to restrict software accessibility~\cite{liu2017protc,yoon2017virtualdrone} exclusively to authorized personnel.

UAV software should be able to run in isolation to prevent adversaries from modifying the execution code from another process. 
Researchers have been utilizing container-based isolation~\cite{chen2019container} for UAV software due to its lightweight nature, as it doesn't require a full OS for each application. 
An alternative approach would be to explore virtual machines (VMs), which come with the advantage that VMs do not share the same OS kernel, unlike containers, hence can potentially offer better security.
Although not thoroughly explored, embedded system software protection techniques~\cite{gelbart2009compiler} can be applied to UAVs since both embedded system software and UAV software use dedicated computing hardware and software to execute real-time functions and tasks. Use of TEEs and TPMs falls under this category. 

Rule-based access control, anomaly detection, or reduced complexity ML models need to be developed for such low-capability UAVs. 

{\bf GCS:}
Our study of the existing literature reveals a notable absence of studies focused on protecting GCS. 
Robust environmental measures and enhanced disaster recovery strategies are necessary to protect the GCS and mitigate its vulnerability to natural disasters. GCS should have the same level of security as data centers or electric substations today (both physical and cyber).  
The GCS facilities should be regularly audited for physical and software security compliance. 

{\bf Network:}
Protecting the communication channel for UAV communication requires achieving the Confidentiality, Integrity, and Availability (CIA) triad. In the literature, confidentiality was achieved via encryption and secure communication protocols, such as SSL. Anomaly and intrusion detection techniques and tamper detection methods such as checksums were used to ensure integrity. Availability was achieved via redundancy and robust network infrastructure. 
One of the main concerns with UAVs is the vulnerability of their ADS-B system. The ADS-B signal is not encrypted by standard and doesn't use authentication mechanisms, which makes it vulnerable to unauthorized tracking and spoofing. UAVs are also exposed to vulnerabilities in sattelite communcations (if used), which can be exploited through various attack vectors, including jamming and interception.
The dependence on GPS signals for navigation creates vulnerabilities related to signal manipulation and DDoS attacks, which are not addressed.

{\bf Cloud:}
Non-UAV measures to protect data in the cloud are equally applicable in the UAV domain. Cryptographic measures have been deployed to protect access to cloud/remote servers and data encryption at third-party locations. While the cloud's distributed storage enhances reliability, it simultaneously expands the attack surface, and data in transit between these locations remains susceptible to interception. Additionally, since cloud providers are third parties, UAS operators typically lack the rights and capabilities to manage these security aspects.
When selecting a cloud provider, it is essential to have clearly articulated Service Level Agreements (SLAs).

\section*{Acknowledgment}
The FAA has sponsored this project through the Alliance for System Safety of UAS through Research Excellence (ASSURE), the Center of Excellence for Unmanned Aircraft Systems.  However, the agency neither endorses nor rejects the findings of this research. The research is also partly funded by the US NSF under grants \#2148358 and \#1914635. The presentation of this information is in the interest of invoking technical community comment on the results and conclusions of the research.

We are grateful to Casey Tran from New Mexico State University for his contribution to editing this manuscript. 

\bibliography{refs}

\appendix
{
\footnotesize 
\begin{longtblr}[
  caption = {Abbreviations},
  label = {tab:Abbreviations}
]{
  colspec = {|X[2,l]|X[6,l]|},
  row{1} = {font=\bfseries},
  rowhead = 1,
}
\hline
Acronym & Definition \\
\hline
AA & Attribute Authority \\
\hline
ABAC & Attribute Based Access Control \\
\hline
ADCA  & Adaptive Delegate Consensus Algorithm \\
\hline
ADS-B & Automatic Dependent Surveillance - Broadcast  \\
\hline
AP & Advanced Prioritize \\
\hline
ASLR & Address Space Layout Randomization \\
\hline
ATO &  Air Traffic Organization \\
\hline
CISA & Cybersecurity and Infrastructure Security Agency \\
\hline
COTS & Commercial-off-the-self \\
\hline
CPS & Cyber-physical System \\
\hline
C-SCRM & Cybersecurity Supply Chain Risk Management \\
\hline
CVE & Common Vulnerabilities and Exposure \\
\hline
CWE & Common Weakness Enumeration \\
\hline
DDoS & Distributed Denial of Service \\
\hline
DEP & Data Execution Prevention \\
\hline
DIAAT & Describe System, Identify Hazard, Analyze Risk, Assess Risk, and Treat Risk \\
\hline
DIDUCE & Dynamic Invariant Detection $\cup$ Checking Engine \\
\hline
DJI & Da-Jiang Innovations \\
\hline
DMA & Direct Memory Access \\
\hline
DRI & Drone Readiness Index \\
\hline
FAA & Federal Aviation Administration \\
\hline
FPGA & Field Programmable Gate Array \\
\hline
GCS & Ground Control Station \\
\hline
HADS & Hardware Anomaly Detection System \\
\hline
HAE & Homomorphic Authenticated Encryption \\
\hline
IAM & Integrated Airspace Management \\
\hline
ICAO & International Civil Aviation Organizations \\
\hline
ICN & Information Centric Networks \\
\hline
ICS & Industrial Control System \\
\hline
IDS & Intrusion Detection System \\
\hline
IKCB & Interest-key-content Binding \\
\hline
IoD & Internet of Drone \\
\hline
MFA  & Multi-factor Authentication \\
\hline
ML & Machine Learning \\
\hline
MVE-PCA & Minimum Volume Elliptical Principal Component Analysis \\
\hline
NDN & Named Data Networking \\
\hline
NIST & National Institute of Standards and Technology \\
\hline
NISTIR & NIST Interagency or Internal Reports \\
\hline
NVD & National Vulnerability Database \\
\hline
PiTM & Person-in-the-middle \\
\hline
PRE & Proxy Re-encryption \\
\hline
RBAC & Role Based Access Control \\
\hline
SADS & Software Anomaly Detection System \\
\hline
SATCOM & Satellite Communication \\
\hline
SEDA & Scalable Embedded Device Attestation \\
\hline
SGX & Software Guard Extension \\
\hline
SHR & Synchronization Header \\
\hline
SI & System Identification \\
\hline
SLA & Service Level Agreement \\
\hline
SMS & Safety Management System \\
\hline
SoC & System on Chip \\
\hline
SPAD  & Service Provider Attack Detection \\
\hline
SRM & Safety Risk Management \\
\hline
SSDF & Secure Software Development Framework \\
\hline
TDMA & Time Division Multiple Access \\
\hline
TEE & Trusted Execution Environments \\
\hline
TPM & Trusted Platform Module \\
\hline
TSAD  & Tenant Specific Attack Detection \\
\hline
UAANET & UAV ad-hoc networks \\
\hline
UAS  & Unmanned Aircraft System \\
\hline
UAV  & Unmanned Aircraft/Aerial Vehicle \\
\hline
WPA & Wi-Fi Protected Access \\
\hline
\end{longtblr}
}


 \begin{IEEEbiography}
 {Sharad Shrestha} (S'22) completed his Bachelor of Engineering at Pulchowk Campus, Nepal, in 2013, followed by a Master of Science in Computer Science from Central Michigan University, USA, in 2016. He is presently a Ph.D. student in computer science with a concentration on cybersecurity at New Mexico State University. His research interests encompass Smart Grid technologies, Unmanned Aerial Vehicles (UAVs), and 5G networks, focusing on enhancing their security and efficiency. \textbf{}
 \end{IEEEbiography}

\begin{IEEEbiography}
    {Mohammed Ababneh} received his B.S. and M.S. degrees in computer science from AL al-bayt University and Yarmouk University in Jordan, respectively. He is currently pursuing his Ph.D. in computer science at New Mexico State University. His areas of focus in research include; layer-2 blockchain protocols, payment channel networks, and cryptographic protocol design.
\end{IEEEbiography}

 \begin{IEEEbiography}
 {Satyajayant Misra}
 (S'05, M'09) is a professor in computer science and electrical and computer engineering at New Mexico State University. He completed his Ph.D. in Computer Science from ASU, AZ, USA, in 2009. His research interests are in wireless networks, the Internet, and smart grid architectures and protocols. He serves on several IEEE/ACM journal editorial boards and conference executive committees including editorship in IEEE Transactions on Mobile Computing and IEEE Transactions on Network Science and Engineering. He has authored over 100 peer-reviewed publications. 
 \end{IEEEbiography}

 \begin{IEEEbiography}
     {Roopa Vishwanathan} is an Assistant Professor in the Dept. of Computer Science at New Mexico State University, Las Cruces, NM. Her current areas of research are applied cryptography, security of blockchain-based applications, and scalability of blockchains (Layer 2 protocols). Her research group's results in this area include the design of pathfinding, routing, and rebalancing protocols for payment channel networks. The other area she is interested in is cryptographic protocol design. Past work in this area includes design of secure revocable chameleon hash schemes, design of revocable attribute-based encryption and signature schemes, and design of redactable or rewritable blockchains.
 \end{IEEEbiography}

\begin{IEEEbiography}
    {Henry M. Cathey, Jr.} is the Aerospace Division Director at the New Mexico State University’s, Physical Science Laboratory and serves as the Director of the FAA approved New Mexico UAS Flight Test Site.  Responsibilities include support for scientific ballooning, Unmanned Aircraft Systems, and other suborbital efforts. His core concentrations focus on research and new technology implementation. He leads research efforts focusing on sUAS DAA requirements, BVLOS, Cybersecurity, Counter UAS, and led FAA STEM minority outreach education using UAV’s as the central learning platform. In July 2015 he was awarded a NASA Exceptional Public Service Medal for his contributions to NASA’s mission.
\end{IEEEbiography}

\begin{IEEEbiography}
    {Matthew Jansen} is a recent graduate from Oregon State University's Computer Science program in Corvallis, OR, USA. His research interests are in host-based intrusion detection, with a focus in graph-based endpoint auditing, detection, and attack investigation. He received his Honors B.S. and M.S. in Computer Science from Oregon State University in 2020 and 2023, respectively. 
\end{IEEEbiography}

\begin{IEEEbiography}
    {Jinhong Choi} is currently pursuing the M.S. degree in computer science at Oregon State University, located in Corvallis, OR, USA. His research interest lies in cyber-physical security for drones, focusing on implications of manipulated data inputs, from sensors and incoming messages. He is also interested in assessing algorithms for drone swarm, utilizing a nano drone platform.
\end{IEEEbiography}

\begin{IEEEbiography}
    {Rakesh B. Bobba} (M'06) is an Associate Professor in the School of Electrical Engineering and Computer Science, Oregon State University, Corvallis, OR, USA. He is an Dr. Bobba's research interests are in the design of secure and trustworthy networked and distributed computer systems, with a current focus on cyber-physical critical infrastructures, real-time systems, trustworthy machine learning, and privacy. He received his B.E.(Hons.) in Electrical and Electronics Engineering from Birla Institute of Technology and Science, Pilani, India in 2000; and  M.Sc. and Ph.D. degrees in Electrical and Computer Engineering from University of Maryland at College Park in 2007 and 2009, respectively.
\end{IEEEbiography}

\begin{IEEEbiography}
    {Yeongjin Jang}  is a Principal Engineer, Software Vulnerability Research at Samsung Research America.  Dr. Jang’s research interests are in trustworthy computing, vulnerability discovery and analysis, side-channel attack and defense, developing new exploit primitives, mobile security, practical applied cryptography, jailbreaking, and building defense mechanisms. He received his B.S. Computer Science from KAIST in 2010; and M.S. and Ph.D. degrees in Computer Science from Georgia Tech. in 2016 and 2017, respectively.
\end{IEEEbiography}
 
\end{document}